\documentclass{aa}  
\usepackage{graphicx}
\usepackage{txfonts}
\usepackage{natbib}
\bibpunct{(}{)}{;}{a}{}{,}

\begin{document}
   \title{Comparison of synthetic maps from truncated jet-formation 
          models with YSO jet observations}

   \author{Matthias Stute\inst{1,2}
	  \and
	  Jos\'e Gracia\inst{3,4,5}
	  \and
	  Kanaris Tsinganos\inst{2}
	  \and
	  Nektarios Vlahakis\inst{2}
          }

   \offprints{Matthias Stute, \\ \email{matthias.stute@ph.unito.it}}

   \institute{
     Dipartimento di Fisica Generale "A. Avogadro", Universit\`a degli Studi di 
     Torino, Via Pietro Giuria 1, 10125 Torino, Italy
     \and
     IASA and Section of Astrophysics, Astronomy and Mechanics, 
     Department of Physics, University of Athens, 
     Panepistimiopolis, 157 84 Zografos, Athens, Greece
     \and
     High Performance Computing Center Stuttgart (HLRS), Universit\"at 
     Stuttgart, 70550 Stuttgart, Germany
     \and
     Max-Planck-Institut f\"ur Kernphysik, Postfach 10 39 80, 69029 Heidelberg,
     Germany
     \and
     School of Cosmic Physics, Dublin Institute of Advanced Studies, 
     31 Fitzwilliam Place, Dublin 2, Ireland
   }

   \date{Received 21 October 2008; accepted 16 April 2010}

  \abstract
   {Significant progress has been made in the last years in the understanding 
    of the jet formation mechanism through a combination of numerical 
    simulations and analytical MHD models for outflows characterized by the 
    symmetry of self-similarity. Analytical radially self-similar models 
    successfully describe disk-winds, but need several improvements. In a 
    previous article we introduced models of truncated jets from disks, i.e. 
    evolved in time numerical simulations based on a radially self-similar MHD 
    solution, but including the effects of a finite radius of the jet-emitting 
    disk and thus the outflow.}
   {These models need now to be compared with available observational data. A 
    direct comparison of the results of combined analytical theoretical models 
    and numerical simulations with observations has not been performed as yet. 
    This is our main goal.}
   {In order to compare our models with observed jet widths inferred from
    recent optical images taken with the Hubble Space Telescope (HST) and 
    ground-based adaptive optics (AO) observations, we use a new set of tools 
    to create emission maps in different forbidden lines, from which we 
    determine the jet width as the full-width half-maximum of the emission.}
   {It is shown that the untruncated analytical disk outflow solution considered
    here cannot fit the small jet widths inferred by observations of several 
    jets. Furthermore, various truncated disk-wind models are examined, whose 
    extracted jet widths range from higher to lower values compared to the 
    observations. Thus, we can fit the observed range of jet widths by tuning 
    our models.}
   {We conclude that truncation is necessary to reproduce the observed jet 
    widths and our simulations limit the possible range of truncation radii. We 
    infer that the truncation radius, which is the radius on the disk mid-plane 
    where the jet-emitting disk switches to a standard disk, must be between 
    around 0.1 up to about 1 AU in the observed sample for the considered 
    disk-wind solution. One disk-wind simulation with an {{\em inner}} truncation 
    radius at about 0.11 AU also shows potential for reproducing the 
    observations, but a parameter study is needed.}

   \keywords{MHD --- methods: numerical --- ISM: jets and outflows --- Stars: 
    pre-main sequence, formation}

   \maketitle

\section{Introduction}

Astrophysical jets and disks \citep{Liv09} seem to be inter-related, notably in 
young stellar objects (YSOs), where jet signatures are well 
correlated with the infrared excess and accretion rate of the circumstellar 
disk \citep{CES90, HEP04}. Disks provide the plasma which is outflowing in the 
jets, while jets in turn provide the disk with the needed angular momentum 
removal so that accretion onto the protostellar object takes place 
\citep{Har09}. On the theoretical front, the most widely accepted description 
of this accretion-ejection phenomenon \citep{Fer07} is based on the interaction 
of a large scale magnetic field with an accretion disk around the central 
object. Then, plasma is channeled and magneto-centrifugally accelerated along 
the open magnetic field lines threading the accretion disk, as first described
in \citet{BlP82}. Several works have extended this study either by semi-analytic
models using radially self-similar solutions of the full magnetohydrodynamics 
(MHD) equations with the disk treated as a boundary condition \citep{VlT98}, 
by self-consistently treating the disk-jet system semi-analytically 
\citep[e.g.][]{Fer97,CaF00}, or, by self-consistently treating numerically the 
disk-jet system \citep[e.g.][]{ZFR07, TFM09}.

The original \citet{BlP82} model, however, has serious limitations for a needed 
meaningful comparison of its predictions with observations. First, 
singularities exist at the jet axis, the outflow is not asymptotically 
super-fast, and most importantly, an intrinsic scale in the disk is lacking with
the result that the jet formally extends to radial infinity, to mention just a 
few. First, the singularity at the axis can be easily taken care of by 
numerical simulations extending the analytical solutions close to this symmetry 
axis \citep[][GVT06 hereafter]{GVT06}. Next, the outflow speed at large 
distances may be tuned to cross the corresponding limiting characteristic, with 
the result that the terminal wind solution is causally disconnected from the 
disk and hence perturbations downstream of the super-fast transition (as 
modified by self-similarity) cannot affect the whole structure of the steady
disk-wind outflow \citep[][V00 hereafter]{VTS00}, a state which has also been 
shown to be structurally stable \citep[][M08 hereafter]{MTV08}. The next step of
introducing a scale in the disk has been done in a previous paper 
\citep[][paper I hereafter]{STV08}, wherein we presented numerical simulations 
of truncated flows whose initial conditions are based on analytical 
self-similar models.

\begin{table*}[!htb]
\caption{List of numerical science models}
\label{tbl_models}
\centering
\begin{tabular}{l l l l l}
\hline\hline
Name & Domain $[ R_0$ x $R_0]$ & Resolution & Description & 
$R_{\rm trunc}$ $[ R_0 ]$  \\
\hline
model SC1a & [0,50] $\times$ [6,100]  & 200 $\times$ 400 & 
$\alpha_{\rm trunc} = 0.4$, external analytical solution $\lambda_1 = 10^3$,
$\lambda_2 = 10^{-3}$ & 5.375 \\
model SC1b & [0,50] $\times$ [6,100]  & 200 $\times$ 400 & 
$\alpha_{\rm trunc} = 0.2$, external analytical solution $\lambda_1 = 10^3$,
$\lambda_2 = 10^{-3}$ & 5.125 \\
model SC1c & [0,50] $\times$ [6,100]  & 200 $\times$ 400 & 
$\alpha_{\rm trunc} = 0.1$, external analytical solution $\lambda_1 = 10^3$,
$\lambda_2 = 10^{-3}$ & 4.875 \\
model SC1d & [0,50] $\times$ [6,100]  & 200 $\times$ 400 & 
$\alpha_{\rm trunc} = 0.01$, external analytical solution $\lambda_1 = 10^3$,
$\lambda_2 = 10^{-3}$ & 3.625 \\
model SC1e & [0,50] $\times$ [6,100]  & 200 $\times$ 400 & 
$\alpha_{\rm trunc} = 0.001$, external analytical solution $\lambda_1 = 10^3$,
$\lambda_2 = 10^{-3}$ & 2.625 \\
model SC2     & [0,50] $\times$ [6,100]  & 200 $\times$ 400 & 
$\alpha_{\rm trunc} = 0.4$, external analytical solution $\lambda_1 = 100$, 
$\lambda_2 = 0.1$ & 5.375 \\
model SC3     & [0,50] $\times$ [6,100]  & 200 $\times$ 400 &  
same as model SC2, but solutions are swapped & 5.375 \\
model SC4     & [0,50] $\times$ [6,100]  & 200 $\times$ 400 &  
$\alpha_{\rm trunc} = 0.4$, external analytical solution $\lambda_1 = 1$, 
$\lambda_2 = 0.1$ & 5.375 \\
model SC5     & [0,50] $\times$ [6,100]  & 200 $\times$ 400 & 
same as model SC4, but solutions are swapped & 5.375 \\
\hline
\hline
model SC1f & [0,50] $\times$ [6,100]  & 200 $\times$ 400 & 
$\alpha_{\rm trunc} = 0.0005$, external analytical solution $\lambda_1 = 10^3$,
$\lambda_2 = 10^{-3}$ & 2.375 \\
model SC1g & [0,10] $\times$ [6,20]  & 200 $\times$ 400 & 
$\alpha_{\rm trunc} = 1\times10^{-5}$, external analytical solution 
$\lambda_1 = 10^3$, $\lambda_2 = 10^{-3}$ & 0.575 \\
\hline
\end{tabular}
\end{table*}

In order to test our truncated models, we will now apply our simulations to 
observations. In recent years, many NIR and optical data have become available 
exploring the morphology and kinematics of the jet launching region 
\citep[e.g.][and references therein]{DCL00,RDB07,Dou08}. Hubble Space Telescope 
(HST) and adaptive optics (AO) observations give access to the innermost 
regions of the wind, where the acceleration and collimation occurs 
\citep{RMD96,DCL00,WRB02,HEP04}. Because YSO jets emit in a number of atomic 
(and molecular) lines, we used a set of tools described in \citet{G??10} to 
create emission maps in different forbidden lines which were used by other 
authors to extract the jet width from images. The observed jet widths will be 
compared with those extracted from our synthetic images. A similar study has 
been done by \citet{CFR99}, \citet{GCF01} and \citet{DCF04} using a different 
set of semi-analytical self-similar disk-jet solutions from \citet{Fer97} and 
\citet{CaF00}. Observed jet widths could be reproduced by manually truncating 
the solutions inside 0.07 AU and outside 1 AU, but the modification of 
flow streamlines induced by truncation was ignored.

The remainder of the paper is organized as follows: we briefly review the 
initial setup of the numerical simulations in Sect. \ref{sec_num_models} and
describe our procedure for the comparison with observations in Sect. 
\ref{sec_obs}. The results of our studies are presented in Sects. 
\ref{sec_untrunc} - \ref{sec_inner}. Finally, we conclude with the 
implications of the results in terms of the structure of the disk and the 
respective launching radii of the jets in YSOs. 

\section{Initial model setup and numerical simulations} \label{sec_num_models}

This work is based on the results of our numerical simulations discussed in our 
paper I and two new models. We solved the MHD equations with the PLUTO 
code\footnote{http://plutocode.to.astro.it/} \citep{MBM07} starting from an
initial condition set according to a steady, radially self-similar solution 
as described in V00, hereafter labelled ADO ({\em analytical disk outflow 
solution}, as in M08), which crosses all three critical surfaces. At the 
symmetry axis, the analytical solution was modified as described in GVT06 and 
M08.

To study the influence of the truncation of the analytical solution, we divided
our computational domain into a jet region and an external region, separated by 
a truncation field line $\alpha_{\rm trunc}$. For lower values of the normalized 
magnetic flux function, i.e. $\alpha < \alpha_{\rm trunc}$ -- or conversely 
smaller cylindrical radii -- our initial conditions are fully determined by the 
solution of V00 and the modification of GVT06 and M08 close to the axis. In the 
outer region we modified all quantities and initialized them with another 
analytical solution, but with modified parameters. From V00, one can show that 
if we start with an arbitrary MHD solution with the variables $\rho$, $p$, 
$\vec v$, $\vec B$, one can easily construct a second solution by using 
two free parameters\footnote{Strictly speaking this is only true when all 
lengths are also scaled, but because only the gravity term explicitly depends
on the length scale and in our case the gravitational force is small compared 
to the other forces, this slight inconsistency is unimportant (see paper I).}
$\lambda_1$ and $\lambda_2$, with $\rho\,' = \lambda_1\,\lambda_2^2\,\rho$, 
$p\,' = \lambda_2^2\,p$, $\vec B\,' = \lambda_2\,\vec B$ and $\vec v\,' = 
\lambda_1^{-1/2}\,\vec v$. Thus some or all quantities are scaled down in the 
external region depending on our choice of parameters. In Table \ref{tbl_models}
we give the parameters of the models (some of which we studied in paper I as 
well as new ones) used in this study.

For further details, we refer the reader to paper I.

\section{Comparison of synthetic emission runs with observations} 
\label{sec_obs}

Numerical simulations and observations cannot be directly compared. While the 
former describe the plasma in terms of physical quantities like density, 
pressure, magnetic field and velocities, the latter observes only photon flux 
as a function of frequency. This comparison can be facilitated by means of 
constructing synthetic maps. However, translating numerical simulations into 
such synthetic maps is very complicated. In general, the local emissivity is 
a function of temperature, electron density and density of the respective ion, 
as e.g., $n_{\rm OII}$ for singly ionized oxygen. The emissivities are 
integrated along a given line-of-sight and projected onto the plane of the sky 
producing an ideal synthetic emission map. Finally, real detectors distort this 
ideal map through their detector response, which needs to be taken into account.

In this paper, we use the set of tools 
OpenSESAMe\footnote{http://homepages.dias.ie/\~\/jgracia/OpenSESAMe/} v0.1 
described in \citet{G??10} to produce 
synthetic observations from our simulations in different consecutive stages. 
The first stage approximates the ionization state of each atom by locally 
solving a chemical network under the assumption of local equilibrium. The second
stage calculates the statistical equilibrium of level populations for each ion 
of interest as a function of temperature and density and yields the emissivity 
for individual transitions of interest. Further stages take care of integration 
along the line-of-sight and projection. Finally, the ideal maps are convolved
with a Gaussian point-spread-function (PSF) to mimic the finite spatial 
resolution of a given instrument. These synthetic emission maps are then
quantitatively analyzed with similar techniques used on real observed maps.
We refer to ``runs'' as runs of OpenSESAMe with different sets of scalings (see 
below) and to ``models'' as different MHD simulations with PLUTO (Table 
\ref{tbl_models}).

\begin{figure}[!htb]
  \centering
  \includegraphics[width=\columnwidth]{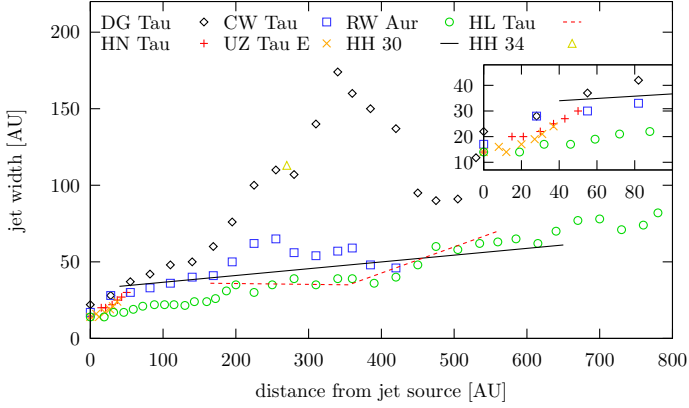}
  \caption{Variation of jet width (FWHM) derived from [SII] and [OI] images as 
    a function of distance from the source. Data points are from CFHT/PUEO and 
    HST/STIS observations of DG Tau (diamonds), HN Tau (plus signs), CW Tau 
    (squares), UZ Tau E (crosses), RW Aur (circles), HH 34 (one triangle), 
    HH 30 (black solid line) and HL Tau (red dashed line); data are taken from
    \citet{RDB07} and references therein.}
  \label{Fig_observations}
\end{figure}

We compare the width of jets measured from HST and AO observations 
\citep[][and references therein]{DCL00,RDB07,Dou08} with the width from 
synthetic emission maps calculated from our MHD models. We convolved the maps 
with a Gaussian with a full-width half-maximum (FWHM) of 15 AU ($\sigma = 6.37$ 
AU) throughout this paper, equivalent to HST's resolution of 0.1'' at a 
distance of 150 pc. We use a sample of eight jets: DG Tau, HN Tau, CW Tau, UZ 
Tau E, RW Aur, HH34, HH30 and HL Tau (Fig. \ref{Fig_observations}). In order to 
determine the width of the jets in our models, we use a method which is as 
close as possible to that applied by the observers. We create convolved 
synthetic maps for the emission in the [SII] $\lambda$6731 and [OI] 
$\lambda$6300 lines for each numerical model and each run of OpenSESAMe and 
determine the jet width from the map's FWHM as a function of distance along the 
axis. 

\subsection{Normalizations} \label{sec_norm}

Throughout the first paper, we used only the dimensionless quantities in which 
PLUTO performs its calculations. In order to compare our results with 
observations, however, i.e. in order to run OpenSESAMe correctly, we have to 
scale them to physical units by providing scaling factors for density $\rho_0$,
pressure $p_0$, velocity $v_0$, magnetic field strength $B_0$, a length scale 
$R_0$ and a mass scale $\mathcal{M}$. However, in terms of the normalizations 
used in the PLUTO code, only three of those quantities are independent. A 
possible choice is the mass of the central object, velocity scale and density
scale, while the remaining factors are calculated from these.

Here we use three different ``coordinate systems'': i) the 
computational grid of cells with indices ($i$,$j$) from (0,0) to (199,399), ii) 
the PLUTO domain from (0,6) to (50,100) and iii) the physical scale of the jet 
in AU, which is simply the PLUTO domain multiplied with a length scale $R_0$. 
In the solution of V00, the length scale $R_0$ is connected to the mass of the 
central object and the velocity normalization via
\begin{equation}
\label{length_scale}
R_0 = \frac{\mathcal{G}\,\mathcal{M}}{4\,\,v_0^2} = 
110.9\,\textrm{AU}\,\left(\frac{v_0}{
\textrm{km s$^{-1}$}}\right)^{-2}\,\left(\frac{\mathcal{M}}{
0.5\,\textrm{M}_{\odot}}\right) \,.
\end{equation}

From the velocity and density normalization directly follow the 
normalizations for the magnetic field and pressure as
\begin{eqnarray}
\label{pressure_scale}
p_0 &=& \rho_0\,v_0^2  = 10^{-11}\,\textrm{g cm$^{-1}$ s$^{-2}$}\,
\left(\frac{\rho_0}{10^{-21}\,\textrm{g cm$^{-3}$}}\right) \,
\left(\frac{v_0}{\textrm{km s$^{-1}$}}\right)^2 \,, \\
\label{mag_scale}
B_0 &=& \sqrt{4\,\pi\,\rho_0\,v_0^2} = 11.21\,\textrm{$\mu$G}\,
\left(\frac{\rho_0}{10^{-21}\,\textrm{g cm$^{-3}$}}\right)^{1/2} \,
\left(\frac{v_0}{\textrm{km s$^{-1}$}}\right) \,.
\end{eqnarray}

The mass of the central object affects only the length scale. The pressure and 
temperature of the jet and thus the synthetic emission maps are affected only 
by the unit density and unit velocity. 

As typical jet velocities in YSOs we assumed values of 100, 300, 600 and 1000 
km s$^{-1}$, as typical masses of T Tauri stars 0.2, 0.5 and 0.8 M$_{\odot}$ 
\citep{HEG95}, and as jet number densities values of 125, 500, 1000 and 
$5\times10^4$ cm$^{-3}$. We adopt the nomenclature for our runs as e.g. 
($n_{\rm jet}$, $v_{\rm jet}$, $M$) with $n_{\rm jet}$ in cm$^{-3}$, $v_{\rm jet}$ 
in km s$^{-1}$ and $M$ in M$_{\odot}$. 

In order to find the normalizations listed above, we needed to have typical 
values of jet density and velocity at a certain position ($R$, $z$) for a given 
mass $M$, and iteratively solved the Eq. (\ref{length_scale}). Details of this 
algorithm and also a graphical picture of this approach are given in Appendix 
\ref{app_norm}. The normalizations vary from numerical model to numerical model,
therefore we used the corresponding different values for each model. 

As another constraint, we require that $R_0$ is small enough that the 
FWHM of the Gaussian of 15 AU is sampled by a reasonable number of pixels. 
Because the resolution of our numerical simulations was four pixels per $R_0$, 
requiring at least two pixels per FWHM requires $R_0 < 30$ AU. This limit 
highly reduces the number of runs, from 576 to 180. The only valid runs are 
\begin{itemize}
\item ($\cdots$,600,0.2), ($\cdots$,600,0.5), ($\cdots$,1000,0.5), 
($\cdots$,1000,0.8) for models ADO, SC1a--b, SC2, SC4
\item ($\cdots$,600,0.2), ($\cdots$,600,0.5), ($\cdots$,1000,0.2), 
($\cdots$,1000,0.5), ($\cdots$,1000,0.8) for models SC1c--f
\item ($\cdots$,600,0.2), ($\cdots$,1000,0.2), ($\cdots$,1000,0.5), 
($\cdots$,1000,0.8) for model SC1g
\item ($\cdots$,100,0.2) for model SC3
\item no valid runs for model SC5.
\end{itemize}

The corresponding values of $R_0$ are listed in Table \ref{tbl_R0}.

\subsection{The artefact of limb-brightening} \label{sec_corrections}

\begin{figure}[!htb]
  \centering 
  \includegraphics[width=\columnwidth]{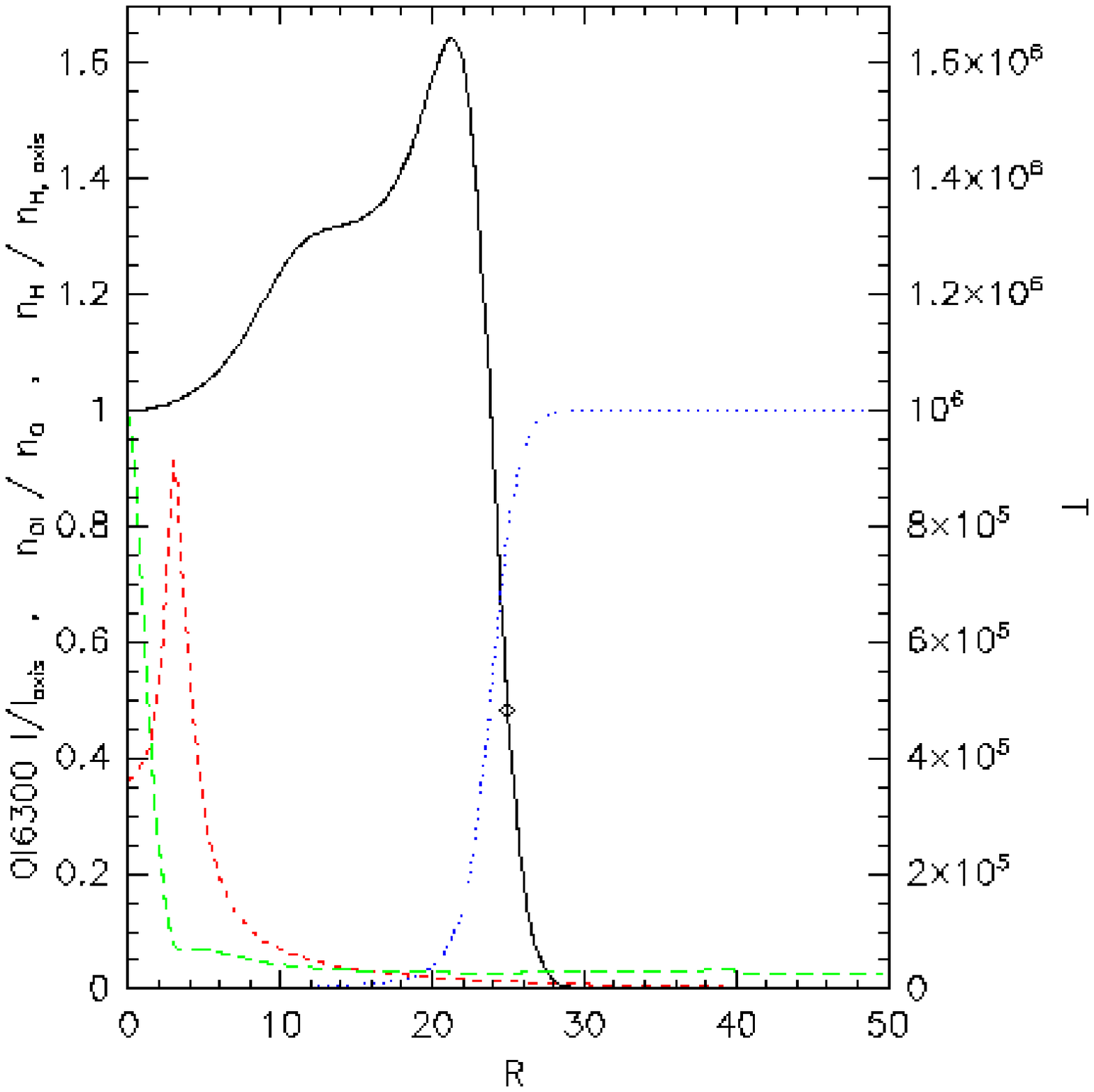}
  \includegraphics[width=\columnwidth]{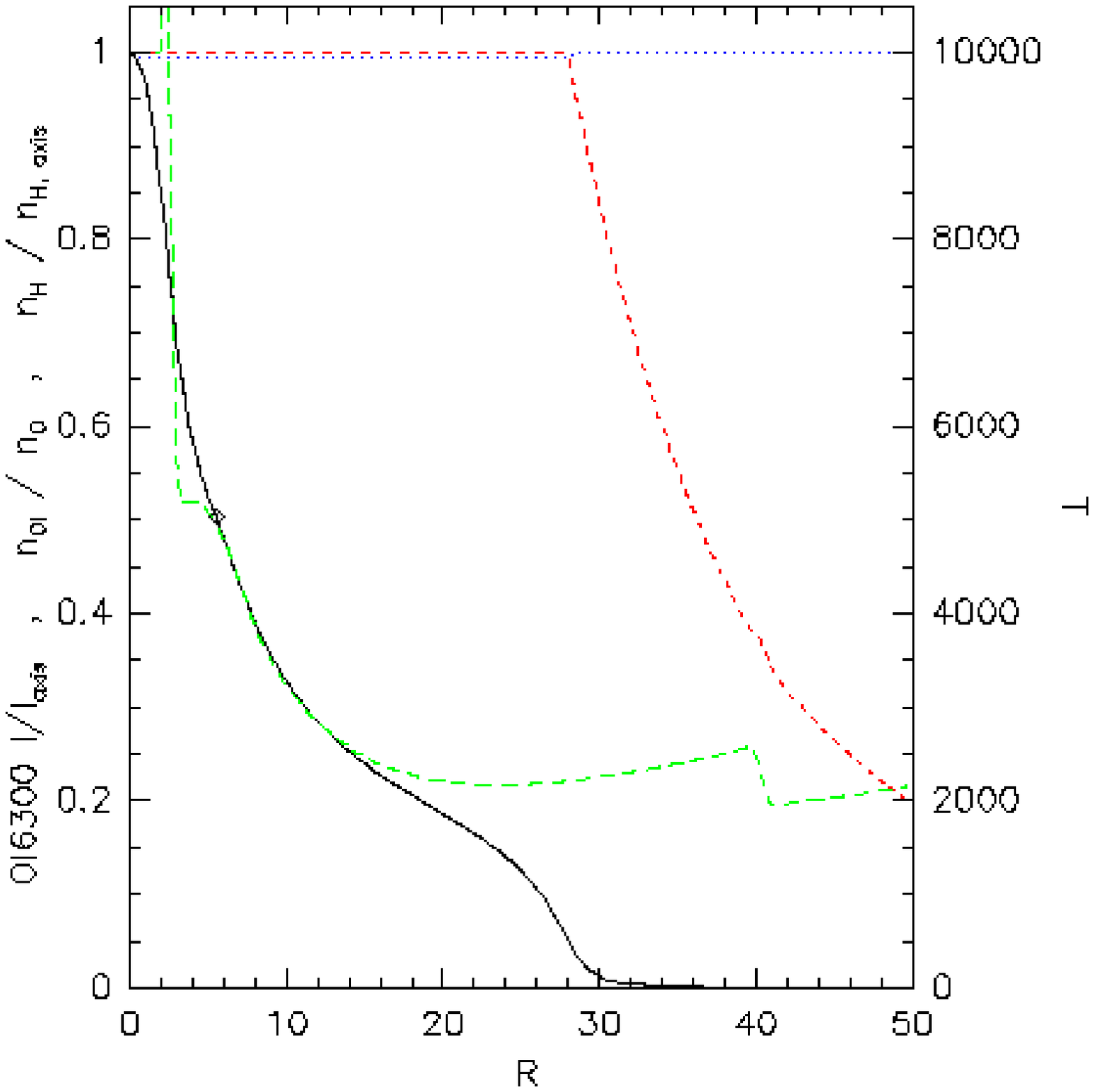}
  \caption{Transverse intensity cuts of the convolved [OI] $\lambda$6300 
    emission at $z = 100$ AU (black, solid), normalized to its value on the 
    axis, for model ADO and run (500,600,0.5); without corrections of density 
    and temperature they show the effect of limb-brightening (top) which 
    vanishes after the corrections (bottom); also plotted are cuts of the 
    normalized hydrogen number density (long-dashed, green), of 
    $n_{OI} / n_{O}$ (blue, dotted) and of the temperature (dashed, red, y-axis
    on the right-hand side). $R_0 = 7.60$ AU.}
  \label{Fig_limb_brightening}
\end{figure}

After plotting transverse intensity cuts through the convolved emission maps, 
it can be seen that the maximum emission does not come from the axis, but from 
a small shell at a finite radius (Fig. \ref{Fig_limb_brightening}, left). This 
effect of limb-brightening, which has been never observed up to now in real 
protostellar jets, is a direct consequence of the density and 
pressure/temperature profiles in the analytical solution and is present in all
our models and runs, the untruncated model ADO as well as our truncated models. 
Details of the physical reasons for this behavior are given in Appendix 
\ref{app_limb}.

Because the analytical solution is not well defined close to the axis and we had
to interpolate it in our numerical simulations, and because it is likely that a
stellar wind resides inside the disk wind \citep{MMT09}, we applied the 
following corrections before running OpenSESAMe:

\begin{itemize}
\item we {\em limited} the temperature to $10^4$ K in the whole domain
\item we limited the density around the axis by setting the density inside 
1 FWHM to its value at 1 FWHM for each $z$.
\end{itemize}

This value of $T$ is typical of that deduced from analyses of line ratios in 
jets \citep[e.g.][]{BE99, LCD00}. The artefact of limb-brightening is removed by
limiting $T$. For numerical reasons, the density shows a steep increase close 
to the jet axis. This artefact has to be corrected to obtain jetwidths above 1 
FWHM.

\section{Do we have to truncate the disk?} \label{sec_untrunc}

\begin{figure*}[!htb]
  \centering 
  \includegraphics[width=0.245\textwidth]{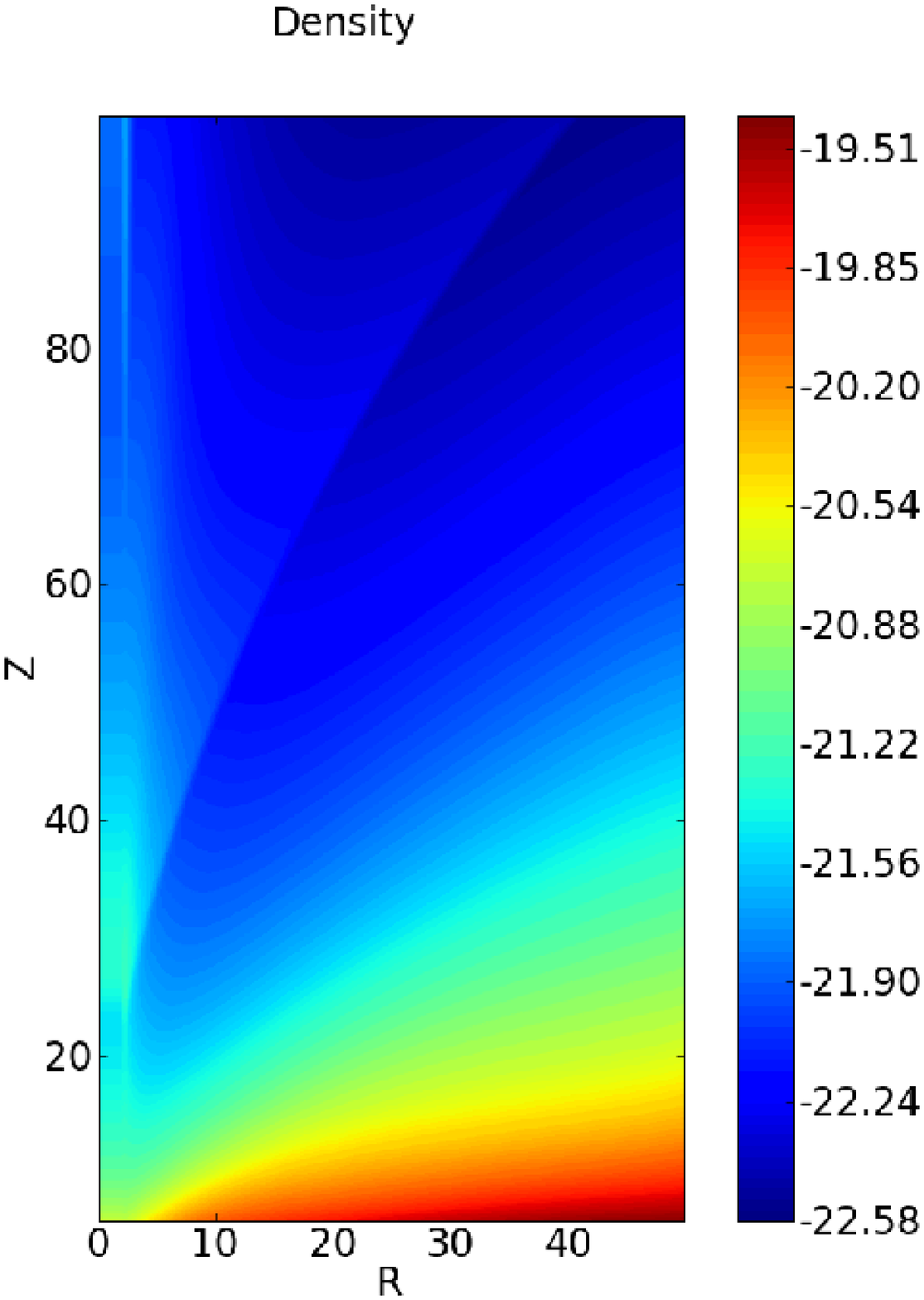}
  \includegraphics[width=0.245\textwidth]{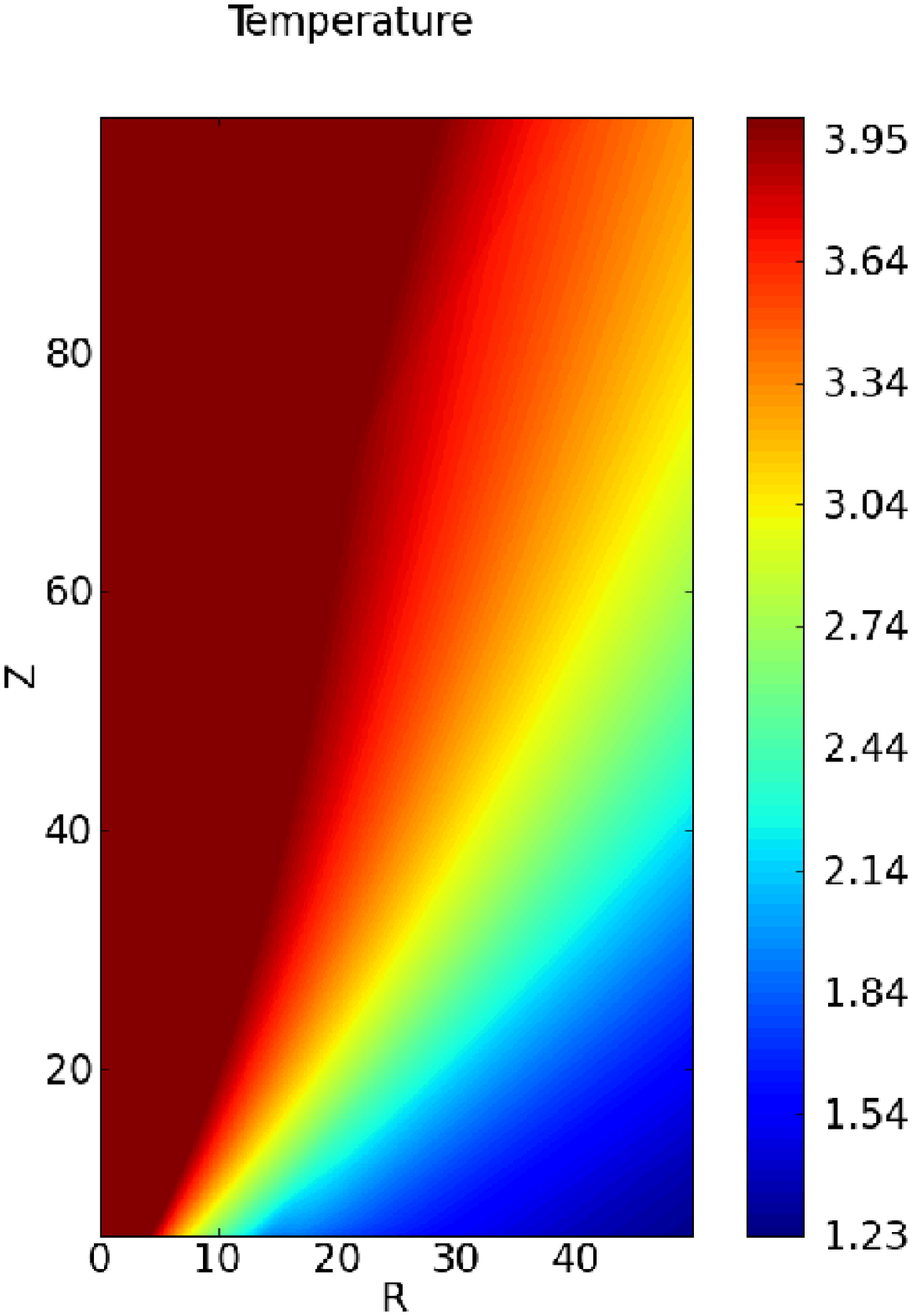} 
  \includegraphics[width=0.245\textwidth]{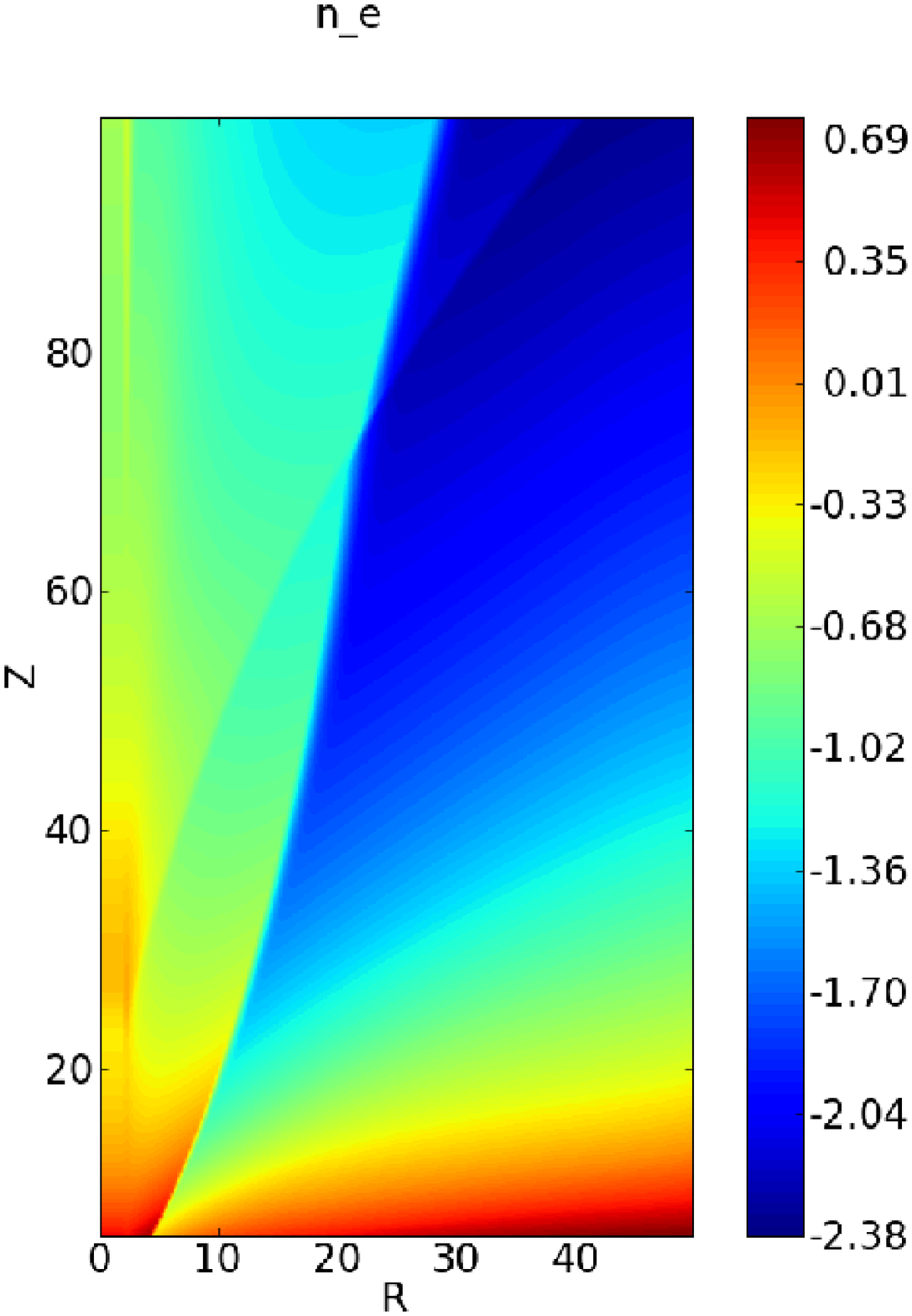}
  \includegraphics[width=0.245\textwidth]{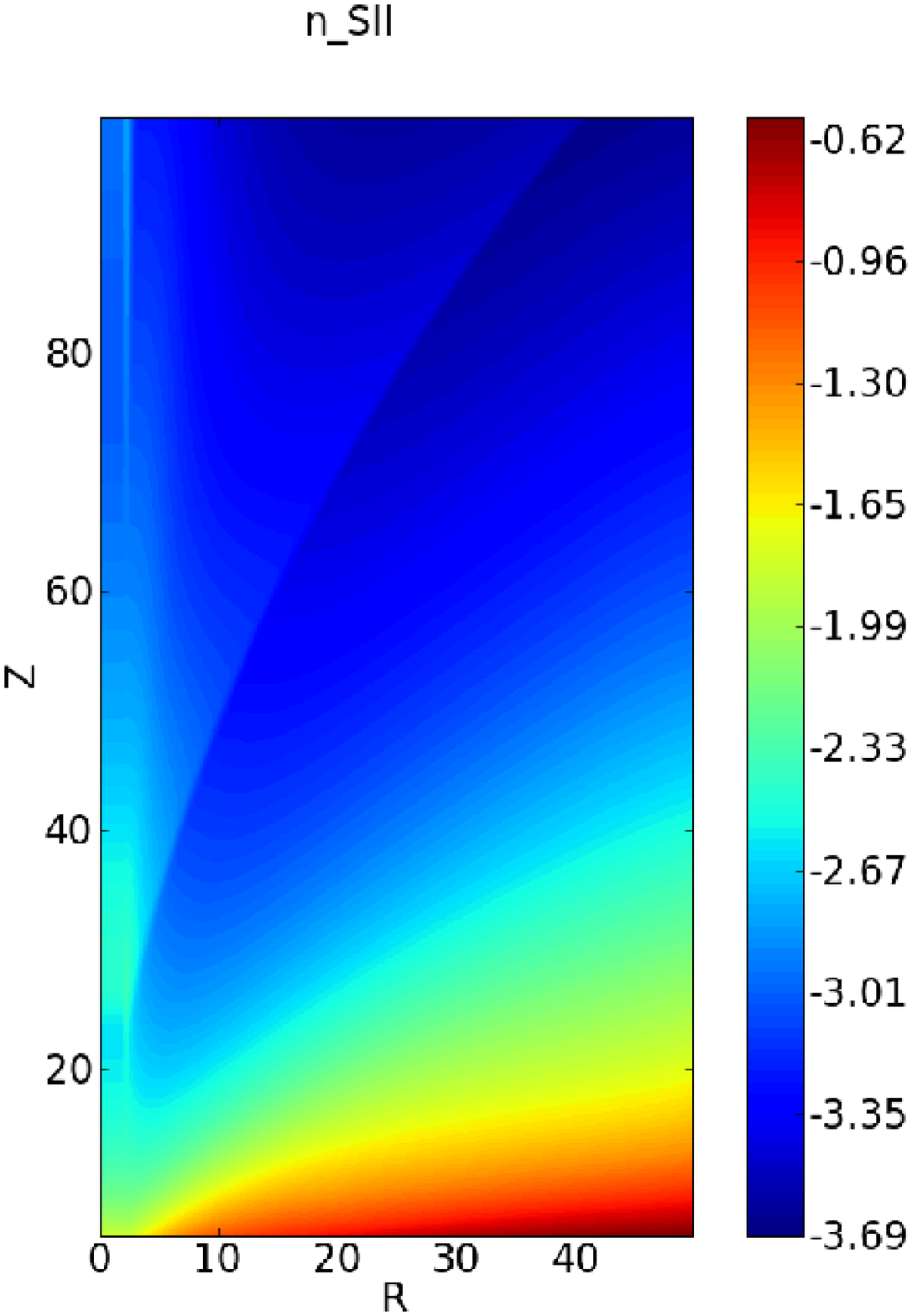}
  \includegraphics[width=0.245\textwidth]{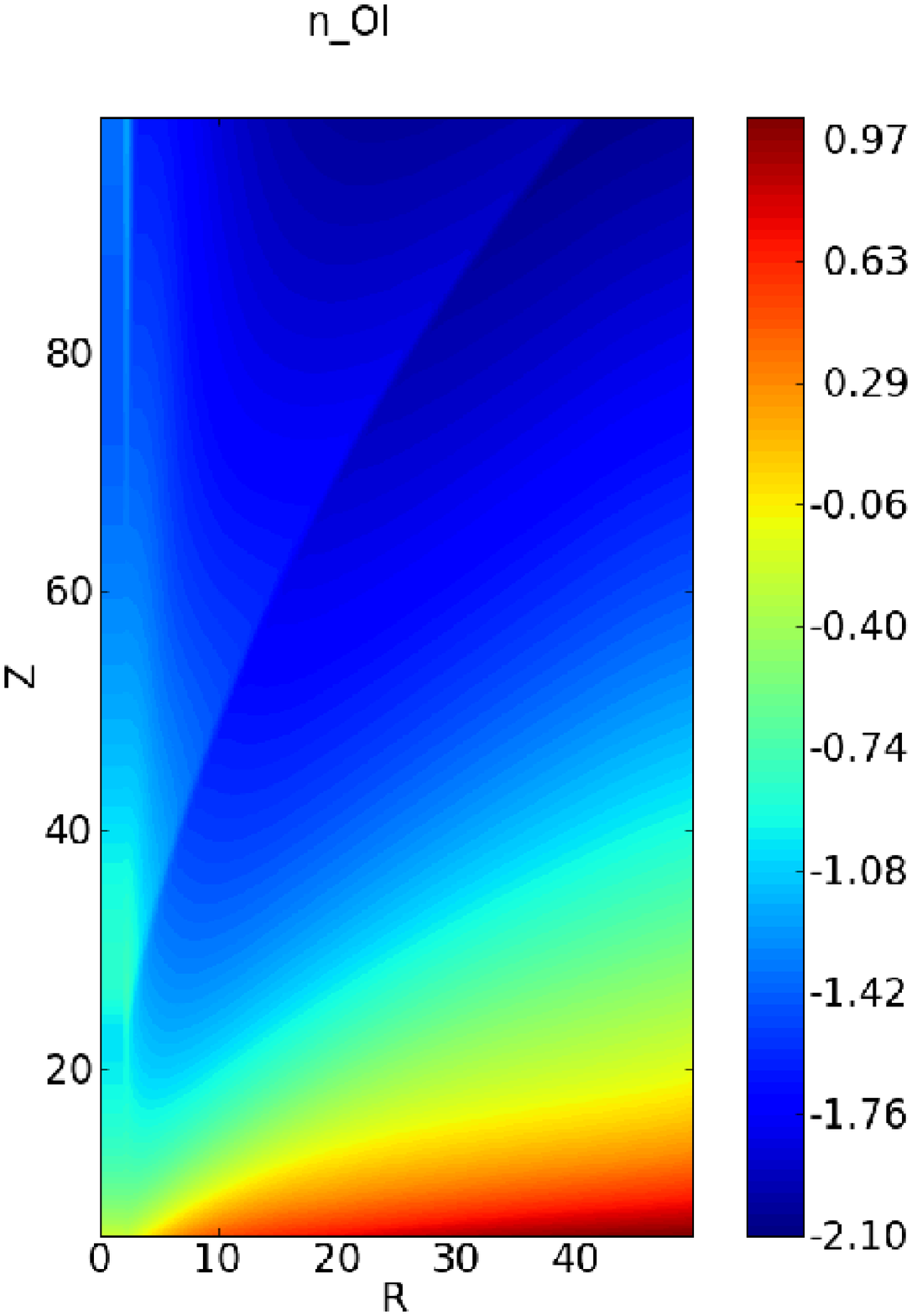} 
  \includegraphics[width=0.245\textwidth]{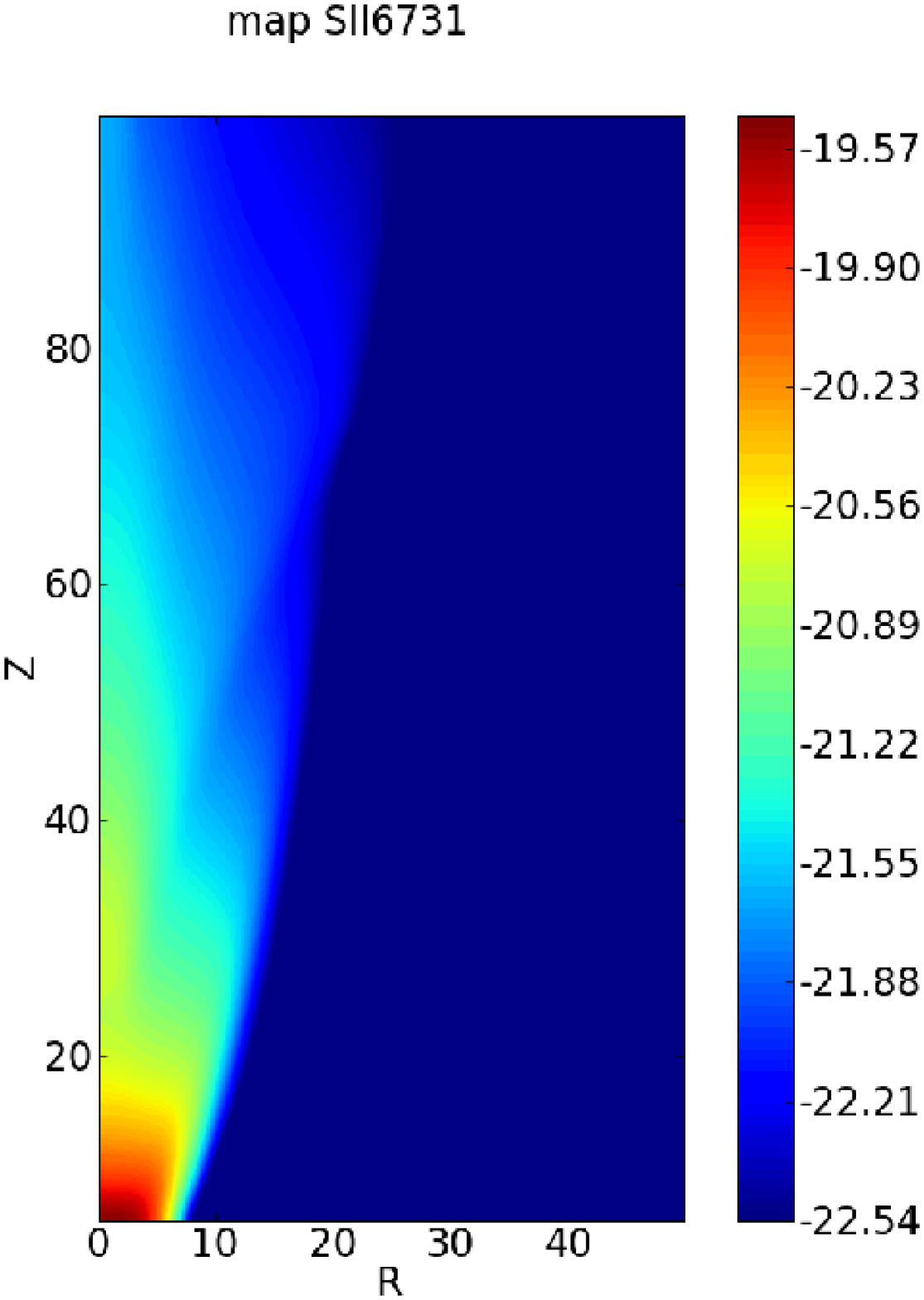}
  \includegraphics[width=0.245\textwidth]{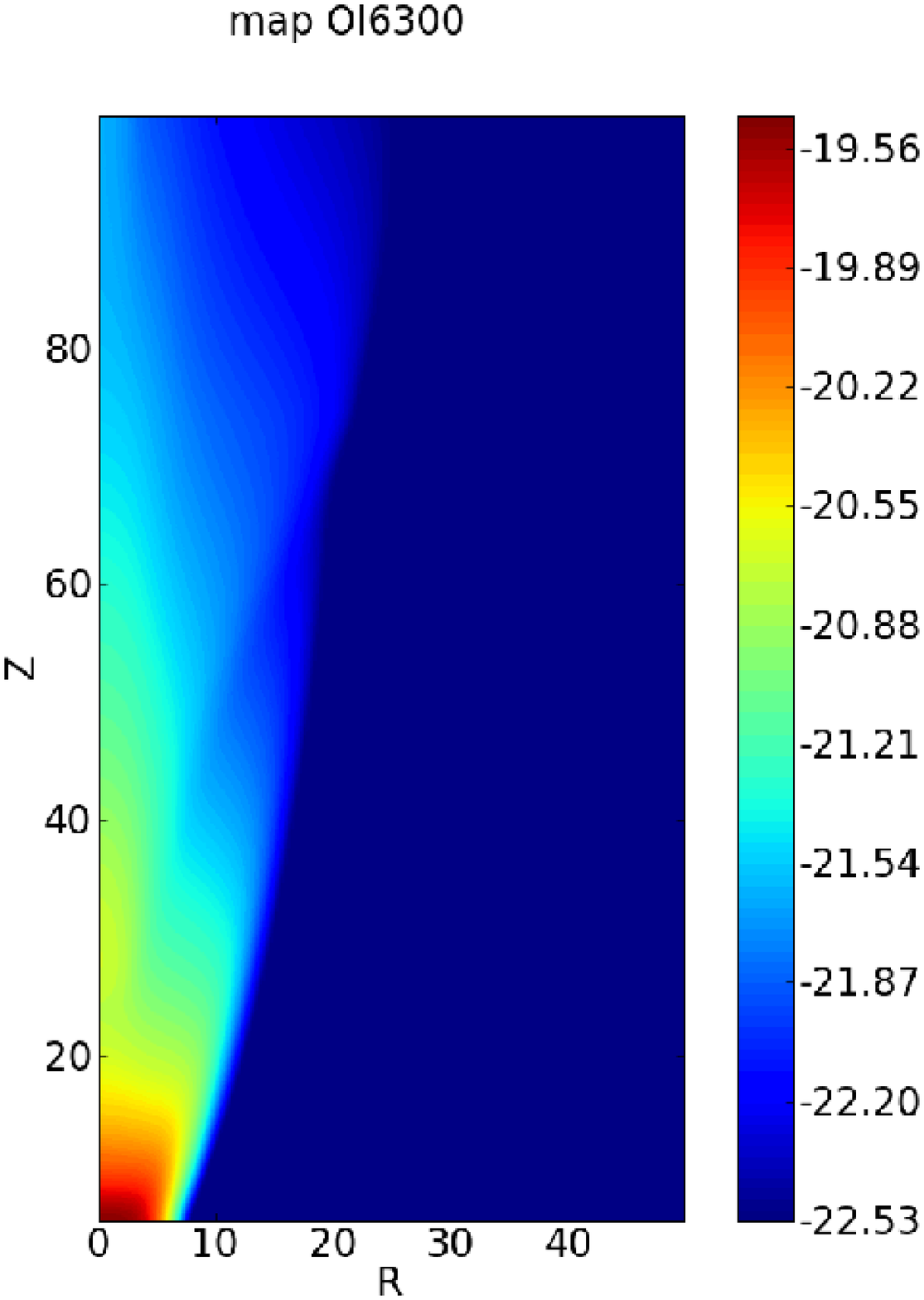}
  \caption{Typical output of OpenSESAMe: cuts of the density and temperature, 
    the electron and SII and OI ion densities and synthetic emission maps of 
    the [SII] $\lambda$6731 and [OI] $\lambda$6300 lines for model ADO and run 
    (500,600,0.5). $R_0 = 7.60$ AU.}
  \label{Fig_pipeline_output_ADO}
\end{figure*}

Before we present the implications of truncating the analytical solution below,
we have to compare the unmodified ADO solution \citep{VTS00} with observations 
to show that the truncation is indeed needed. Figure 
\ref{Fig_pipeline_output_ADO}
shows the typical output of OpenSESAMe: cuts of the density and temperature, 
the electron and SII and OI ion densities and synthetic emission maps of the 
[SII] $\lambda$6731 and [OI] $\lambda$6300 lines, all for model ADO and run 
(500,600,0.5). The electron density shows a steep cut-off along a field line 
anchored at about $5\,R_0$ where the temperature drops below $10^4$ K, while 
the [SII] and [OI] emission maps show a somewhat shallower cut-off, which is 
located at a larger radius (about $8\,R_0$). Synthetic emission maps of the 
[OI] $\lambda$6300 line for all models and runs are shown in Appendix 
\ref{app_emiss}.

\begin{figure*}[!htb]
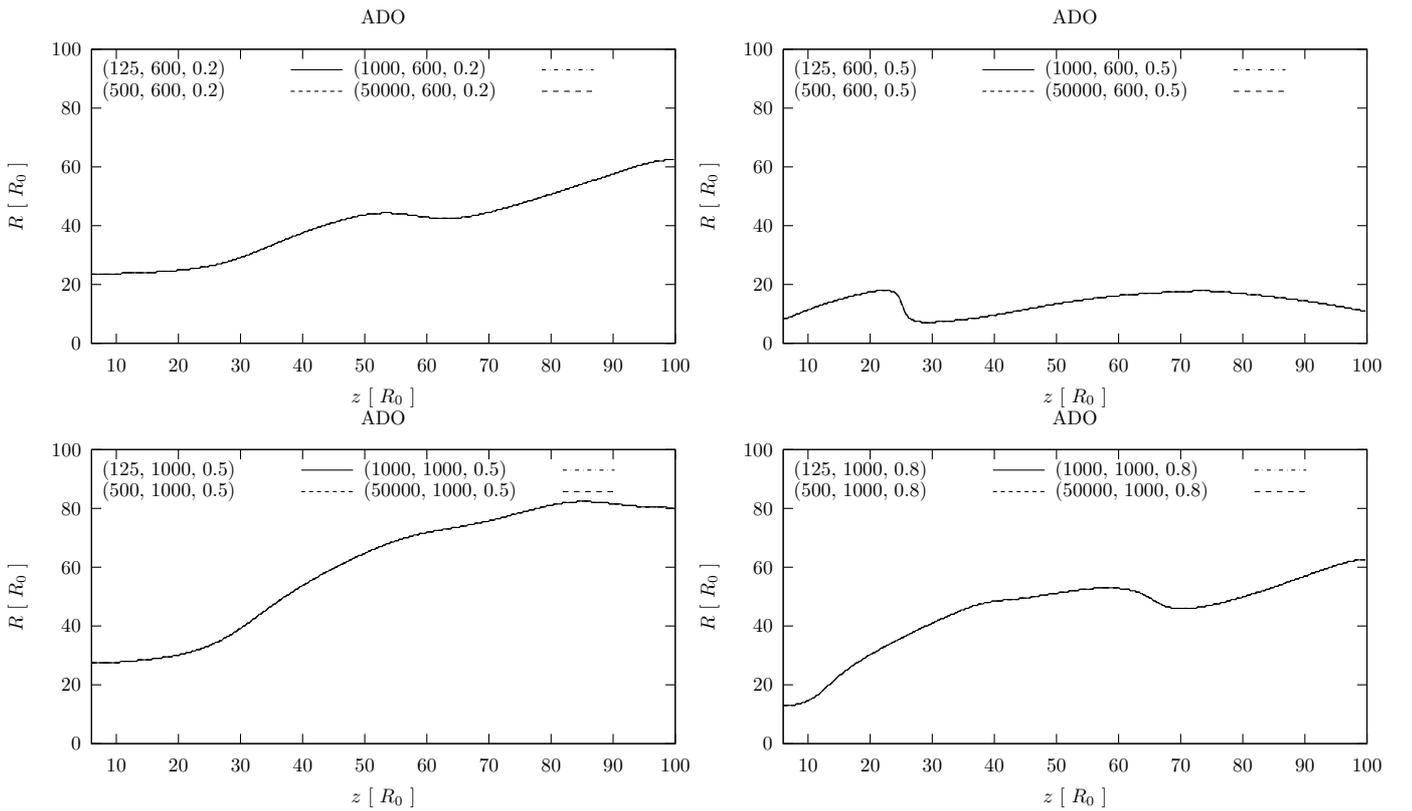

  \centering
  \includegraphics[width=0.49\textwidth]{11204f4a.eps}
  \includegraphics[width=0.49\textwidth]{11204f4b.eps}\\
  \includegraphics[width=0.49\textwidth]{11204f4c.eps}
  \includegraphics[width=0.49\textwidth]{11204f4d.eps}
  \caption{Dimensionless jet width derived from synthetic [OI] images as a 
    function of distance from the source in model ADO.}
  \label{jet_width_ADO_OI}
\end{figure*}

The extracted jet widths in normalized $R_0$ units are plotted in Fig. 
\ref{jet_width_ADO_OI}. In each plot we combine runs with equal velocities and 
masses, i.e. only varying the density, which result in identical jet widths as 
expected. Because we extracted the jet width from a {\em ratio} of intensities 
by using the FWHM (we divided the maps by the intensity on the axis and check 
where the ratio is 0.5), the factor $\rho^2$ cancels out. 

Much larger changes are present when we compare runs where $\rho_0$ is 
identical, but $v_0$ varies (top right and bottom left plot) and when we look 
for the influence of the mass (plots in each row). The jet widths derived from 
synthetic [OI] are identical to those derived from synthetic [SII] images, thus 
we show only results based on [OI] images.

The jet widths in $R_0$ increase with increasing velocity.  A higher mass 
reduces the normalized jet width considerably. In run (500,600,0.5), as seen in 
Fig. \ref{Fig_pipeline_output_ADO}, the extracted jet width does not 
follow any directly apparent feature in the emissivity maps nor a specific 
contour line.

After rescaling the jet widths with the appropriate values of $R_0$, we compared
the modeled and observed jet widths in AU (Fig. \ref{jet_width_ADO_obs}). The 
jet widths are now rescaled with $R_0$, which is proportional to the mass and 
anti-proportional to $v_0$ (eq. \ref{length_scale}). Because 
$f(i) \equiv v_{\rm jet}/v_0$ is almost constant in our range of $R_0$, 
$v_{\rm jet}$ and $v_0$ can be interchanged. This behavior of $R_0$ is now 
dominant, thus the jet width increases with increasing mass and decreases with 
increasing {\em jet} velocity, although the second effect seems to be of minor 
importance.

The untruncated ADO model on average gives too large jet widths compared to the 
observations of T Tauri jets. The first five data points within 120 AU from the 
jet source in DG Tau can be best approximated with run (500,600,0.2), 
unfortunately our model does not provide results farther out for such a small 
$R_0$. At distances above 200 AU, the run (500,600,0.5) gives a jet width in 
the observed range but only after the first bump, which is intrinsic for 
our analytical model. \citet{HEG95} give a mass of DG Tau of 0.67 $M_\odot$, 
i.e. higher than in both models. Although this difference in mass might not be 
meaningful, we conclude that to be able to reproduce all jets in our 
sample, we need an additional effect which reduces the derived jet width.

\clearpage


\clearpage

\begin{figure}[!htb]
  \centering
  \includegraphics[width=\columnwidth]{11204f5.eps}
  \caption{Jet widths in AU derived from synthetic [OI] images as a 
    function of distance from the source in model ADO; overlaid are the data 
    points of Fig. \ref{Fig_observations}.}
  \label{jet_width_ADO_obs}
\end{figure}

\section{Effects of outer truncation} \label{sec_outer}

\begin{figure*}[!bht]
  \centering
  \includegraphics[width=0.24\textwidth]{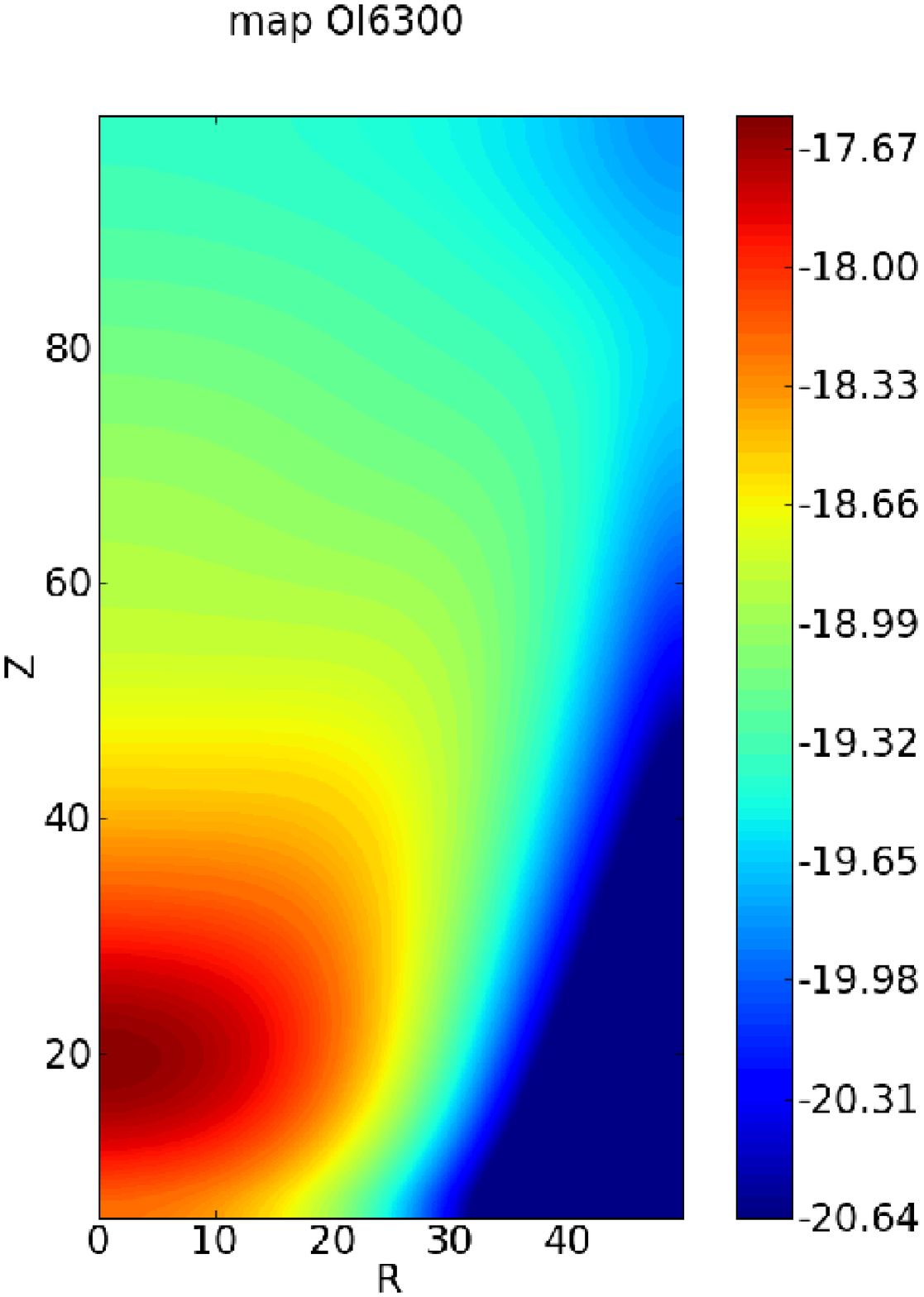}
  \includegraphics[width=0.24\textwidth]{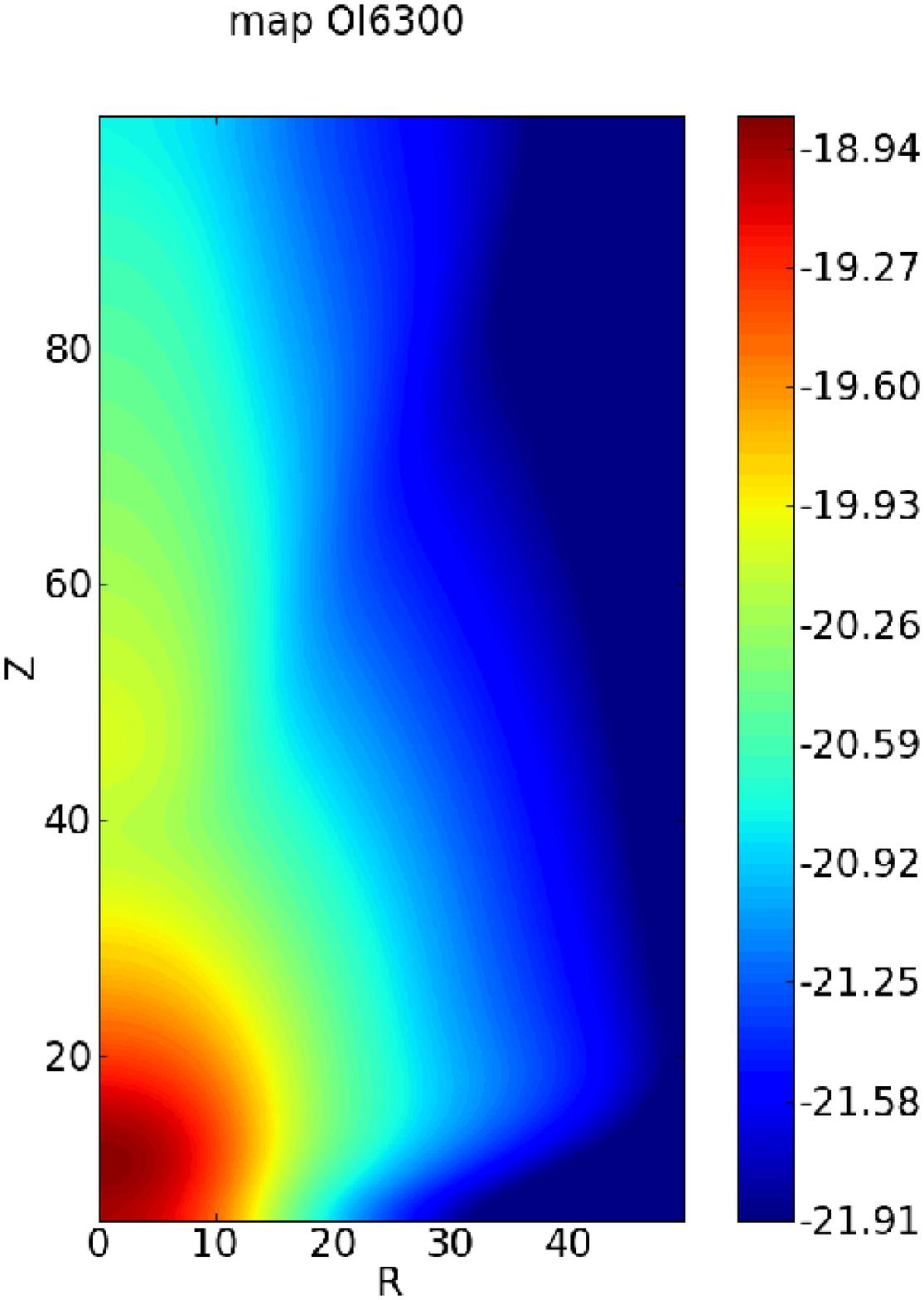}
  \includegraphics[width=0.24\textwidth]{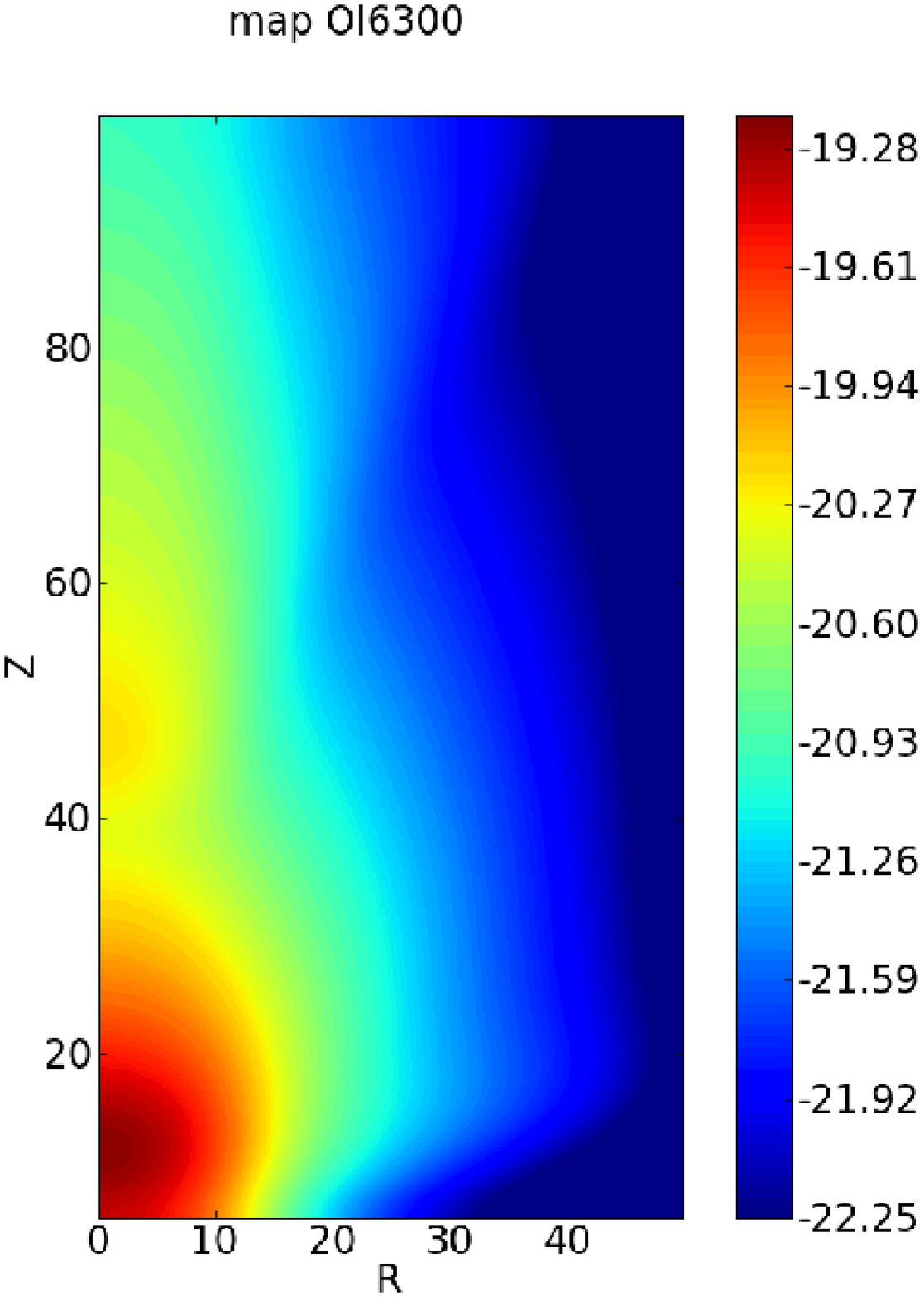}
  \includegraphics[width=0.24\textwidth]{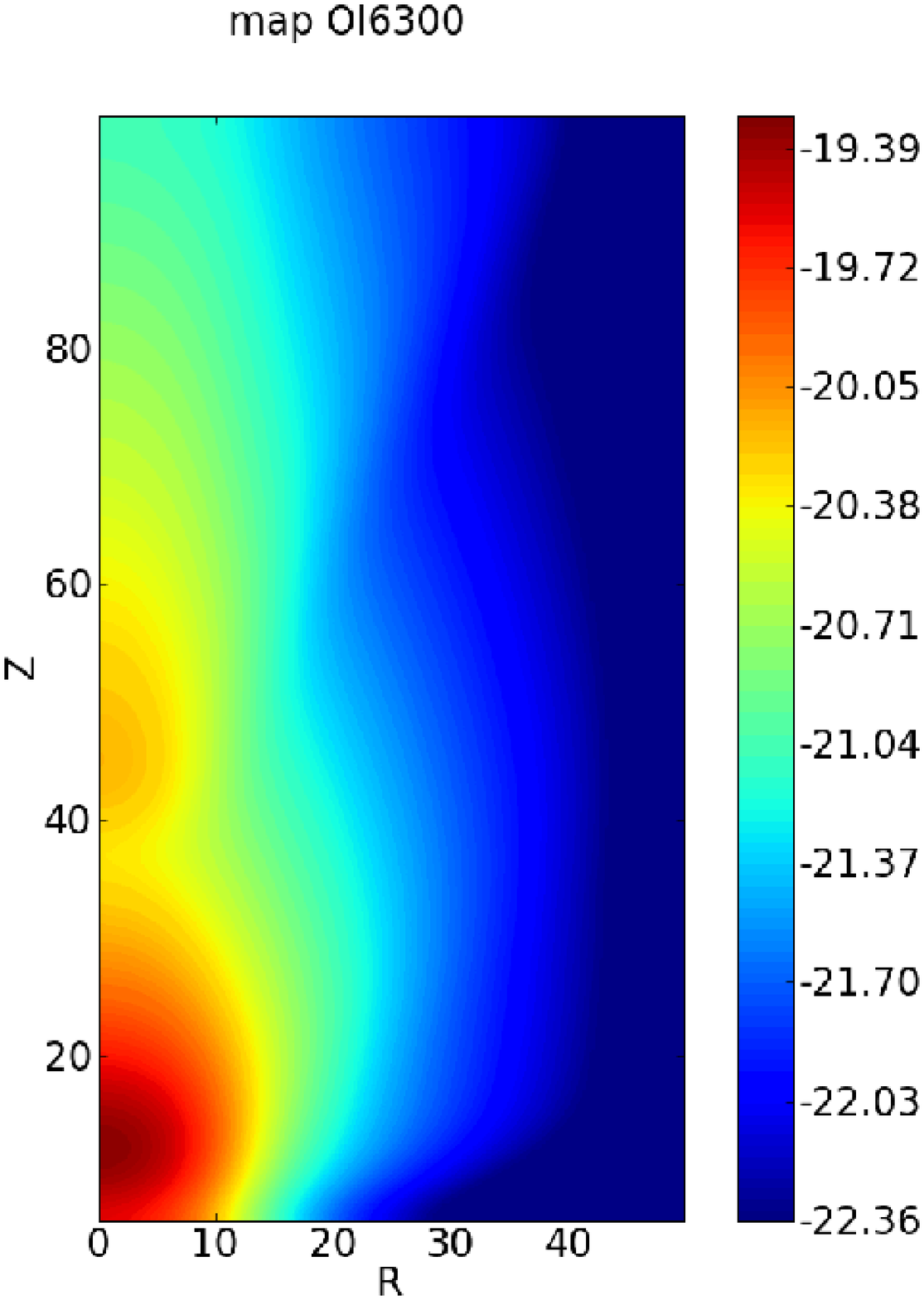}
  \includegraphics[width=0.24\textwidth]{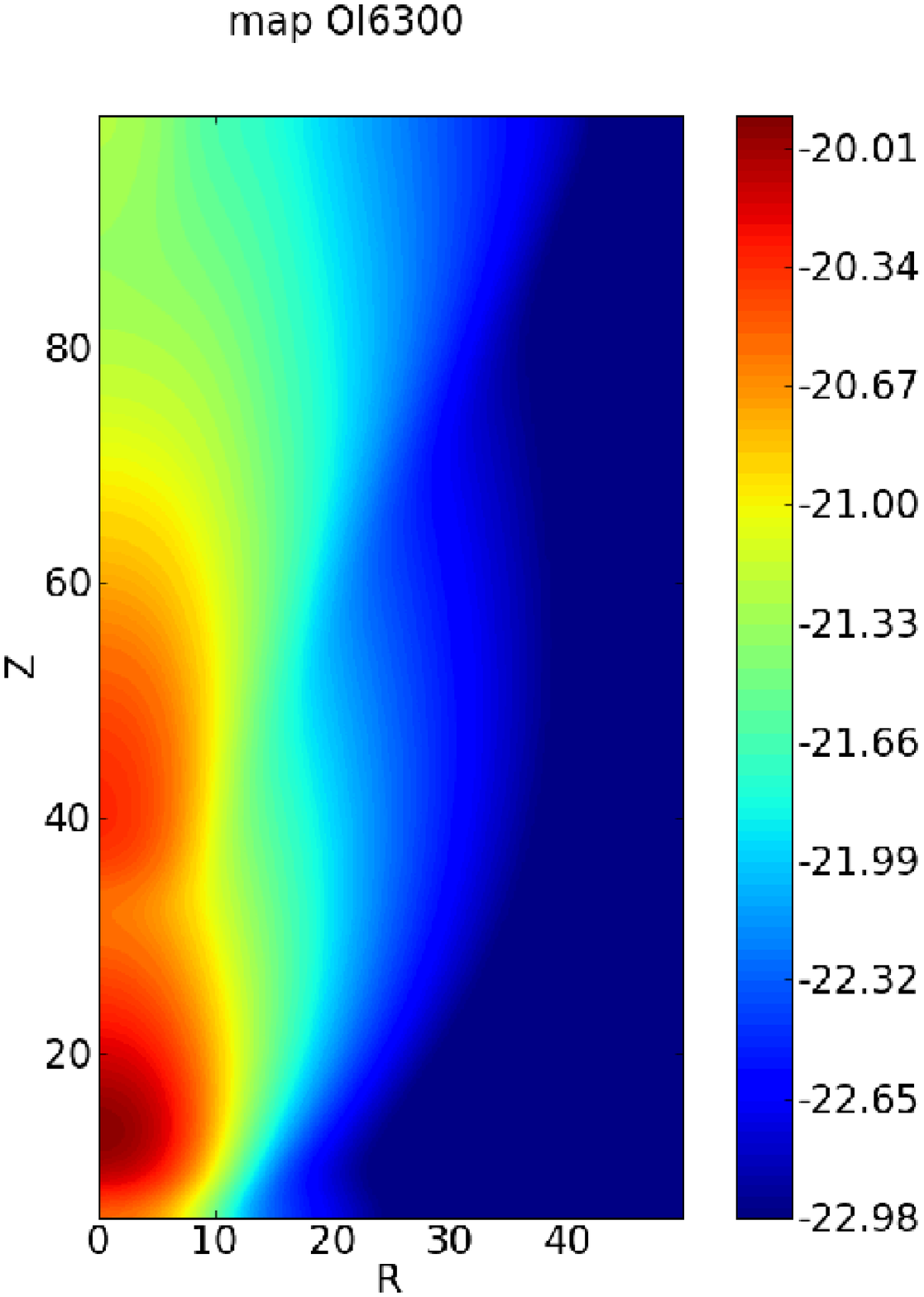}
  \includegraphics[width=0.24\textwidth]{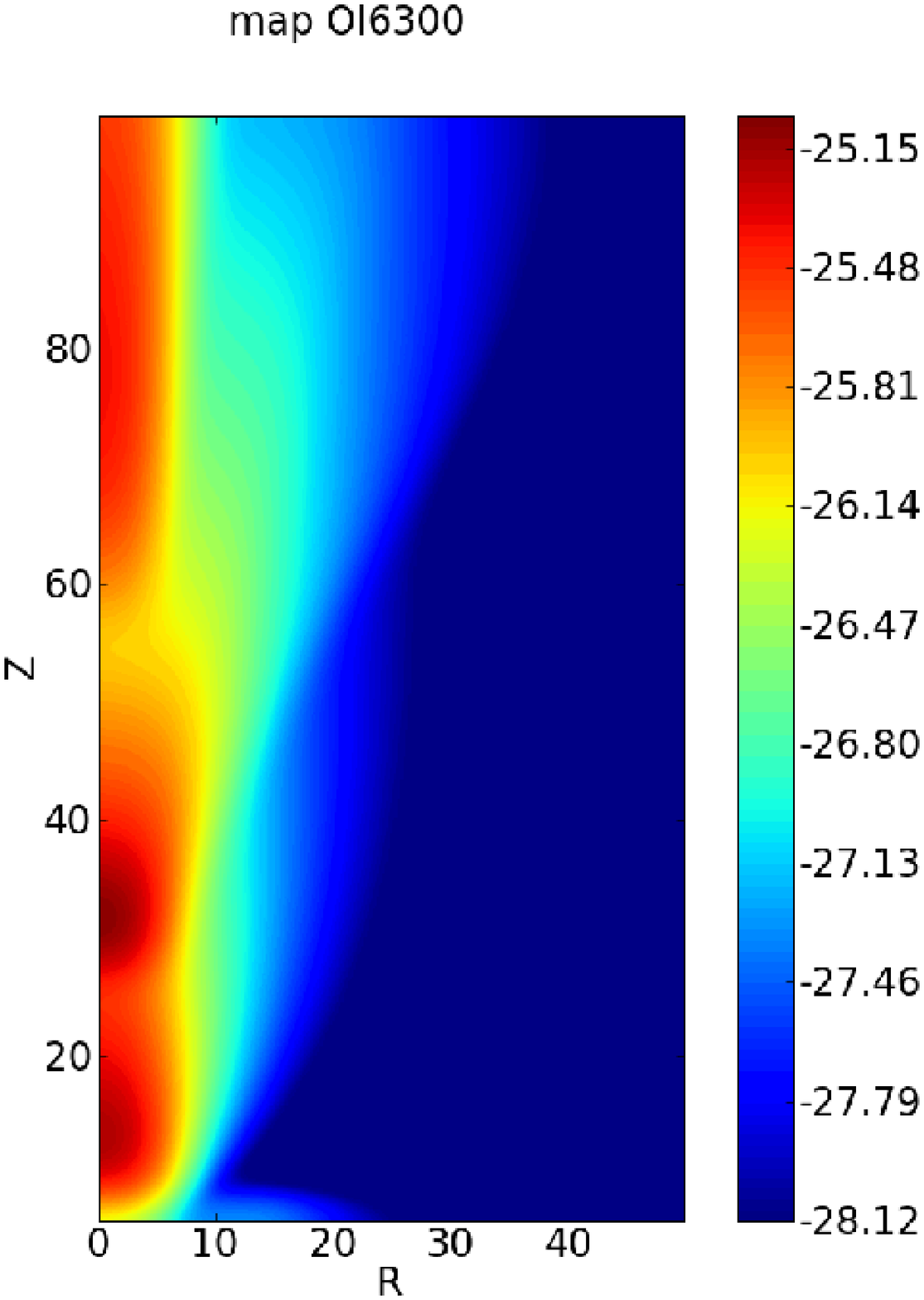}
  \includegraphics[width=0.24\textwidth]{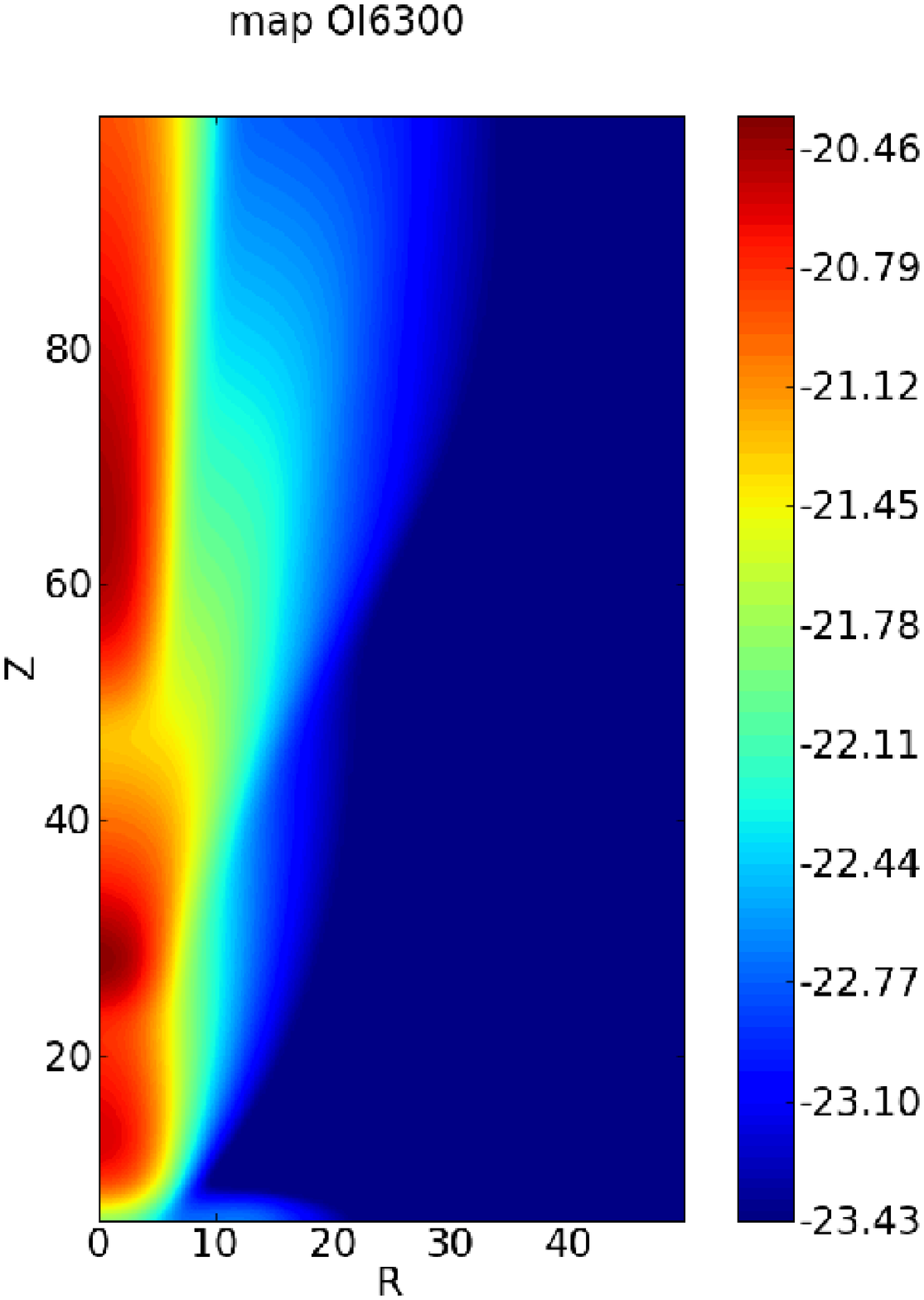}
  \includegraphics[width=0.24\textwidth]{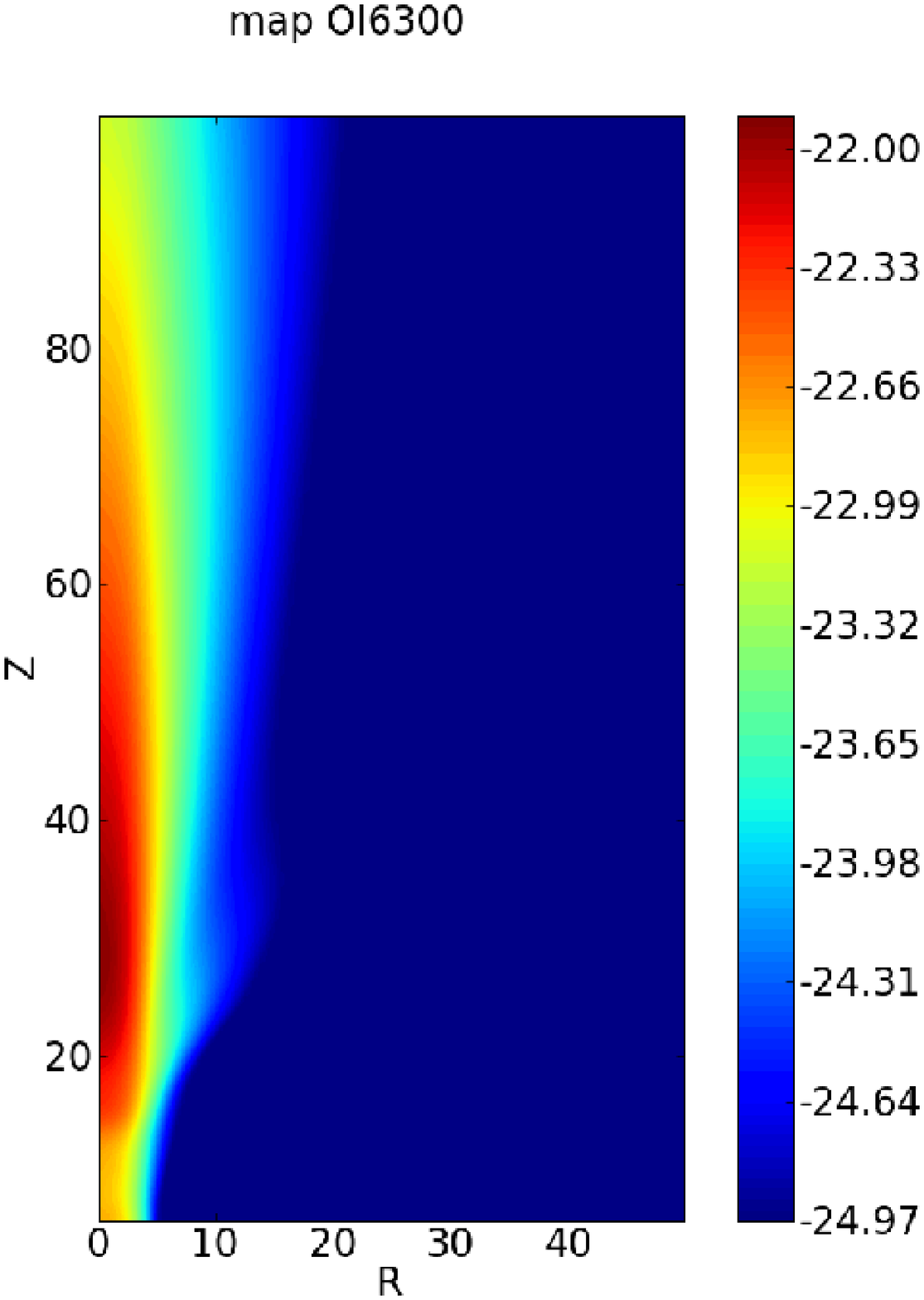}
  \caption{Synthetic emission maps of the [OI] $\lambda$6300 line, convolved 
    with a Gaussian PSF with a FWHM of 15 AU, for run (500,1000,0.5). 
    Top row: models ADO ($R_0 = 1.04$ AU), SC1a ($R_0 = 1.98$ AU), 
      SC1b ($R_0 = 1.98$ AU), SC1c ($R_0 = 2.19$ AU); bottom row: models 
      SC1d ($R_0 = 2.61$ AU), SC1e ($R_0 = 4.22$ AU), SC1f ($R_0 = 4.86$ AU), 
      SC1g ($R_0 = 7.23$ AU).}
  \label{Fig_emissmaps_outertrunc}
\end{figure*}

First we investigate the effects of outer truncation. Convolved synthetic maps 
for the emission in [OI] $\lambda$6300 for some numerical models and run 
(500,1000,0.5) are given in Fig. \ref{Fig_emissmaps_outertrunc}. Truncation 
leads to collimation of the emission region with respect to the model ADO 
without any truncation.

Again we extracted the jet width from emission maps like these. The resulting 
widths derived from the synthetic [OI] images and scaled to AU are presented in 
Fig. \ref{jet_widths_modelSC}. We found similarities in behavior in the 
truncated models to the untruncated model ADO. The jet widths show again no 
dependency on the density, as described for model ADO in the previous section. 
Surprisingly, in models SC1a-c, SC2 and SC4 the runs (500,600,0.2) and 
(500,1000,0.5) and also (500,1000,0.8) lead to almost similar physical jet 
widths. The first two also almost coincide in models SC1d-e. As in model ADO, 
also in the truncated models the run (500,600,0.5) has the smallest jet widths 
(after the first bump). In principle, we can reproduce even smaller values than 
the observed ones. 

\begin{figure*}[!htb]
  \centering
  \includegraphics[width=0.45\textwidth]{11204f7a.eps}
  \includegraphics[width=0.45\textwidth]{11204f7b.eps}
  \includegraphics[width=0.45\textwidth]{11204f7c.eps}
  \includegraphics[width=0.45\textwidth]{11204f7d.eps}
  \includegraphics[width=0.45\textwidth]{11204f7e.eps}
  \includegraphics[width=0.45\textwidth]{11204f7f.eps}
  \includegraphics[width=0.45\textwidth]{11204f7g.eps}
  \includegraphics[width=0.45\textwidth]{11204f7h.eps}
  \includegraphics[width=0.45\textwidth]{11204f7i.eps}
  \includegraphics[width=0.45\textwidth]{11204f7j.eps}
  \caption{Jet widths in AU derived from synthetic [OI] images as a function of 
    distance from the source in models ADO and SC1a--g, SC2 and SC4 and for 
    runs (500,600,0.2), (500,600,0.5), (500,1000,0.5), (500,1000,0.8); overlaid 
    are the data points of Fig. 1. \label{jet_widths_modelSC}}
\end{figure*}

\section{Effects of inner truncation} \label{sec_inner}

\begin{figure}[!hbt]
  \centering
  \includegraphics[width=\columnwidth]{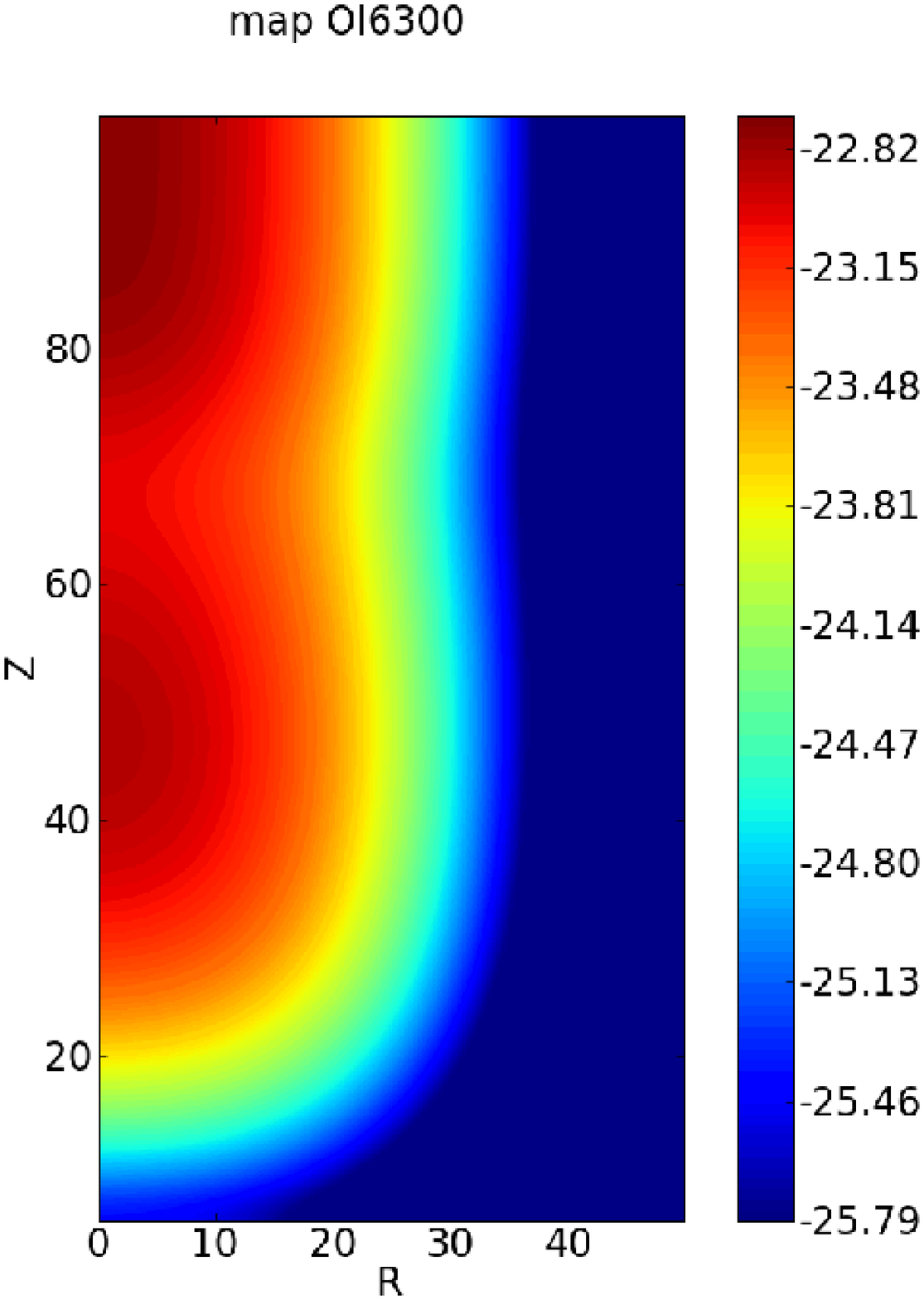}
  \caption{Synthetic emission maps of the [OI] $\lambda$6300 line, convolved 
    with a Gaussian PSF with a FWHM of 15 AU, for model SC3 and run 
    ($\cdots$, 100, 0.2). $R_0 = 0.97$ AU.\label{Fig_emissmaps_innertrunc}}
\end{figure}

In paper I we also performed numerical simulations, in which we truncated the 
analytical solution in the interior, i.e. at an inner truncation radius. The 
physical picture behind this scenario is a stellar magnetosphere truncating the 
jet-emitting disk. We showed that inner truncation leads to a decrease of the 
jet radius and compression of the material in the inner region. 
Unfortunately, only one run met our scaling requirements (Sect. \ref{sec_norm}):
model SC3 and run ($\cdots$, 100, 0.2). For this model, a convolved synthetic 
map for the emission in [OI] $\lambda$6300 is given in Fig. 
\ref{Fig_emissmaps_innertrunc}. 

After rescaling the derived jet width to AU, we found an almost constant width 
in the range of the observed values. Note that our model does not provide 
results farther out than 100 AU due to a small $R_0$.

\begin{figure}[!htb]
  \centering
  \includegraphics[width=\columnwidth]{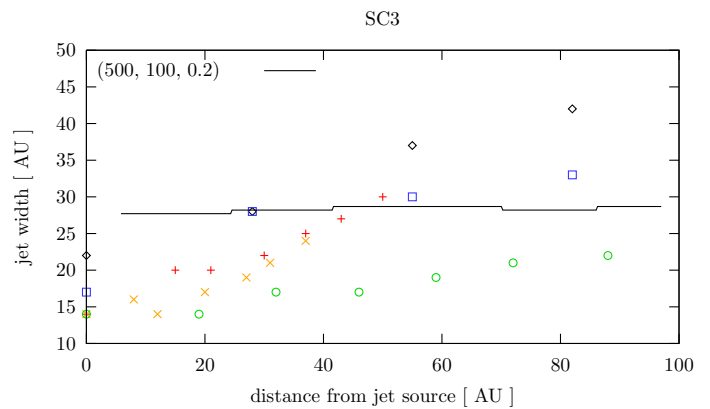}
  \caption{Jet widths in AU derived from synthetic [OI] images as a function of 
    distance from the source in model SC3 and for run (500,100,0.2); overlaid 
    are the data points of Fig. 1. \label{jet_widths_modelSC35}}
\end{figure}

\section{Discussion}

We studied the jet widths derived from synthetic emission maps in different 
forbidden lines as the full-width half-maximum of the emission.

We found that the untruncated model ADO of \citet{VTS00} cannot account for the
small jet widths found in recent optical images taken with HST and AO. The 
density normalization is not important for the resulting measured jet width as 
long as we are far from the critical regime. 

We investigated different effects for reducing the deriving jet width: by 
imposing an outer radius of the launching region of the underlying accreting 
disk and thus also of the outflow on the observable structure of the jet and by 
imposing an inner radius of the underlying accretion disk due to interactions 
with the stellar magnetosphere.

\subsection{Outer truncation}

We created synthetic images based on our simulations of truncated disk winds 
\citep{STV08} as well as new simulations and found that the extracted jet widths
in the truncated models decrease for models SC1a--1g, compared to those of the 
untruncated model ADO, as naively expected. 

In the present paradigm, jets are emitted only by the inner part of the disk. 
Hence in the other parts the disk can be described by a standard accretion 
disk (SAD), in the inner parts by a jet-emitting disk (JED). \citet{ALK03} 
showed that one can estimate the launching region as
\begin{eqnarray}
R_0 &=& 0.24\,\textrm{AU}\,
\left( \frac{R_\infty}{70\,\textrm{AU}} \right)^{2/3}\,
\left( \frac{v_{\phi,\infty}}{10\,\textrm{km s$^{-1}$}} \right)^{2/3}
\nonumber \\
&& \times \,
\left( \frac{v_{P,\infty}}{500\,\textrm{km s$^{-1}$}} \right)^{-4/3}\,
\left( \frac{\mathcal{M}}{0.5\,\textrm{M$_\odot$}} \right)^{1/3} \,.
\nonumber
\end{eqnarray}

This transition was constrained observationally with measured 
jet rotation velocities and using the equation above and radii of the order of 
0.1--1 AU are used in several theoretical studies as e.g. \citet{CoF08}.

Our results can be used to infer the ``real'' value of the truncation radius 
$R_{\rm trunc}$ in the observed sample of jets and interpret it as the transition
radius of the JED to the SAD, assuming the specific model of V00 applies. 

At the lower boundary in our simulations, the truncation radii are 
given in Table \ref{tbl_models}.  They vary from 5.375 $R_0$ in model SC1a to 
0.575 $R_0$ in model SC1g. However, these radii are set at $z = 6\,R_0$ (the 
lower boundary), not in the equatorial plane. Those can be calculated by 
extrapolating the field line, i.e.
\begin{equation}
R_{\rm trunc} |_{z = 0} = R_{\rm trunc} |_{z = 6} \, 
\frac{G ( \pi / 2 )}{G ( \theta_{\rm trunc})} \,,
\end{equation}
with $\theta_{\rm trunc} = \arctan ( R_{\rm trunc} |_{z = 6} / 6 )$ and $G$ taken
from the analytical solution of V00. This gives the following results:
\begin{itemize}
\item in models SC1a, SC2--4 $R_{\rm trunc} |_{z = 0} = 0.11316\,R_0$, 
\item in model SC1b $R_{\rm trunc} |_{z = 0} = 0.09099\,R_0$, 
\item in model SC1c $R_{\rm trunc} |_{z = 0} = 0.07275\,R_0$, 
\item in model SC1d $R_{\rm trunc} |_{z = 0} = 0.0213\,R_0$, 
\item in model SC1e $R_{\rm trunc} |_{z = 0} = 0.00722\,R_0$, 
\item in model SC1f $R_{\rm trunc} |_{z = 0} = 0.00546\,R_0$ and
\item in model SC1g $R_{\rm trunc} |_{z = 0} = 0.00069\,R_0$. 
\end{itemize}
The last step is a multiplication with $R_0$ of the best-fit model (Table 
\ref{tbl_R0}). The resulting values of $R_{\rm trunc} |_{z = 0}$ in AU for each 
model and run are given in Table \ref{tbl_Rtrunc}.
\begin{table*}[!htb]
\caption{$R_0$ in AU of the different OpenSESAMe runs}
\label{tbl_R0}
\centering
\begin{tabular}{c | c c c c c}
\hline\hline
model & ($\cdots$,600,0.2) & ($\cdots$,600,0.5) & ($\cdots$,1000,0.2) & 
($\cdots$,1000,0.5) & ($\cdots$,1000,0.8) \\
\hline
ADO  & 1.20 &  7.60 & ---  & 1.04 &  2.47 \\
SC1a & 2.32 &  7.60 & ---  & 1.98 &  2.99 \\
SC1b & 2.17 &  7.59 & ---  & 1.98 &  3.33 \\
SC1c & 2.40 &  7.60 & 0.98 & 2.19 &  3.47 \\
SC1d & 2.91 &  8.87 & 1.34 & 2.61 &  4.10 \\
SC1e & 4.54 & 13.96 & 2.06 & 4.22 &  5.86 \\
SC1f & 5.29 & 14.40 & 2.17 & 4.86 &  6.91 \\
SC1g & 8.03 & ---   & 3.04 & 7.23 & 13.69 \\
\hline
SC2  & 2.31 &  7.60 & ---  & 1.96 &  2.98 \\
SC4  & 2.05 &  7.60 & ---  & 2.08 &  2.97 \\
\hline\hline
model & \multicolumn{5}{c}{($\cdots$,100,0.2)} \\
\hline
SC3 & \multicolumn{5}{c}{0.97} \\
\hline
\end{tabular}
\end{table*}
\begin{table*}[!htb]
\caption{Truncation radius at the equator $R_{\rm trunc} |_{z = 0}$ in AU of the 
different OpenSESAMe runs}
\label{tbl_Rtrunc}
\centering
\begin{tabular}{c | c c c c c}
\hline\hline
model & ($\cdots$,600,0.2) & ($\cdots$,600,0.5) & ($\cdots$,1000,0.2) & 
($\cdots$,1000,0.5) & ($\cdots$,1000,0.8) \\
\hline
SC1a & 0.263 & 0.860 & ---   & 0.224 & 0.338 \\
SC1b & 0.197 & 0.691 & ---   & 0.180 & 0.303 \\
SC1c & 0.175 & 0.553 & 0.071 & 0.159 & 0.252 \\
SC1d & 0.062 & 0.189 & 0.029 & 0.056 & 0.087 \\
SC1e & 0.033 & 0.101 & 0.015 & 0.030 & 0.042 \\
SC1f & 0.029 & 0.079 & 0.012 & 0.027 & 0.038 \\
SC1g & 0.006 & ---   & 0.002 & 0.005 & 0.009 \\
\hline
SC2  & 0.261 & 0.860 & ---   & 0.222 & 0.337 \\
SC4  & 0.232 & 0.860 & ---   & 0.235 & 0.336 \\
\hline\hline
model & \multicolumn{5}{c}{($\cdots$,100,0.2)} \\
\hline
SC3 & \multicolumn{5}{c}{0.11} \\
\hline
\end{tabular}
\end{table*}

We found best-fit models for the jets in the observed sample. Note that we 
always ignored the first bump in the synthetic jet widths and focussed on larger
distances from the source:
\begin{itemize}
\item The observed mass of DG Tau (diamonds in Fig. \ref{jet_widths_modelSC}) 
is 0.67 $M_\odot$ \citep{HEG95}, therefore we have to focus on the runs 
(500,1000,0.8), and perhaps also runs (500,600,0.5) and (500,1000,0.5). The 
best-fit model is between ADO and SC1a, thus the truncation radius is larger 
than 0.22 AU. 
\item HN Tau (plus signs) has a mass of 0.72 $M_\odot$ \citep{HEG95}, thus 
again the runs (500,1000,0.8) are favored. Because we ignored the first bump, we
interpolated the jet shape at larger distances and found a best-fit model 
between ADO and SC1a, thus the truncation radius is again larger than 0.34 AU. 
However,this result is highly uncertain.
\item The mass of CW Tau (squares) is the highest in our sample, 1.03 $M_\odot$ 
\citep{HEG95}. Using the runs (500,1000,0.8), the best-fit model is SC1b or 
SC1c. The truncation radius is thus between 0.25 -- 0.3 AU.
\item UZ Tau E (crosses) has the lowest mass in our sample, only 0.18 $M_\odot$ 
\citep{HEG95}, we use the runs (500,600,0.2). Again we had to interpolate the 
jet width from larger distances and choose model SC1a as best-fit model. The
truncation radius is about 0.26 AU.
\item The measured mass of RW Aur (circles) is 0.85 $M_\odot$ \citep{HEG95}, 
thus we have to focus on runs (500,1000,0.8). We need a very high degree of 
truncation as in models SC1e-g, thus a truncation radius of the order of 0.04 
AU.
\end{itemize}

\subsection{Inner truncation}

Because the jet-emitting accretion disk is thought to be truncated by the 
stellar magnetosphere, we also investigated our models in which the analytical 
solution is truncated at an inner radius. We found that inner truncation can 
also reduce the extracted jet widths. The jet width from our model SC3 and run 
($\cdots$,100,0.2) is about 30 AU for the inner 100 AU of the jet. This is well 
in the observed range of 15--45 AU. Because we did not vary the inner 
truncation radius, we can only claim -- based on our results of paper I -- that 
the larger the truncation radius, the higher is the compression of the 
resulting jet. In our models we chose a radius of 5.375 $R_0$ at $z = 6$. For 
model SC3 and run ($\cdots$,100,0.2), this corresponds to an inner truncation 
radius of 0.11 AU, which is only slightly higher than the inner hole in T Tauri 
disks \citep[0.02-0.07 AU,][]{NCG07}. A parameter study varying the inner 
truncation radius is needed.

\section{Conclusions}

We showed as a proof of concept that jet widths derived from numerical 
simulations extending analytical MHD jet formation models can be very helpful 
for understanding recently observed jet widths from observations with adaptive 
optics and space telescopes. However, further aspects have to be investigated 
in more detail.

An intrinsic feature in all our models with outer truncation is the first bump 
in the extracted jet width, which complicates the comparison of synthetic and 
observed jet widths. Only for DG Tau and CW Tau, we could unambiguously find 
models with outer truncation which fitted the observed jet widths at larger 
distances. For HN Tau and UZ Tau E, we have no observed jet widths at larger 
distances, only in the region which is contaminated by the bump. RW Aur shows 
very small observed jet widths at all scales, which cannot be reproduced by any 
of our models and runs. The derived truncation radii ($>0.25$ AU) are several 
times larger than the inner radius of the gaseous disk in T Tauri stars 
\citep[0.02-0.07 AU,][]{NCG07}, thus our use of a self-similar disk-wind 
solution is consistent. 

The model with inner truncation gives a synthetic jet width, which is constant 
and is within the range of observed widths. Another advantage of this model is
that its flow velocities of the order of 100 km s$^{-1}$ are closer to 
observed values. All models with outer truncation have very high velocities 
($> 600$ km s$^{-1}$ at our scaling point of $R_{\rm jet} = 10$ AU and 
$Z_{\rm jet} = 100$ AU), several times higher than in that with inner 
truncation.

Because we originally focused only on the effect of outer truncation, we kept
the inner truncation radius constant in our models SC3 and SC5. In order to 
further explore the ability of inner truncation to reproduce the observations, 
we have to vary the inner truncation radius in a parameter study. Naively, one 
would expect that the derived jet widths are not decreasing for decreasing
truncation radii as for the outer truncation, but are {\em increasing}.
This, however, has to be tested with new simulations.

In our study, we assumed an inclination of 90$^\circ$ of the jet, thus 
projection effects may slightly change our results. We will present the results
of such a study in a forthcoming paper.

\citet{GCF01} showed in their Fig. 1 that the measured jet widths 
are mainly characterized by the ejection index $\xi$, defined by \citet{Fer97}. 
This is related to the model parameter $x$ of the solution of V00, 
$\xi = 2\,(x - 3/4)$, and because in our simulations $x = 0.75$, we get 
$\xi = 0$, which is intrinsic for a standard disk. In the solution of 
\citet{Fer97}, $\xi$ also controls the opening of the field lines, because it is
connected to the lever arm $\lambda$ by $\lambda = 1 + 1 / (2\,\xi)$. 
\citet{GCF01} favored cold solutions with values of $\lambda$ between 50--70. If
heating along streamlines is allowed, the relation is broken and also warm 
solutions of e.g. \citet{CaF00b} with smaller $\lambda$ values of 8 and opening 
of streamlines (the ratio of the maximum radius to the initial launch radius) 
of about 30 can reproduce the observation. In the solution of V00, we have 
$\lambda = G ( \pi / 2 )^{-2} \approx 39$ and the maximum opening 
$G_{\rm max} / G ( \pi / 2 ) \approx 893$. The use of other solutions will 
therefore highly influence our results in terms of the outer truncation radius.

\appendix

\section{Details of the normalization algorithm} \label{app_norm}

In order to find the reference values for the distance $R_0$, pressure $p_0$ 
and magnetic field $B_0$ listed in Eqs. (\ref{length_scale})-(\ref{mag_scale}), 
we require some typical values of jet density and velocity at a certain 
position -- we define the jet as the region at the physical position 
\mbox{$(R_{\rm jet},z_{\rm jet}) = (10,100)$ AU} -- and solve iteratively  
Eq. (\ref{length_scale}):
\begin{enumerate}
\item we start the iteration with the choice $R_0 = 1$ AU, namely in the 
grid cell (40,398)
\item we find $v_0$ by dividing the required jet velocity by the velocity in 
PLUTO units inside the grid cell
\item we calculate $R_0$ with equation (\ref{length_scale})
\item we calculate the physical position of the grid cell using this new $R_0$ 
\item if $R$ ($z$) of the grid cell is larger than 10 AU (100 AU), we move to 
another grid cell 10 \% closer to the jet axis (equatorial plane); if $R$ ($z$) 
of the grid cell is smaller than 10 AU (100 AU), we move to another grid cell 
10 \% further away
\item we stop the iteration when the grid cell does not change anymore between
two steps.
\end{enumerate}

\begin{figure}[!htb]
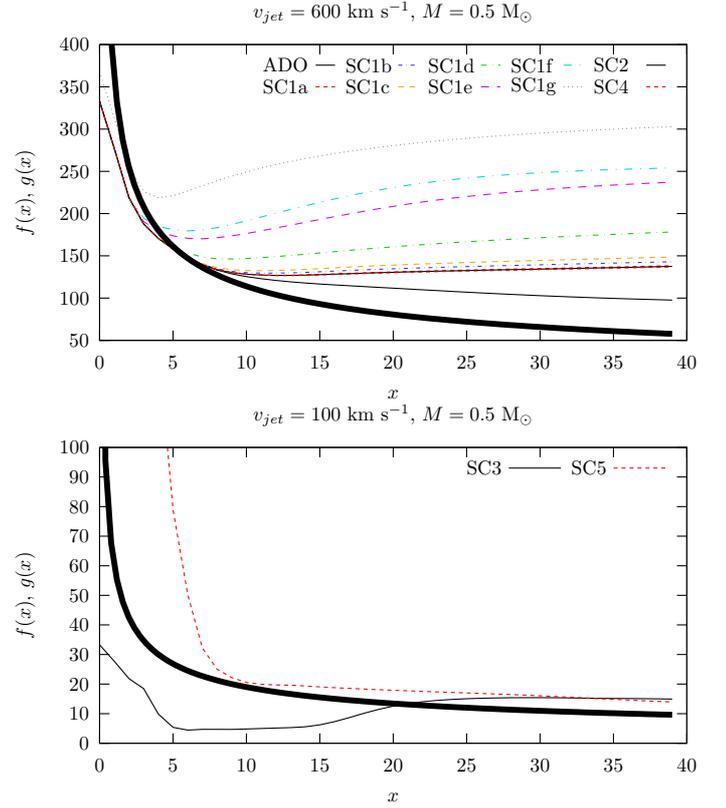

  \centering
  \includegraphics[width=\columnwidth]{11204fA1a.eps}
  \includegraphics[width=\columnwidth]{11204fA1b.eps}
  \caption{Shape of the functions $f(i)$ for all models and of the function 
    $g(i)$ (thick curve) for $v_{\rm jet}= 600$ km s$^{-1}$ (100 km s$^{-1}$ 
    for the models with inner truncation in the bottom panel) and $M = 0.5$ 
    M$_{\odot}$.}
  \label{Fig_normalization}
\end{figure}

The same results, perhaps more intuitively, can be achieved using the following 
graphical picture. We assume that only grid cells along a cone with indices 
($i$, $10\,i$) are important. Using Eq. (\ref{length_scale}) and the 
relation $i=4\,R/R_0$, we have to search for the intersection point
\begin{equation}
f(i) \equiv v(i,10\,i) = \frac{v_{\rm jet}}{v_0} = 4\,v_{\rm jet}\,
\sqrt{\frac{R_{\rm jet}}{G\,M}}\,\frac{1}{\sqrt{i}} \equiv g(i) \,.
\end{equation}
Figure \ref{Fig_normalization} shows $f(i)$ for all models and $g(i)$ for 
$v_{\rm jet}= 600$ km s$^{-1}$ ($v_{\rm jet}= 100$ km s$^{-1}$ for comparison with
models SC3 and SC5) and $M = 0.5$ M$_{\odot}$. After an initial drop 
close to the origin, $f(i)$ converges to an almost constant normalized velocity
for $i > 7$ (i.e. $R_0 < 5.7$ AU for $R_{\rm jet} = 10$ AU) in the models ADO, 
SC1a-c and SC5 and for $i> 20$ (i.e. $R_0 < 2$ AU) in the others. Its value 
increases from the untruncated model ADO (100) monotonically with increasing 
degree of truncation (150 in SC1a up to 300 in SC1g). The models SC2 and SC4 
have values close to that of model SC1a.

Reducing the required mass or increasing the required jet velocity shifts 
the function $g(i)$ upwards and thus increases $i$ of the intersection point 
and decreases $R_0$, until $g(i)$ does not intersect with $f(i)$ anymore 
inside the computational domain. For example, we could not find acceptable 
sets of normalizations for the runs ($\cdots$,1000,0.2) in models ADO, SC1a--b, 
SC2 and SC4; in the models SC1c--g, normalizations were found for all runs.
However, most of these runs were later excluded due to our 
requirement of $R_0 < 30$ AU.

In the inner truncation models SC3 and SC5, the final value of $f(i)$ is much 
lower, only about 15, therefore both functions only intersect for small 
velocities. Again, most of the runs found were later excluded due to our 
requirement of $R_0 < 30$ AU.

\section{Reasons for the artefact of limb-brightening} \label{app_limb}

If the number density is below the critical density, the emissivity of e.g. the 
[SII] $\lambda$6731 line can be written as
\begin{equation} \label{eq_emiss1}
\epsilon ( T ) = n_{\rm e}\,n_{\rm SII}\,f ( T ) = 
n_{\rm e}\,n_{\rm S}\,g ( T )\,f ( T )\, ,
\end{equation}
with $n_{\rm e}$ and $n_{\rm SII}$ the electron and ion density, respectively. 
The latter is a function of the total sulfur density $n_{\rm S}$ and the 
ionization fraction $g ( T ) = n_{\rm SII} / n_{\rm S}$. $f ( T )$ usually has 
the form
\begin{equation}\label{eq_emiss2}
f ( T ) = T^{-C}\,\exp{-D / T}
\end{equation}
with a powerlaw and an exponential cut-off. One can show that $\epsilon ( T )$ 
has a maximum at temperatures of about $4\times10^4$ K. For lower 
temperatures, $f ( T )$ decreases steeply. For temperatures higher than about 
$6\times10^4$ K, $g ( T )$ decreases steeply. The latter cut-off is due to the 
assumption of ionization equilibrium used in OpenSESAMe for calculating the 
level population and chemical composition, while the former is due 
to the exponential temperature dependence in Eq. (\ref{eq_emiss2}), intrinsic to
collisional excitation of a level of upper energy $D$. In eq. (\ref{eq_emiss1}),
$n_{\rm e}$ is also a very steep function of $T$ in ionization equilibrium.

Thus the main emission comes only from a thin shell of the jet, in which the 
temperature is exactly in the range of $(4 - 6)\times10^4$ K. By changing the 
temperature normalization, we can move the emitting shell on the axis, but
the resulting jet widths are of the order of the FWHM of the Gaussian PSF. 
Therefore we have to have a constant temperature profile towards the axis.

\section{Emission maps of the [OI] $\lambda$6300 line for all models and runs} 
\label{app_emiss}

Here we present the emission maps of the [OI] $\lambda$6300 line for all models 
and runs (500, $\cdots$, $\cdots$) for the sake of completeness. 

\begin{figure*}[!bt]
  \centering
  \includegraphics[width=0.24\textwidth]{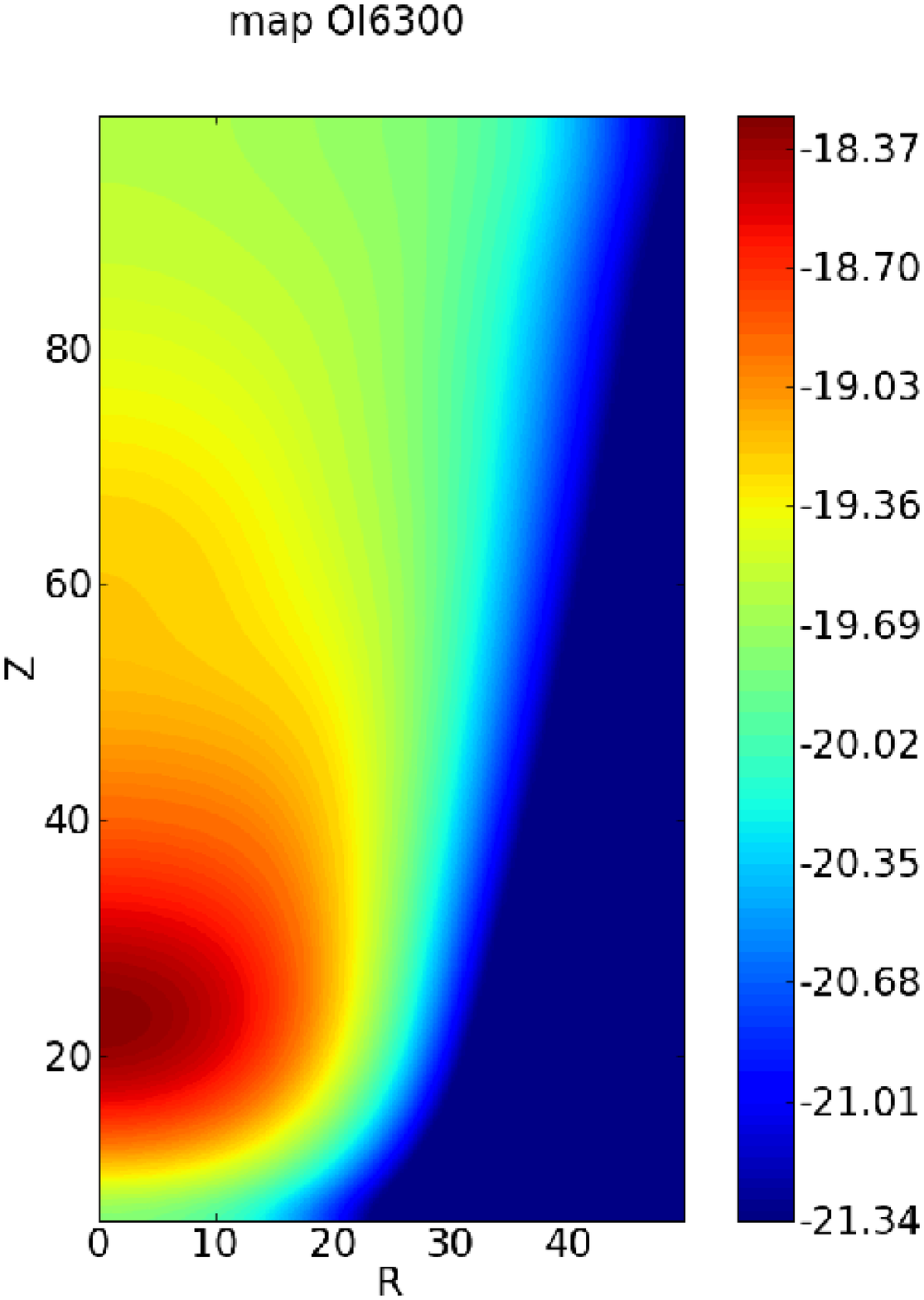}
  \includegraphics[width=0.24\textwidth]{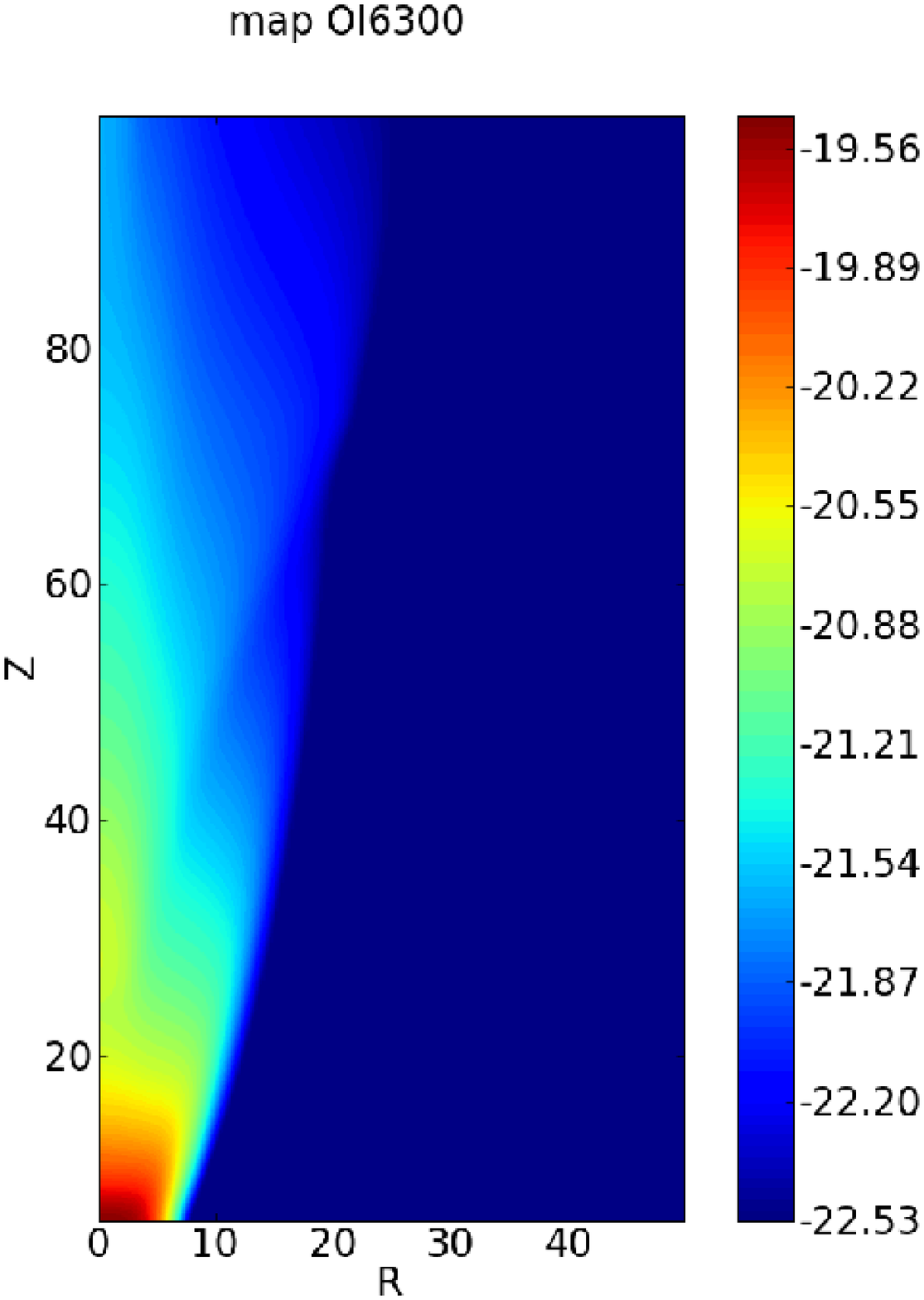}
  \includegraphics[width=0.24\textwidth]{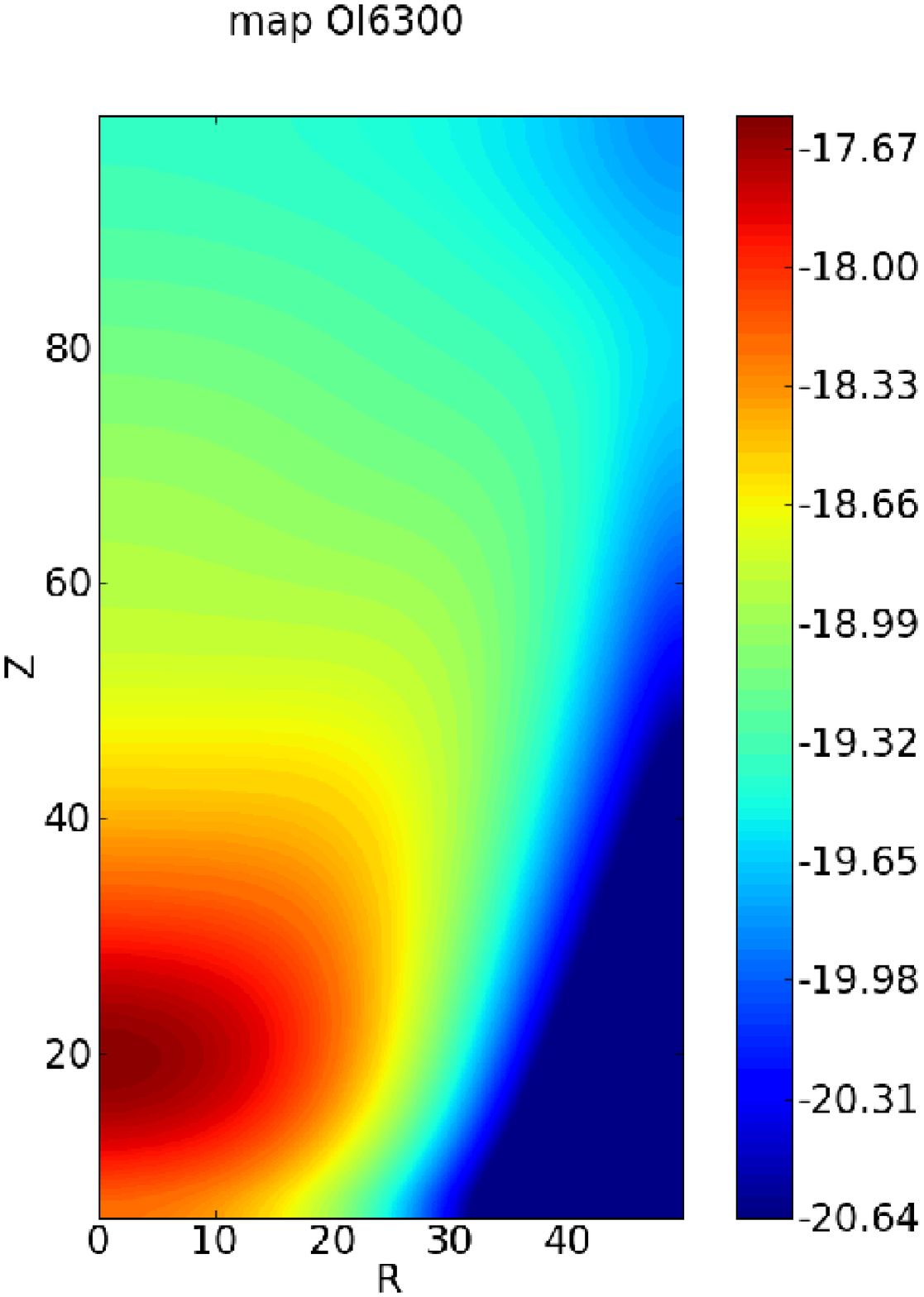}
  \includegraphics[width=0.24\textwidth]{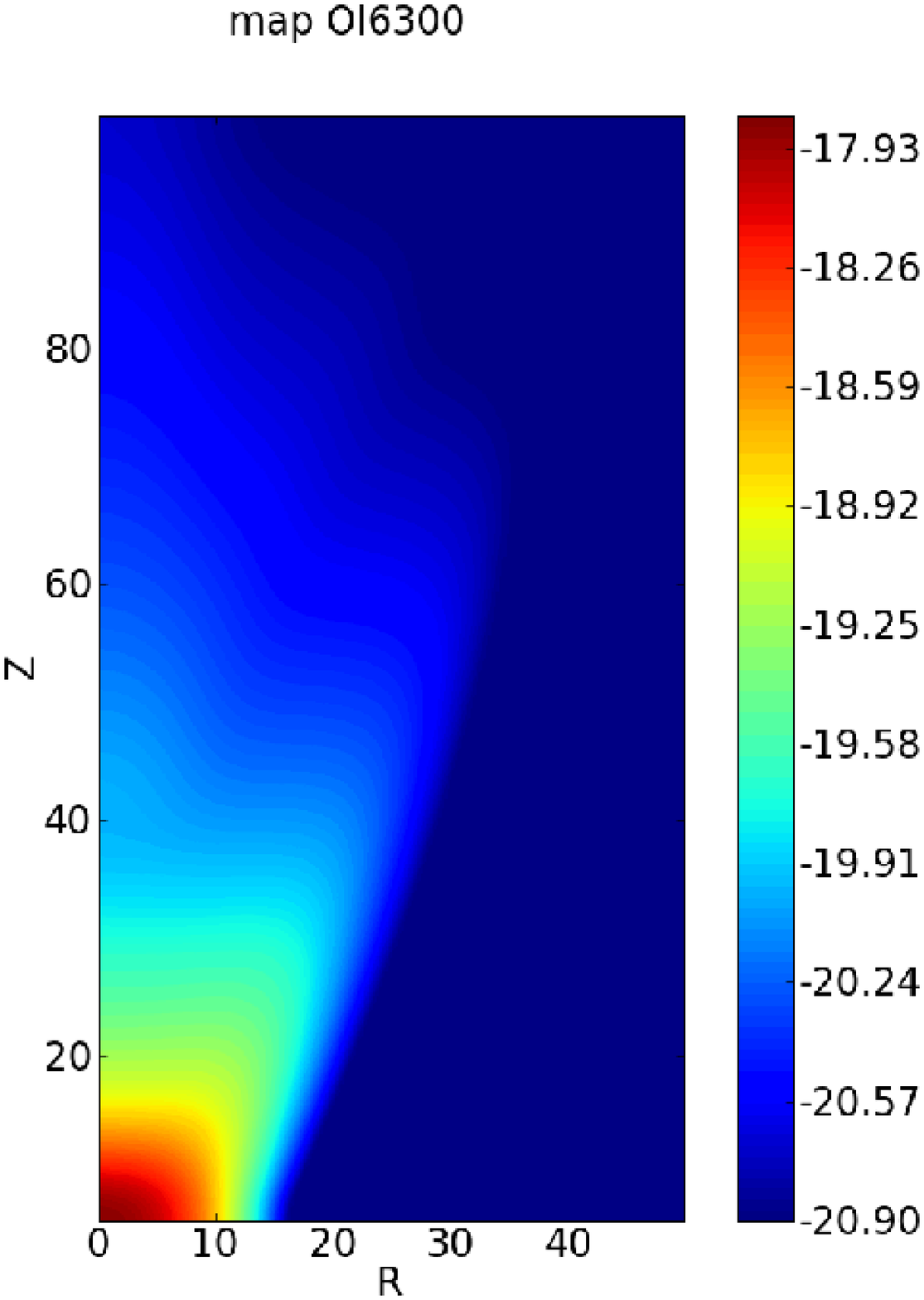}
  \caption{Synthetic emission maps of the [OI] $\lambda$6300 line, convolved 
    with a Gaussian PSF with a FWHM of 15 AU, for model ADO and runs 
    (500, 600, 0.2), (500, 600, 0.5), (500, 1000, 0.5), (500, 1000, 0.8).}
  \label{Fig_emissmaps_all1}
\end{figure*}

\begin{figure*}[!bt]
  \centering
  \includegraphics[width=0.24\textwidth]{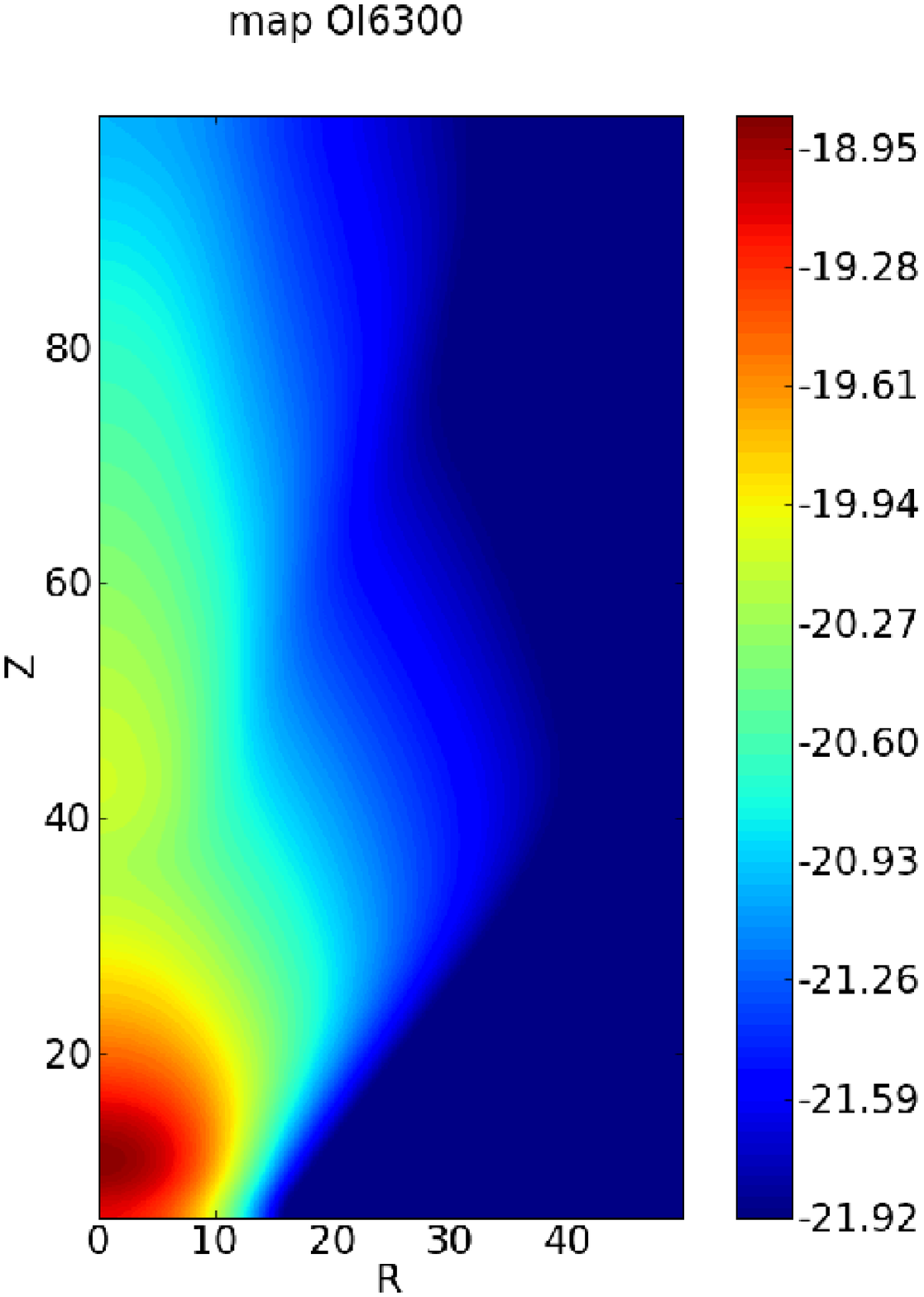}
  \includegraphics[width=0.24\textwidth]{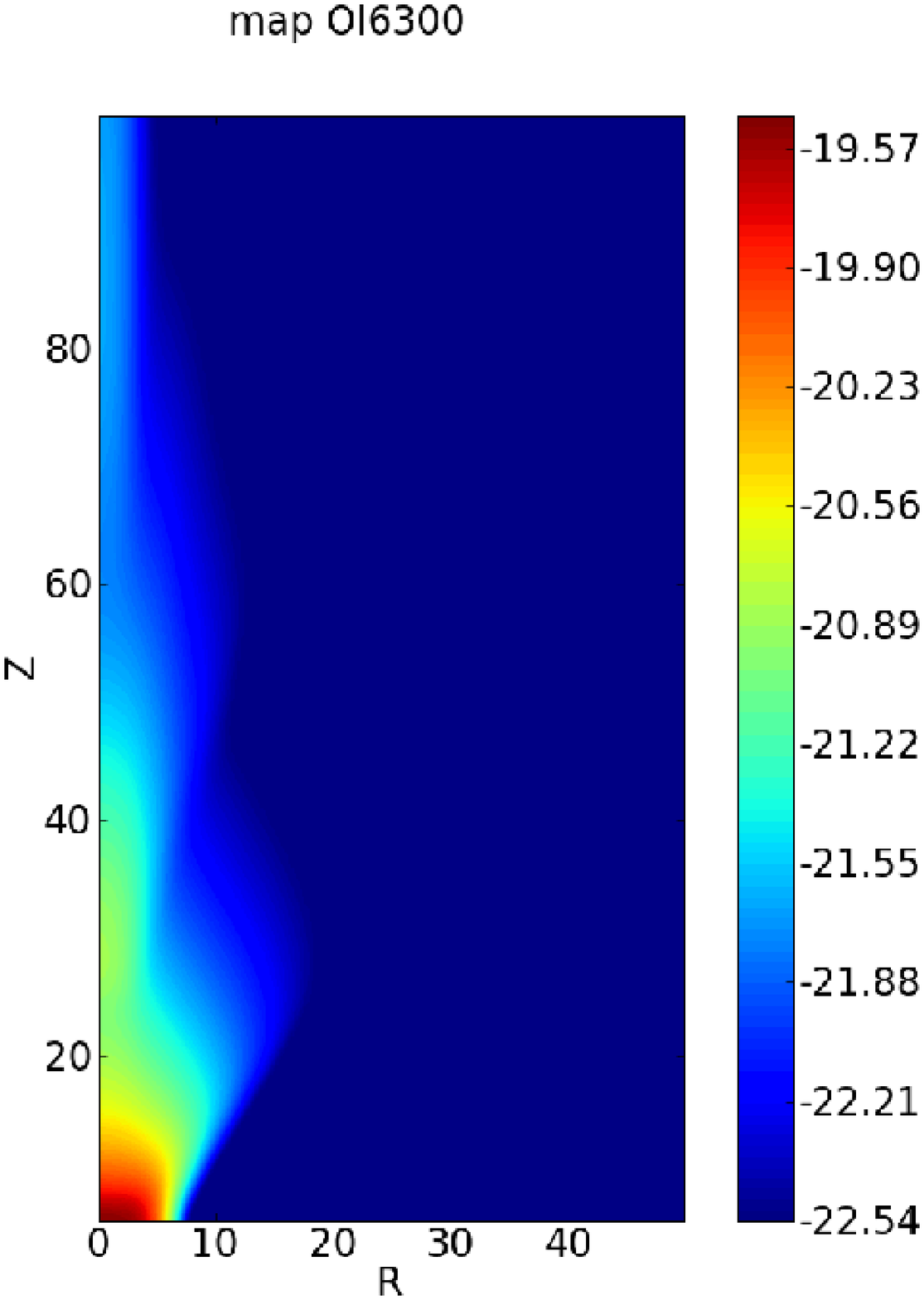}
  \includegraphics[width=0.24\textwidth]{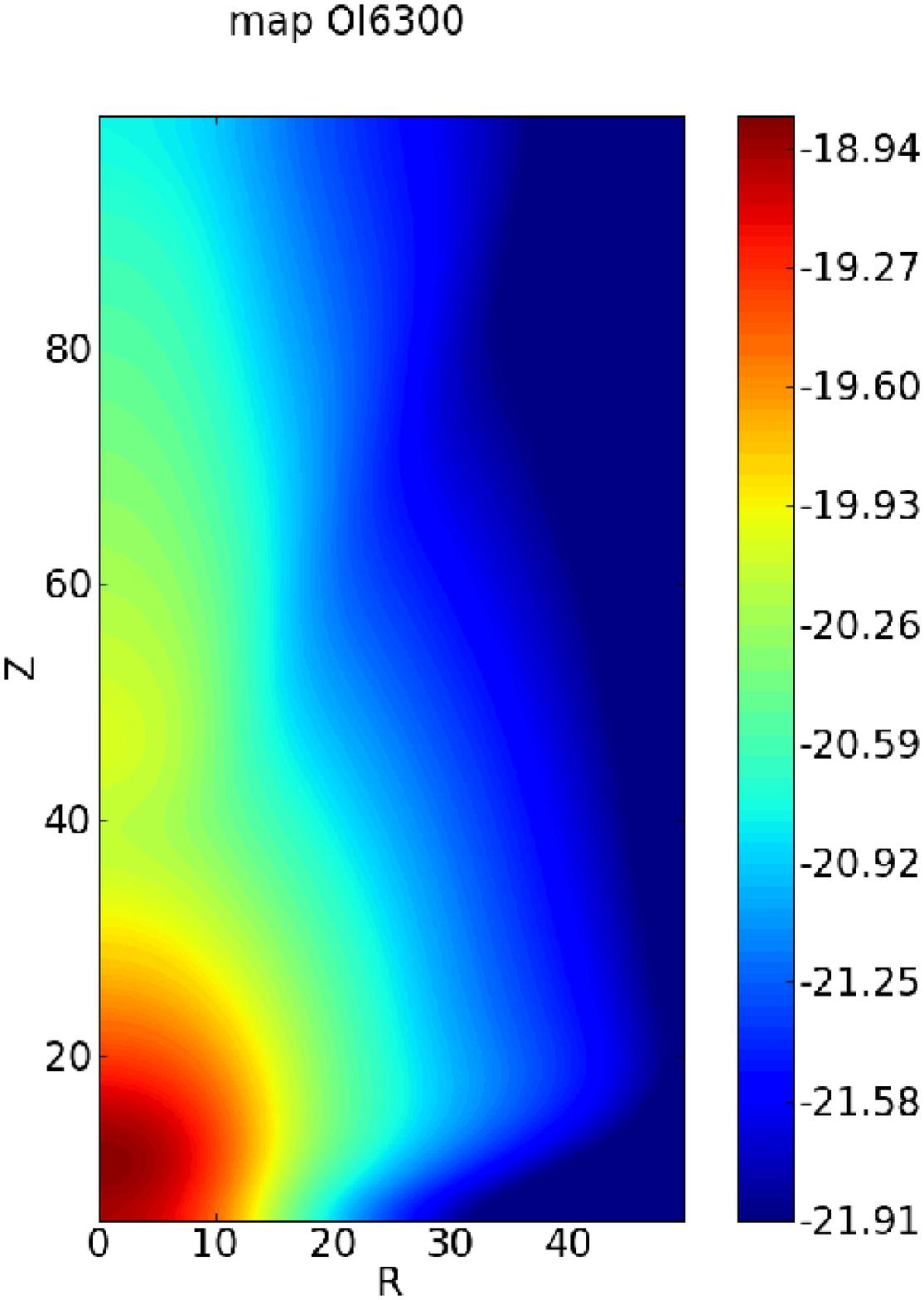}
  \includegraphics[width=0.24\textwidth]{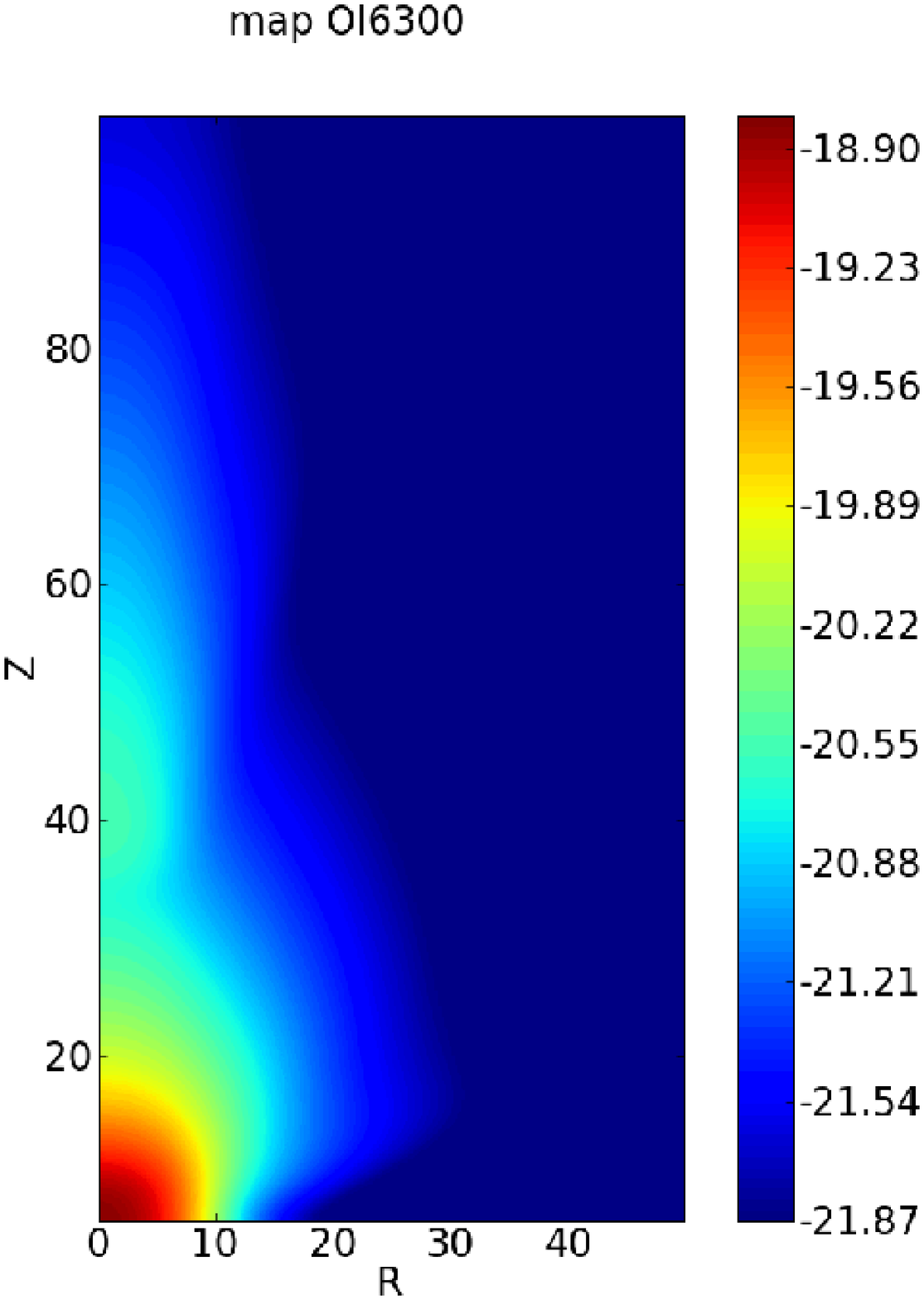}
  \caption{Synthetic emission maps of the [OI] $\lambda$6300 line, convolved 
    with a Gaussian PSF with a FWHM of 15 AU, for model SC1a and runs 
    (500, 600, 0.2), (500, 600, 0.5), (500, 1000, 0.5), (500, 1000, 0.8).}
  \label{Fig_emissmaps_all2}
\end{figure*}
\begin{figure*}[!bt]
  \centering
  \includegraphics[width=0.24\textwidth]{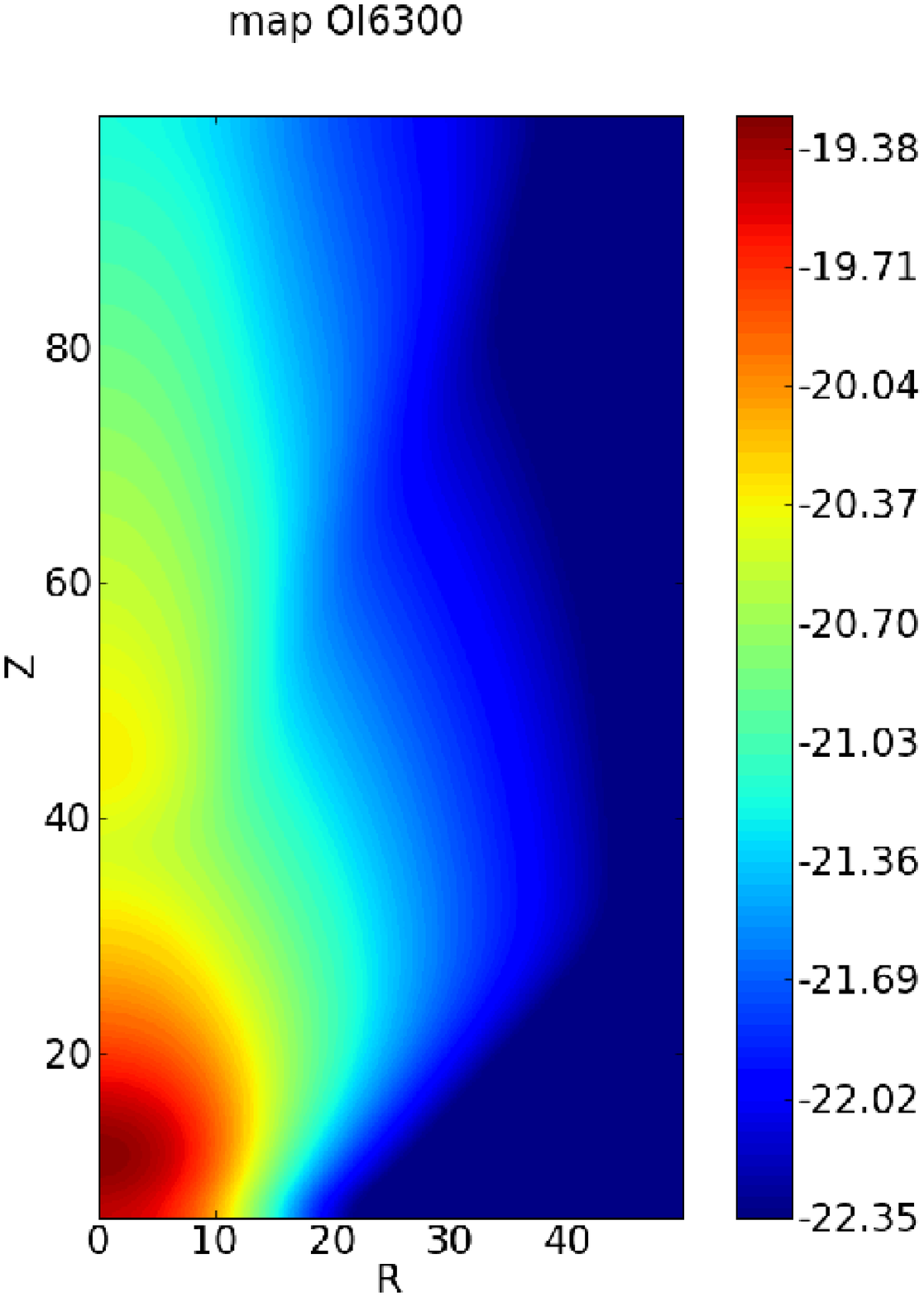}
  \includegraphics[width=0.24\textwidth]{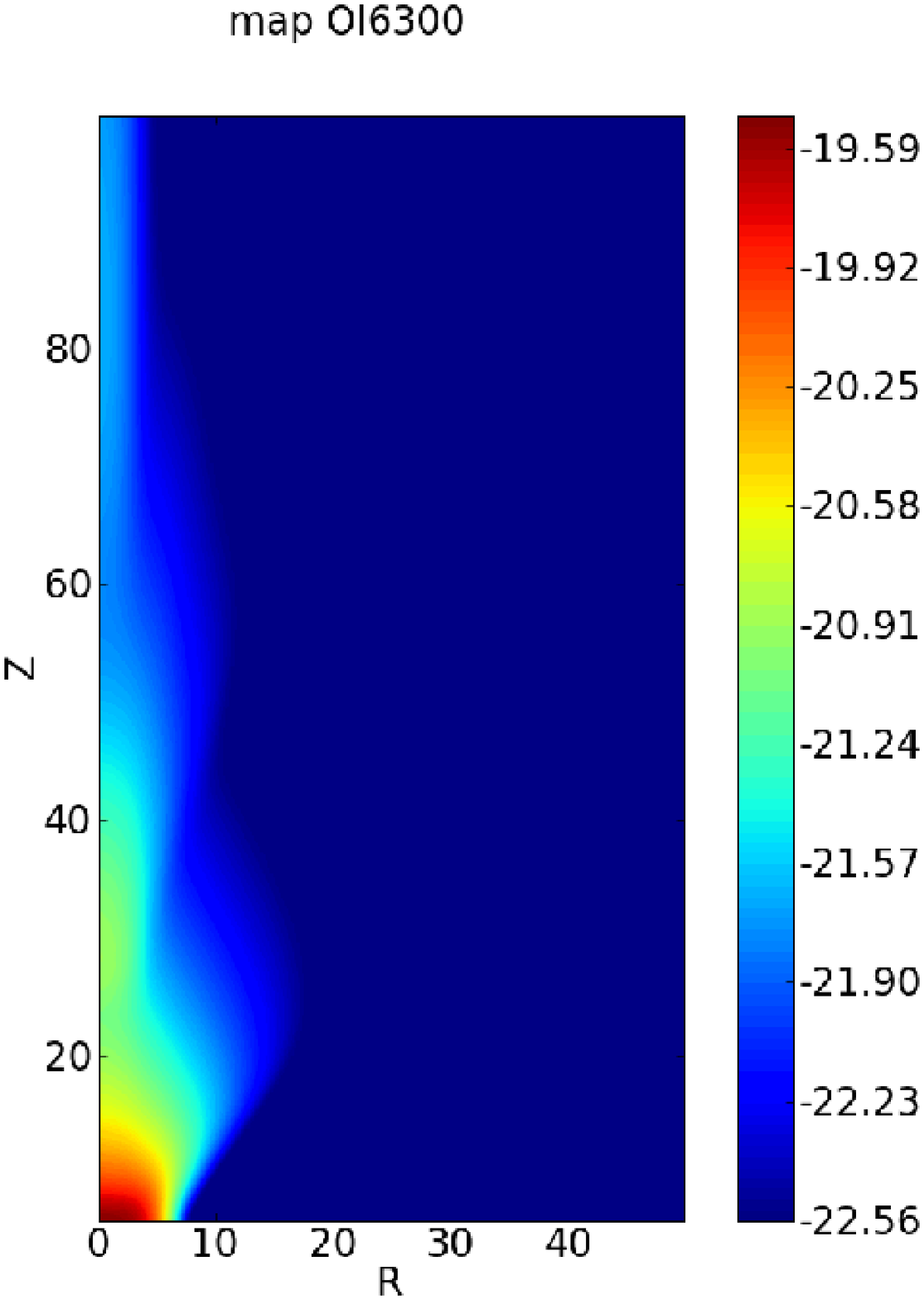}
  \includegraphics[width=0.24\textwidth]{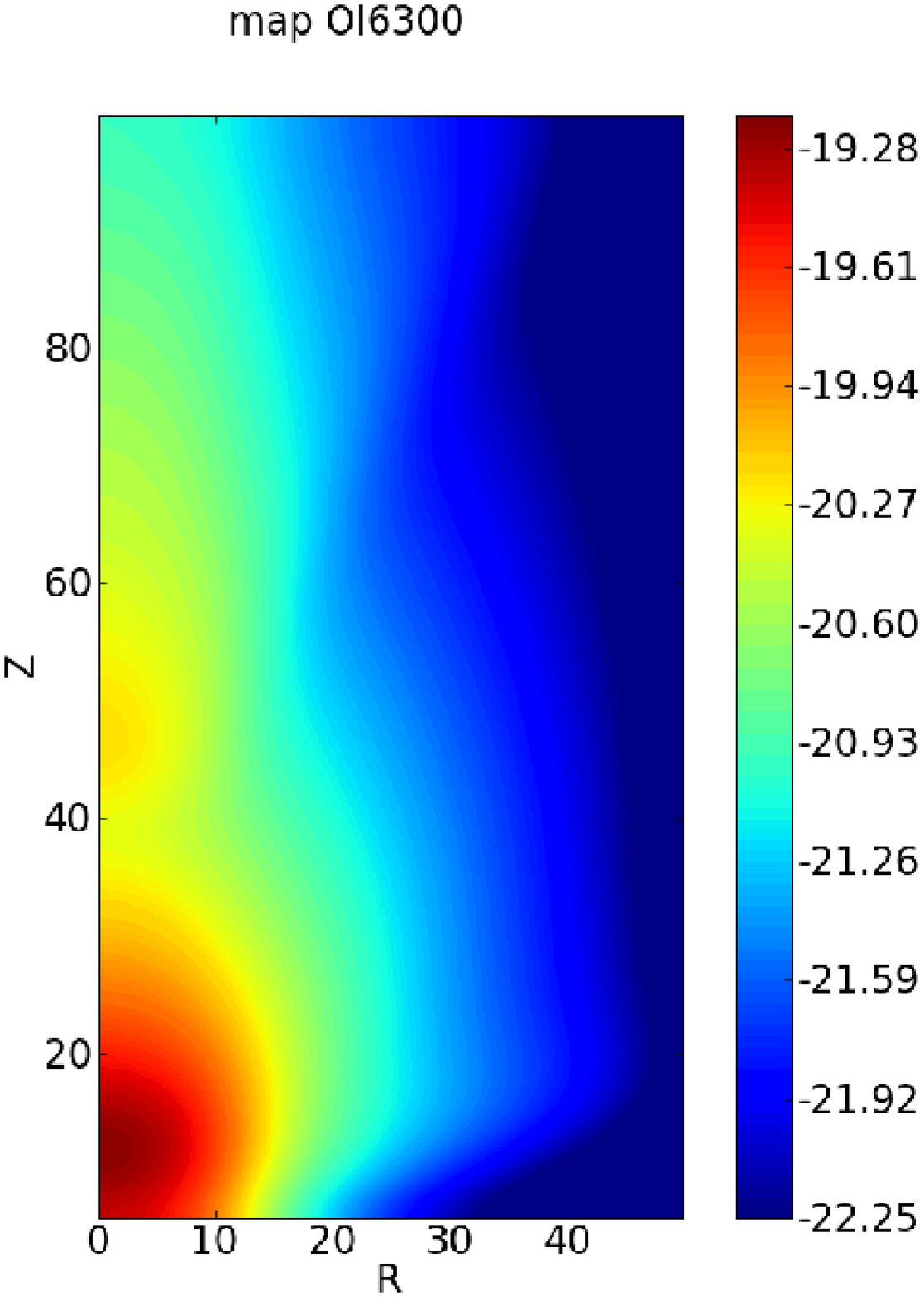}
  \includegraphics[width=0.24\textwidth]{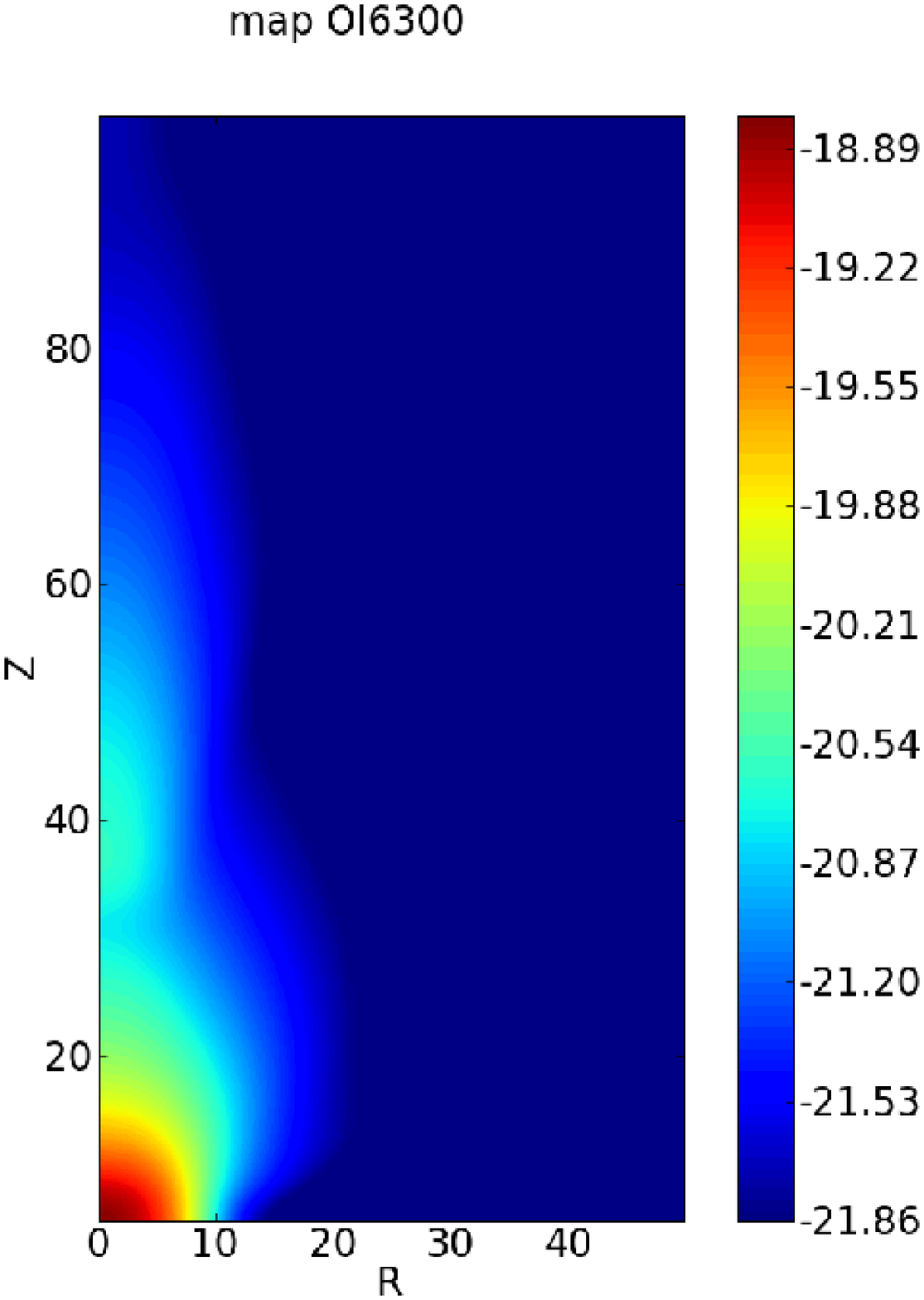}
  \caption{Synthetic emission maps of the [OI] $\lambda$6300 line, convolved 
    with a Gaussian PSF with a FWHM of 15 AU, for model SC1b and runs 
    (500, 600, 0.2), (500, 600, 0.5), (500, 1000, 0.5), (500, 1000, 0.8).}
  \label{Fig_emissmaps_all3}
\end{figure*}
\begin{figure*}[!bt]
  \centering
  \includegraphics[width=0.19\textwidth]{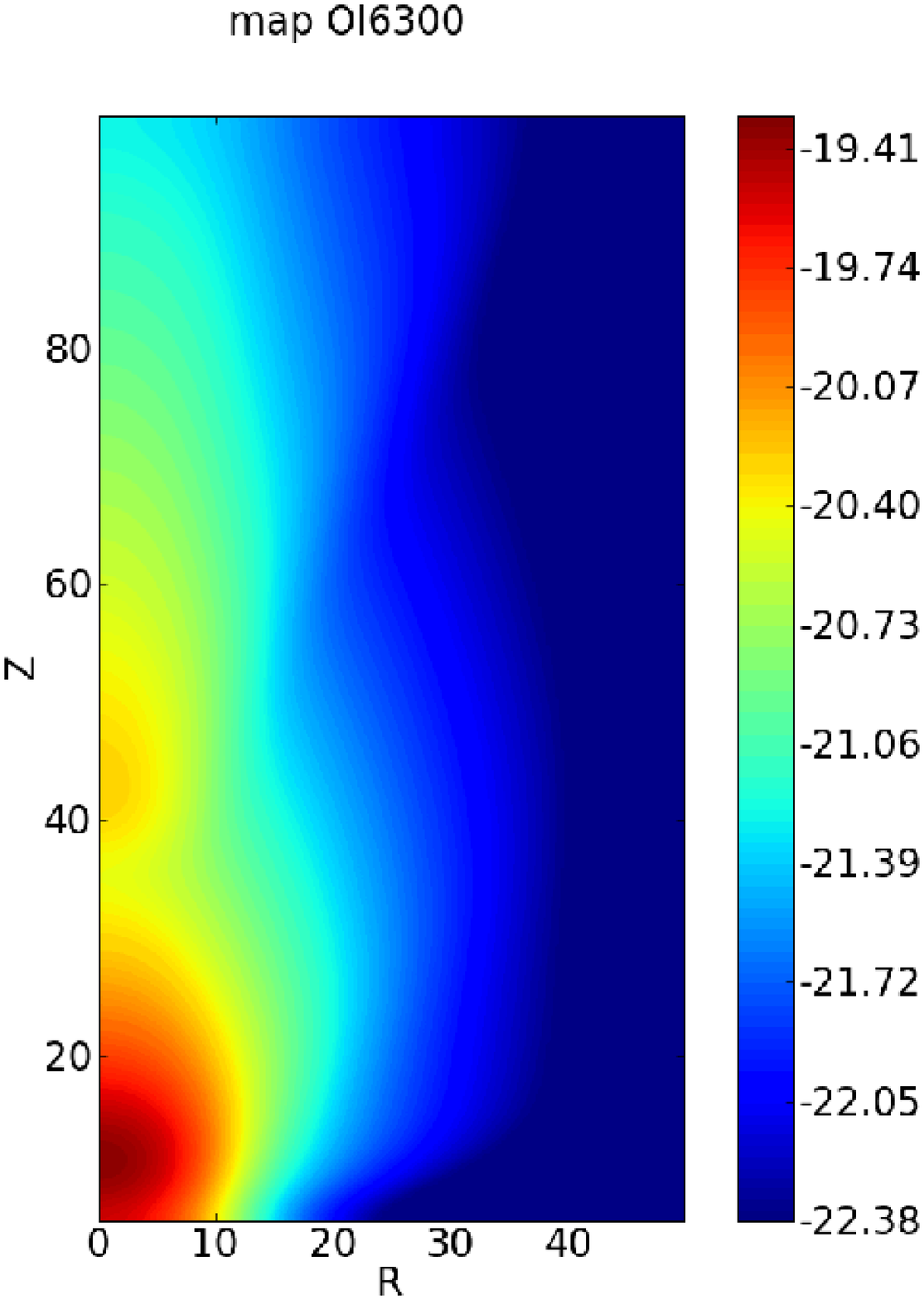}
  \includegraphics[width=0.19\textwidth]{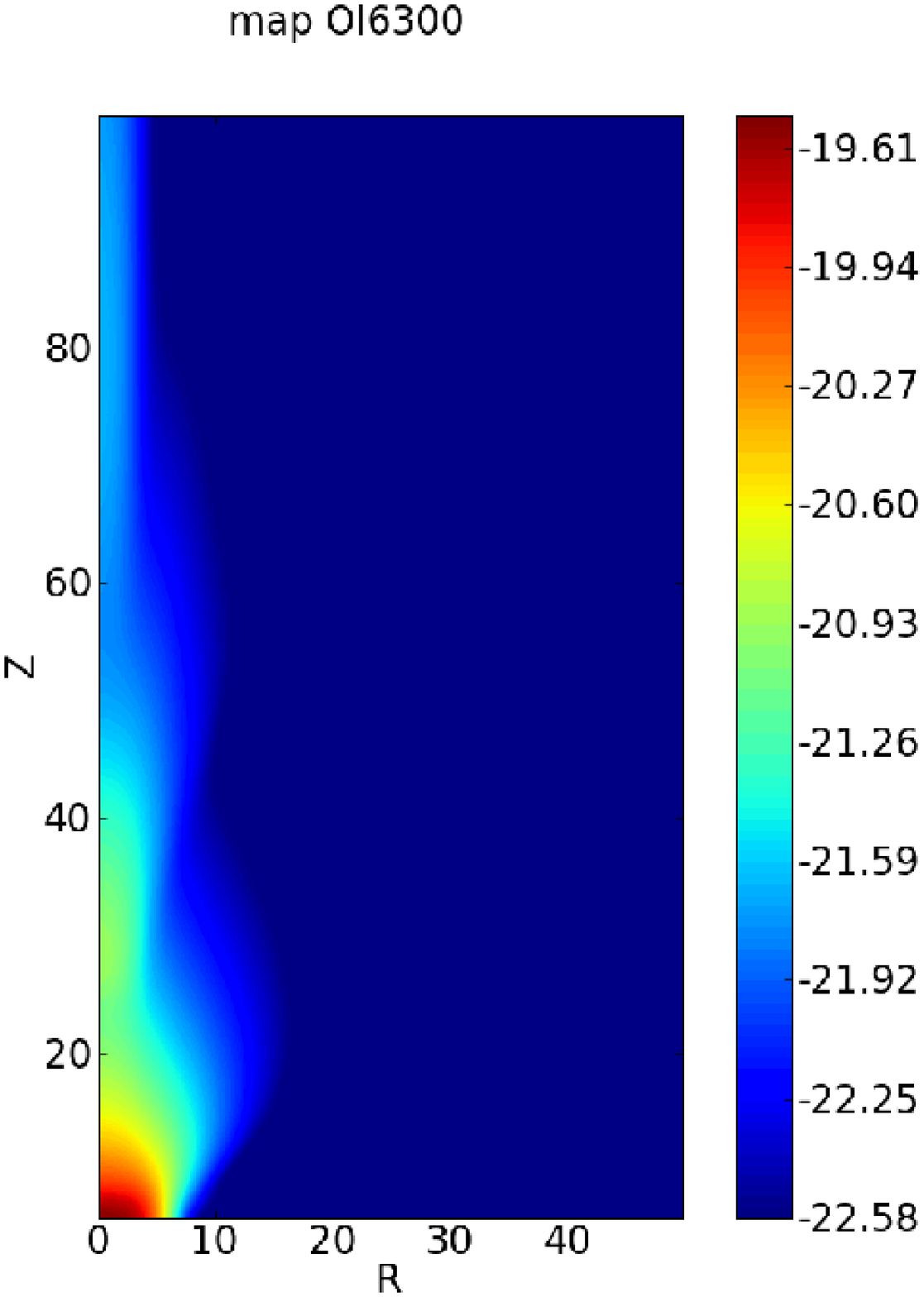}
  \includegraphics[width=0.19\textwidth]{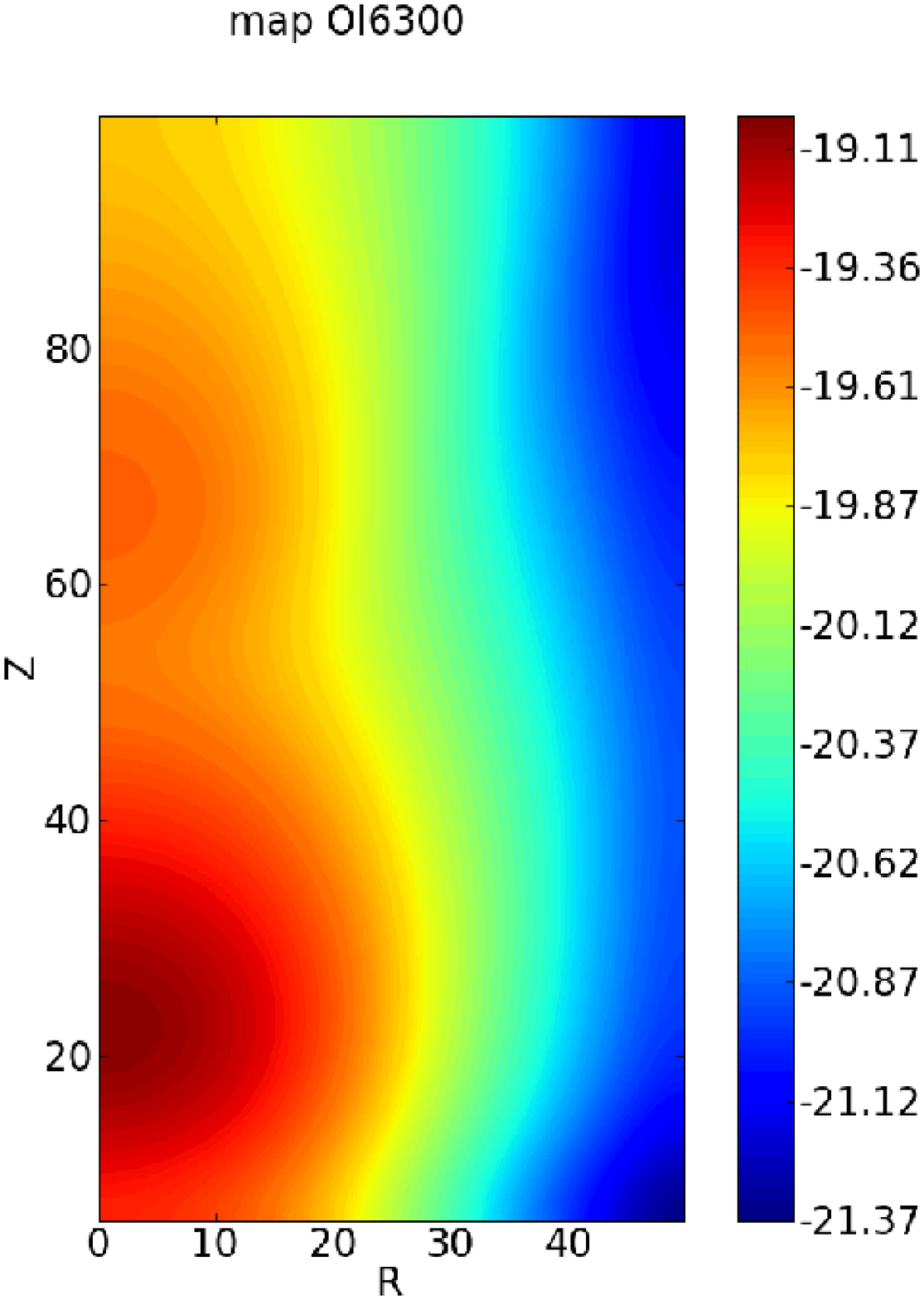}
  \includegraphics[width=0.19\textwidth]{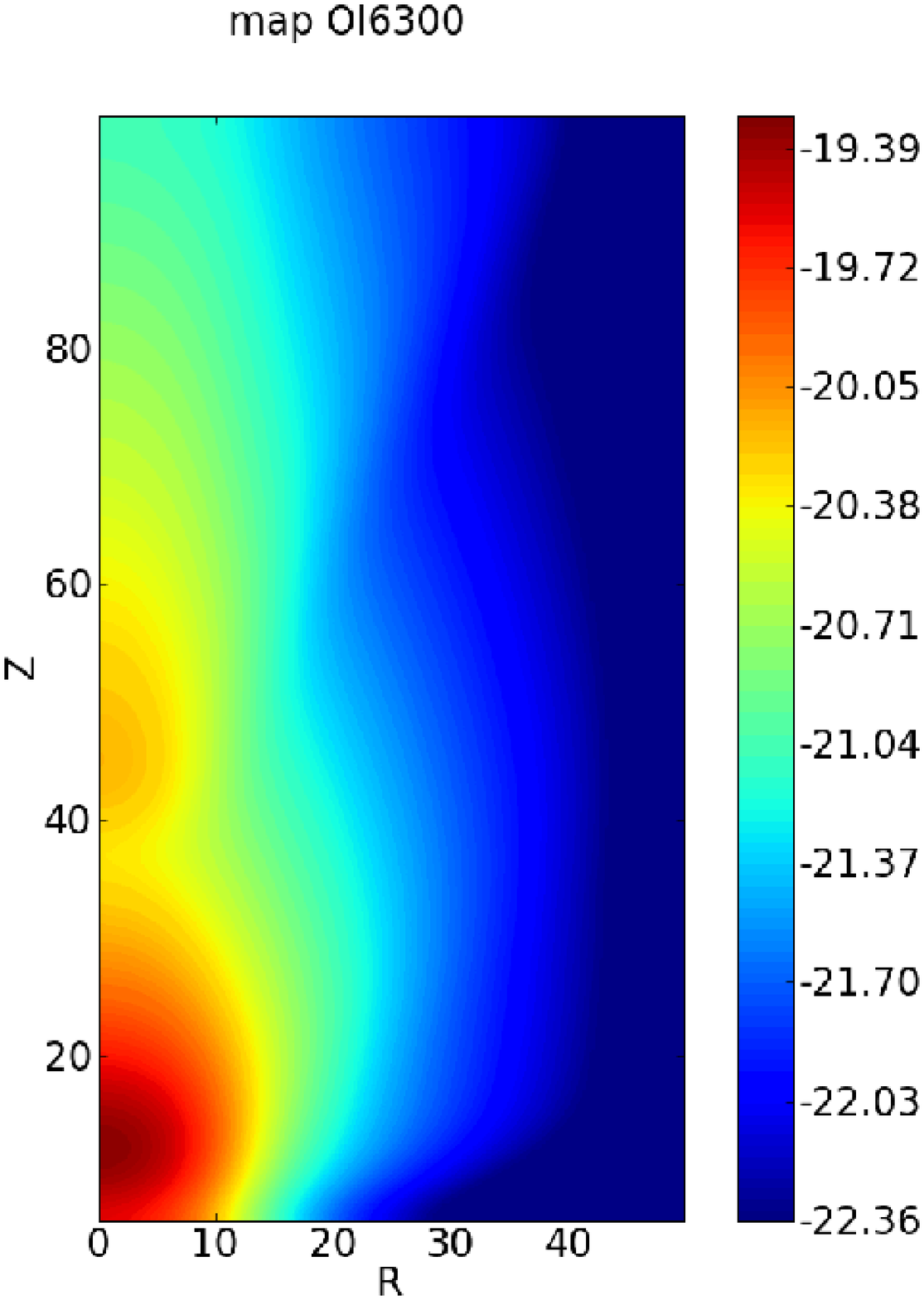}
  \includegraphics[width=0.19\textwidth]{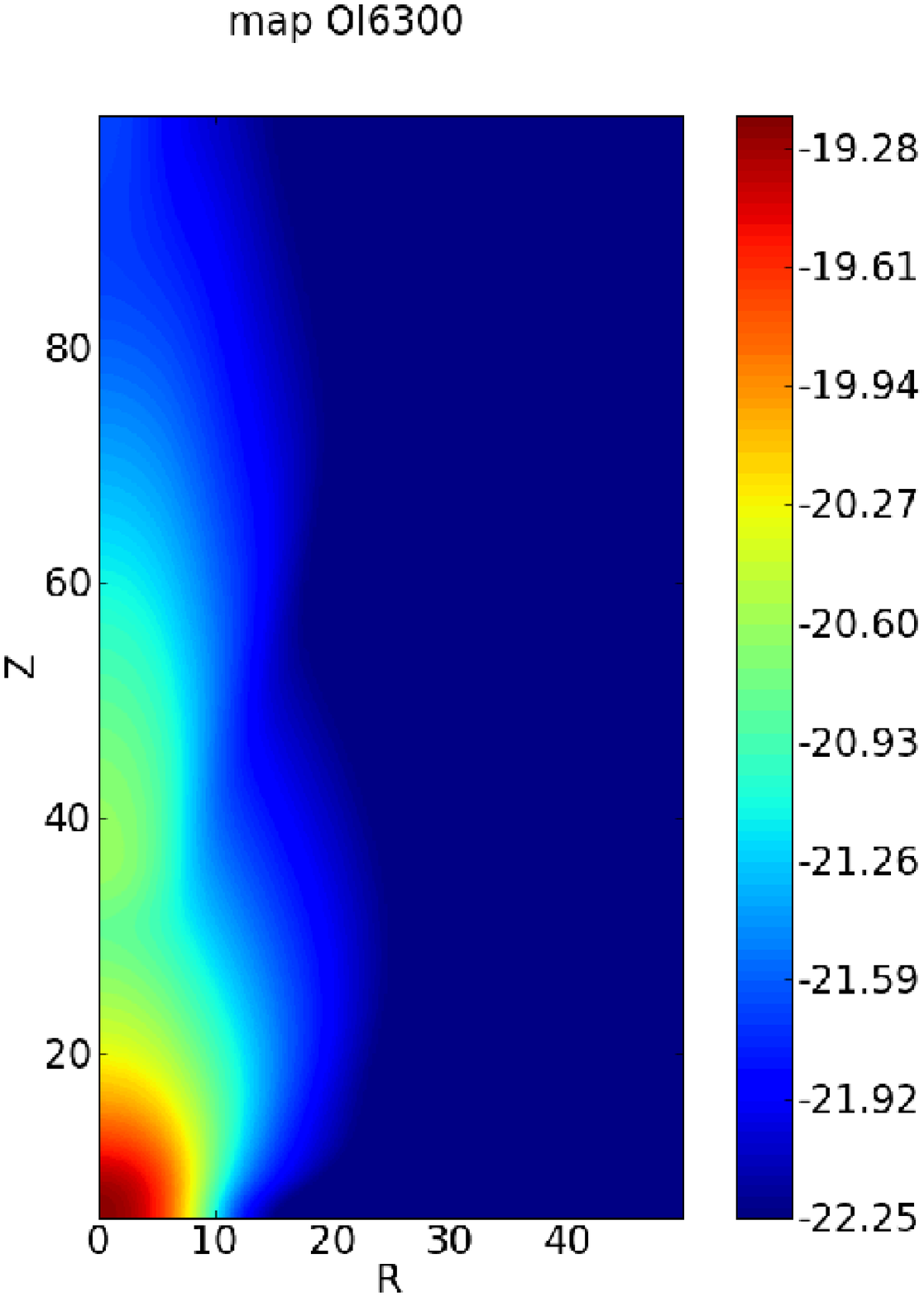}
  \caption{Synthetic emission maps of the [OI] $\lambda$6300 line, convolved 
    with a Gaussian PSF with a FWHM of 15 AU, for model SC1c and runs 
    (500, 600, 0.2), (500, 600, 0.5), (500, 1000, 0.2), (500, 1000, 0.5), 
    (500, 1000, 0.8).}
  \label{Fig_emissmaps_all4}
\end{figure*}
\begin{figure*}[!bt]
  \centering
  \includegraphics[width=0.19\textwidth]{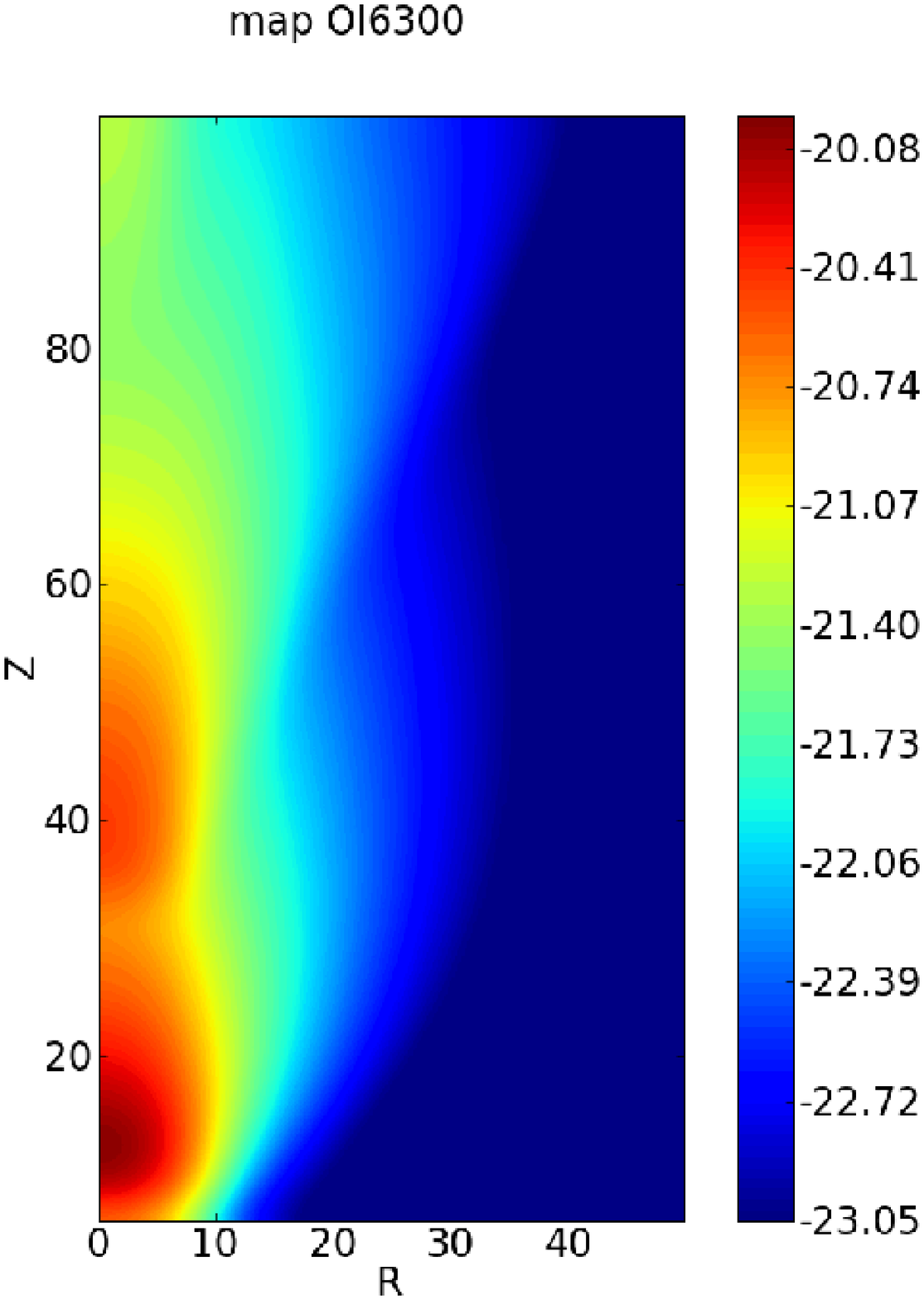}
  \includegraphics[width=0.19\textwidth]{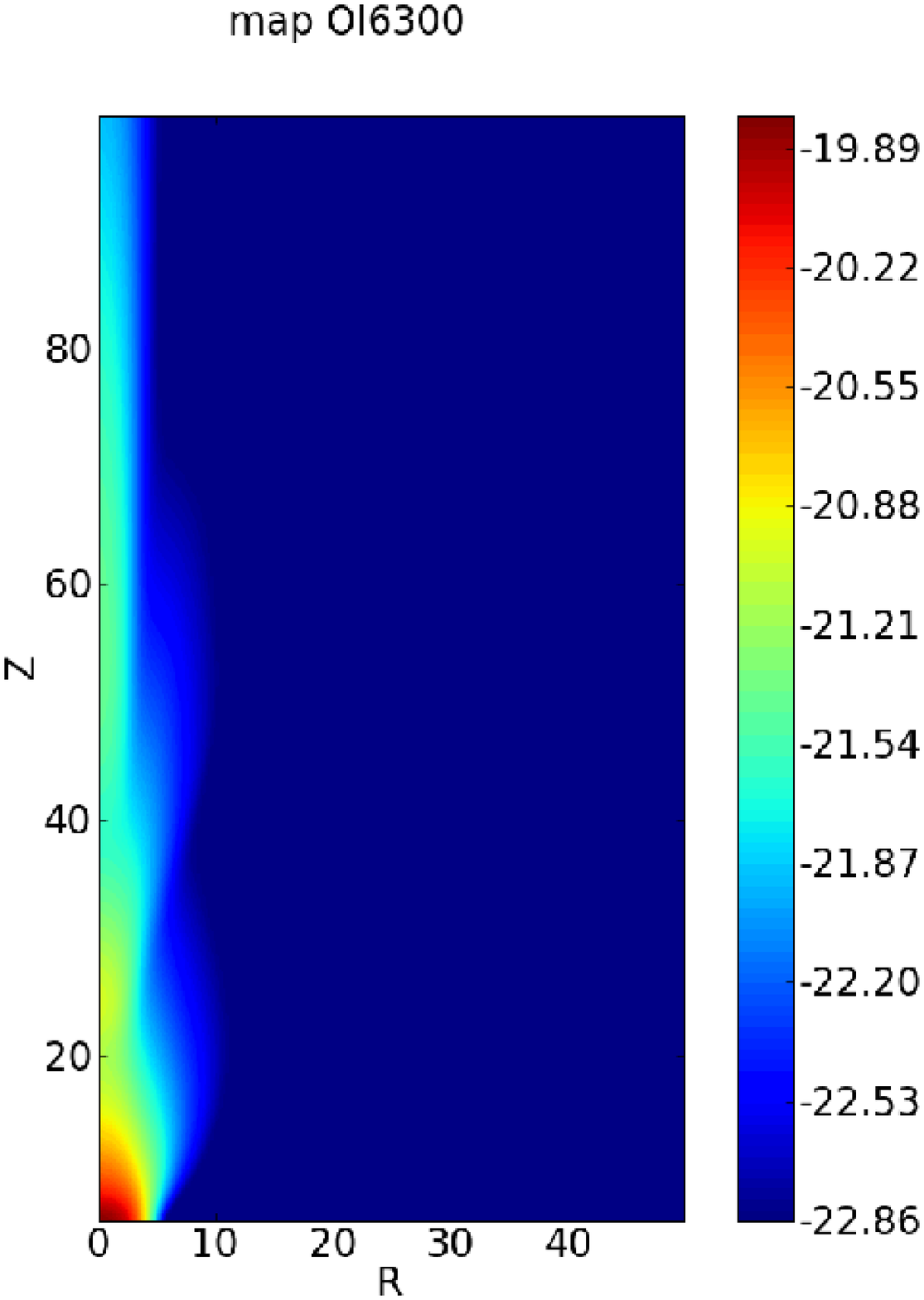}
  \includegraphics[width=0.19\textwidth]{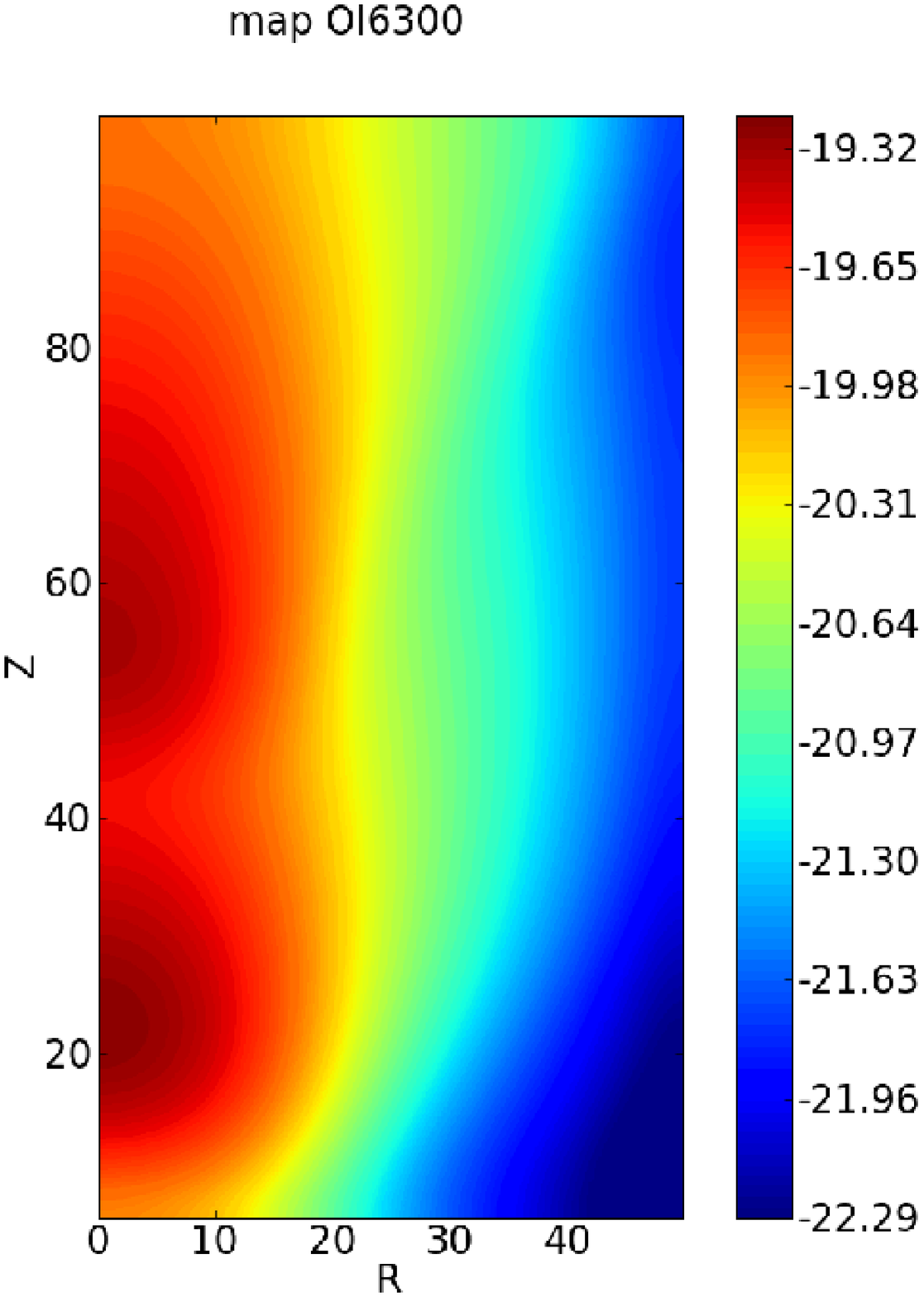}
  \includegraphics[width=0.19\textwidth]{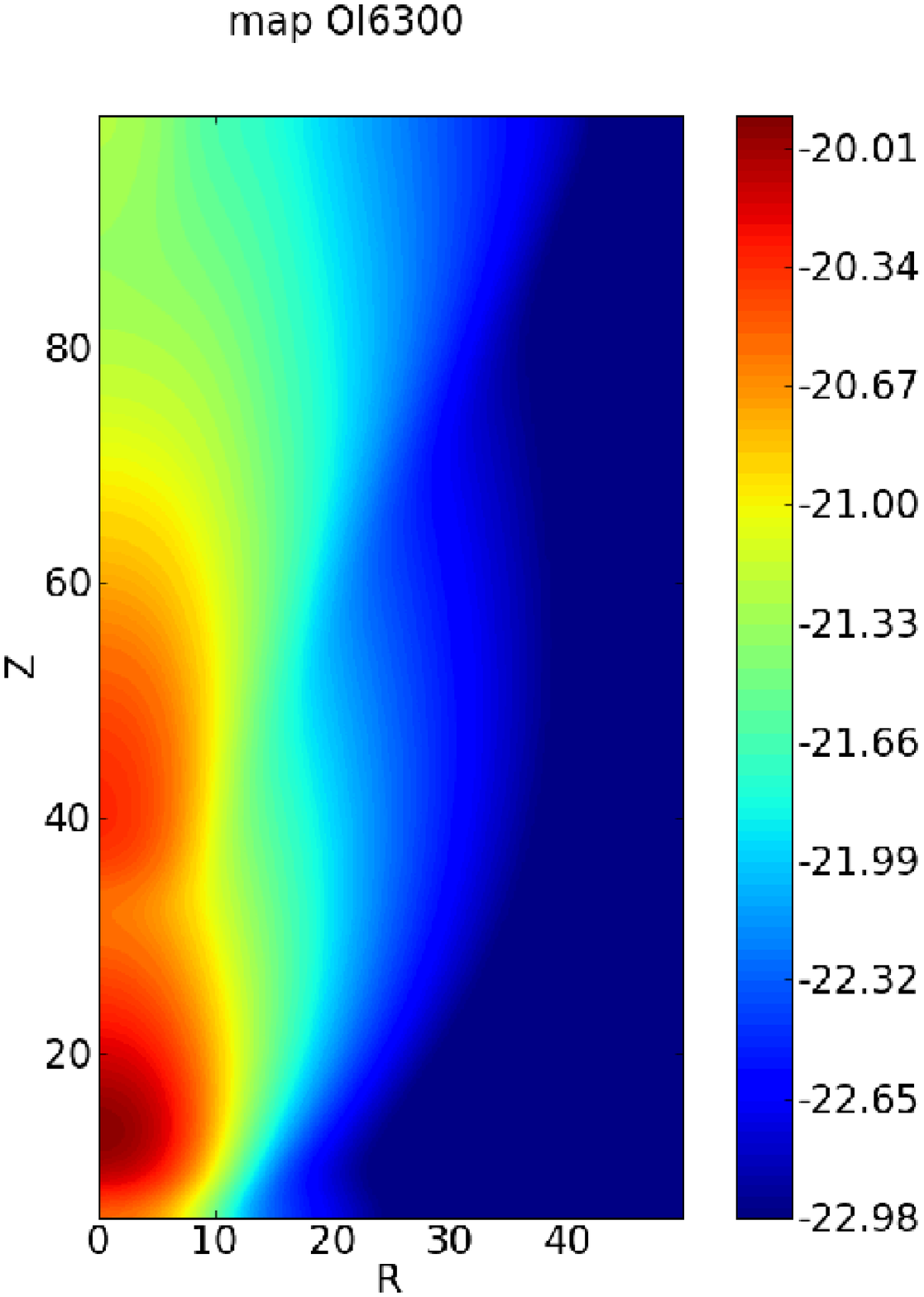}
  \includegraphics[width=0.19\textwidth]{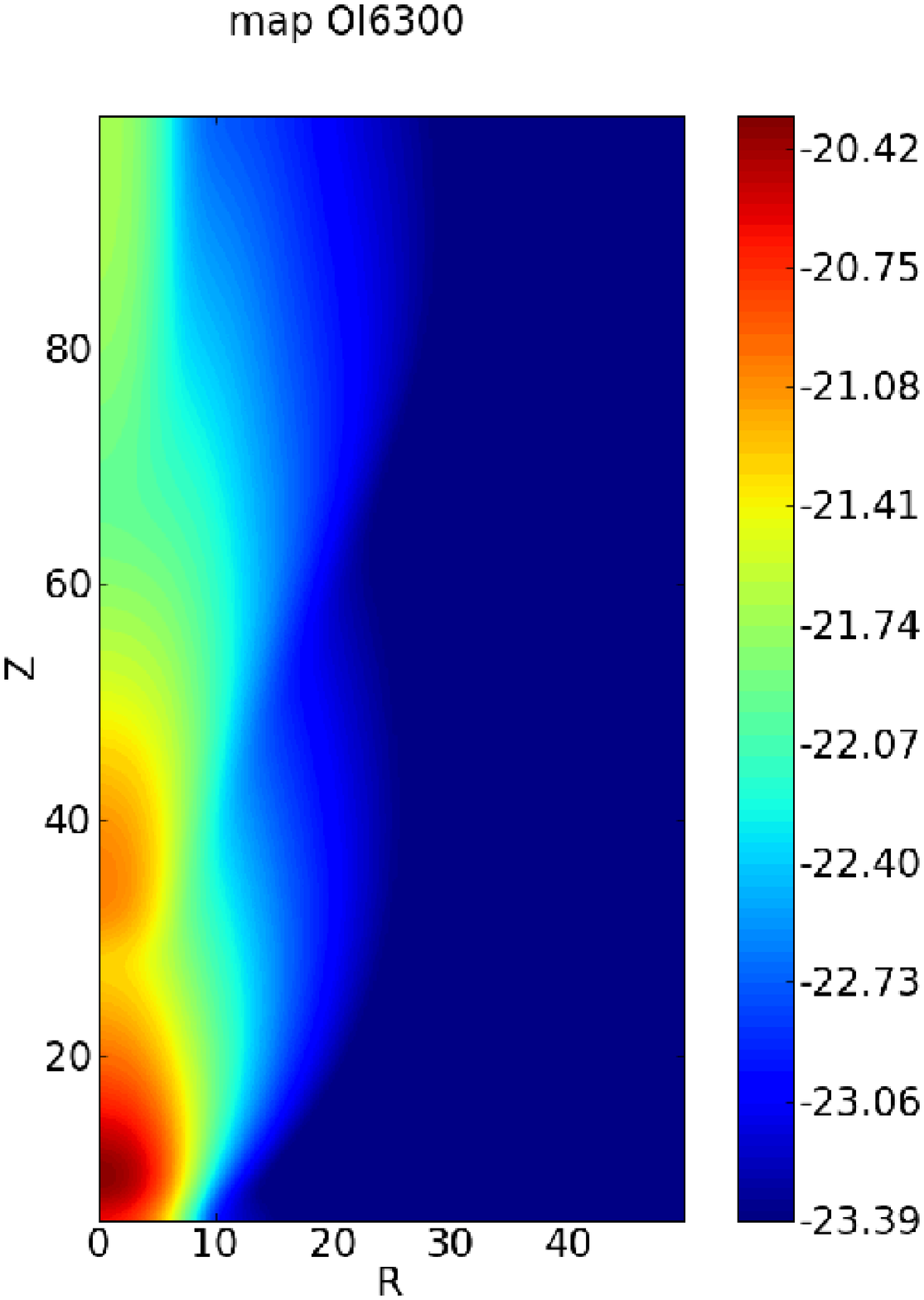}
  \caption{Synthetic emission maps of the [OI] $\lambda$6300 line, convolved 
    with a Gaussian PSF with a FWHM of 15 AU, for model SC1d and runs 
    (500, 600, 0.2), (500, 600, 0.5), (500, 1000, 0.2), (500, 1000, 0.5), 
    (500, 1000, 0.8).}
  \label{Fig_emissmaps_all5}
\end{figure*}
\begin{figure*}[!bt]
  \centering
  \includegraphics[width=0.19\textwidth]{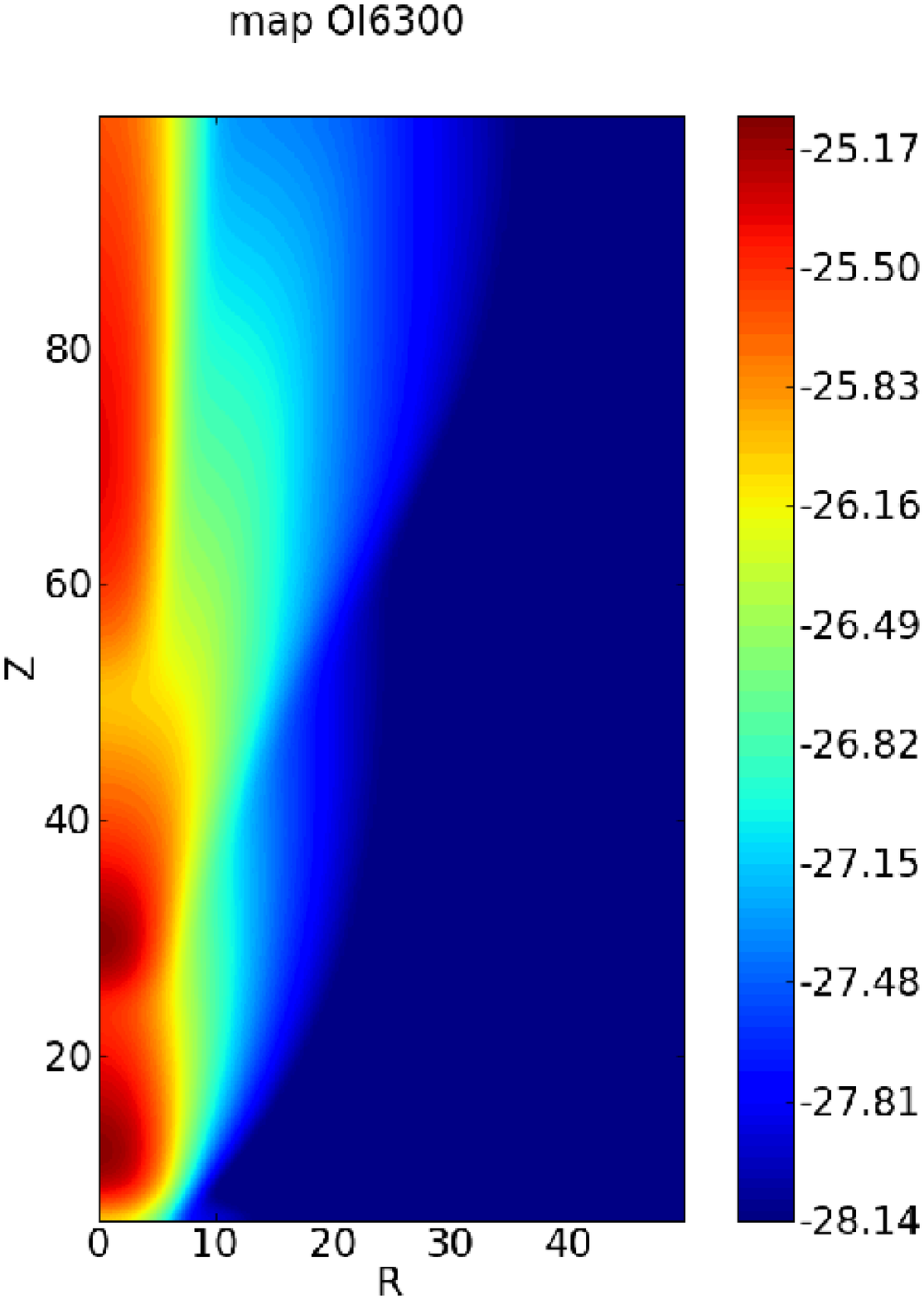}
  \includegraphics[width=0.19\textwidth]{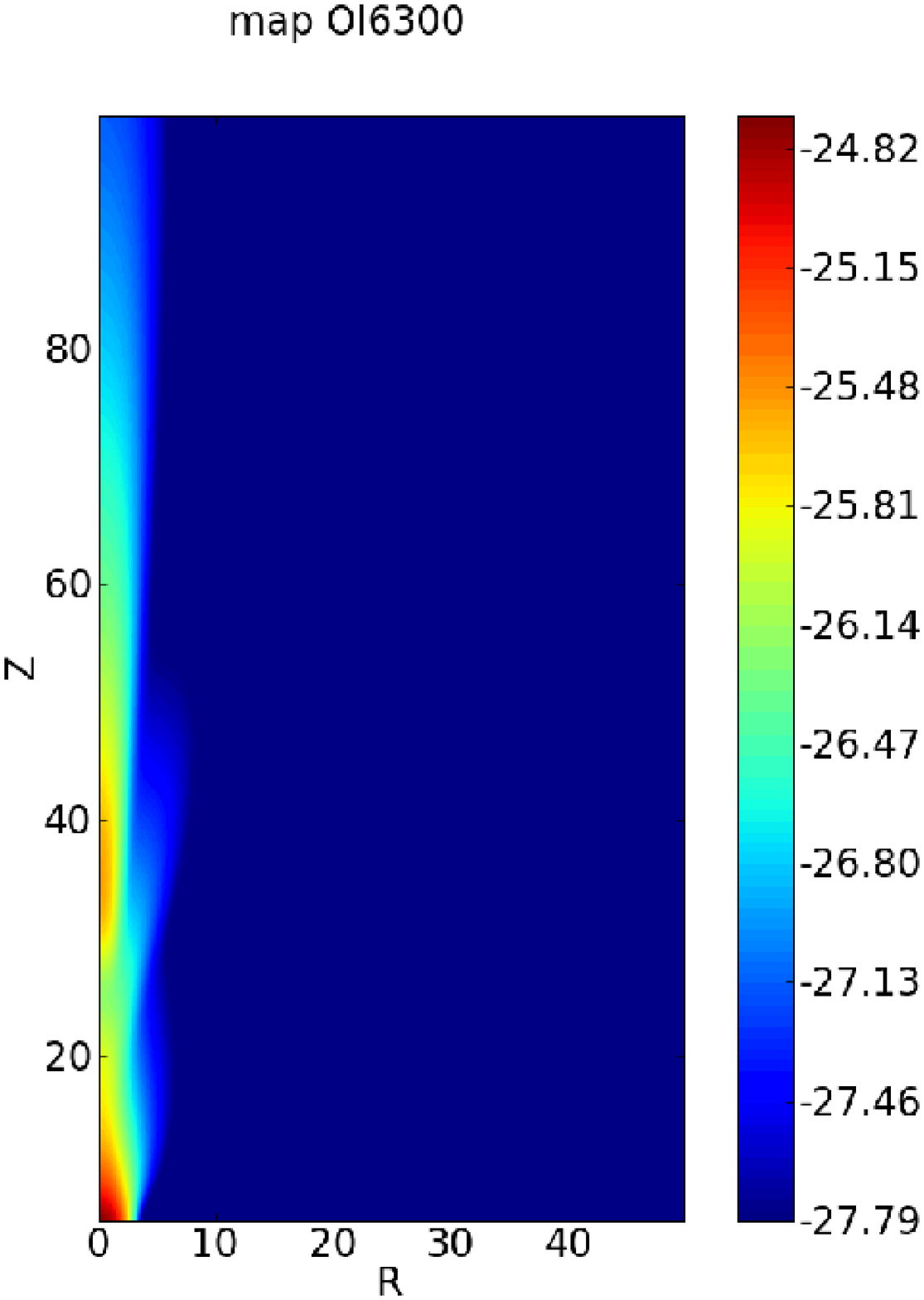}
  \includegraphics[width=0.19\textwidth]{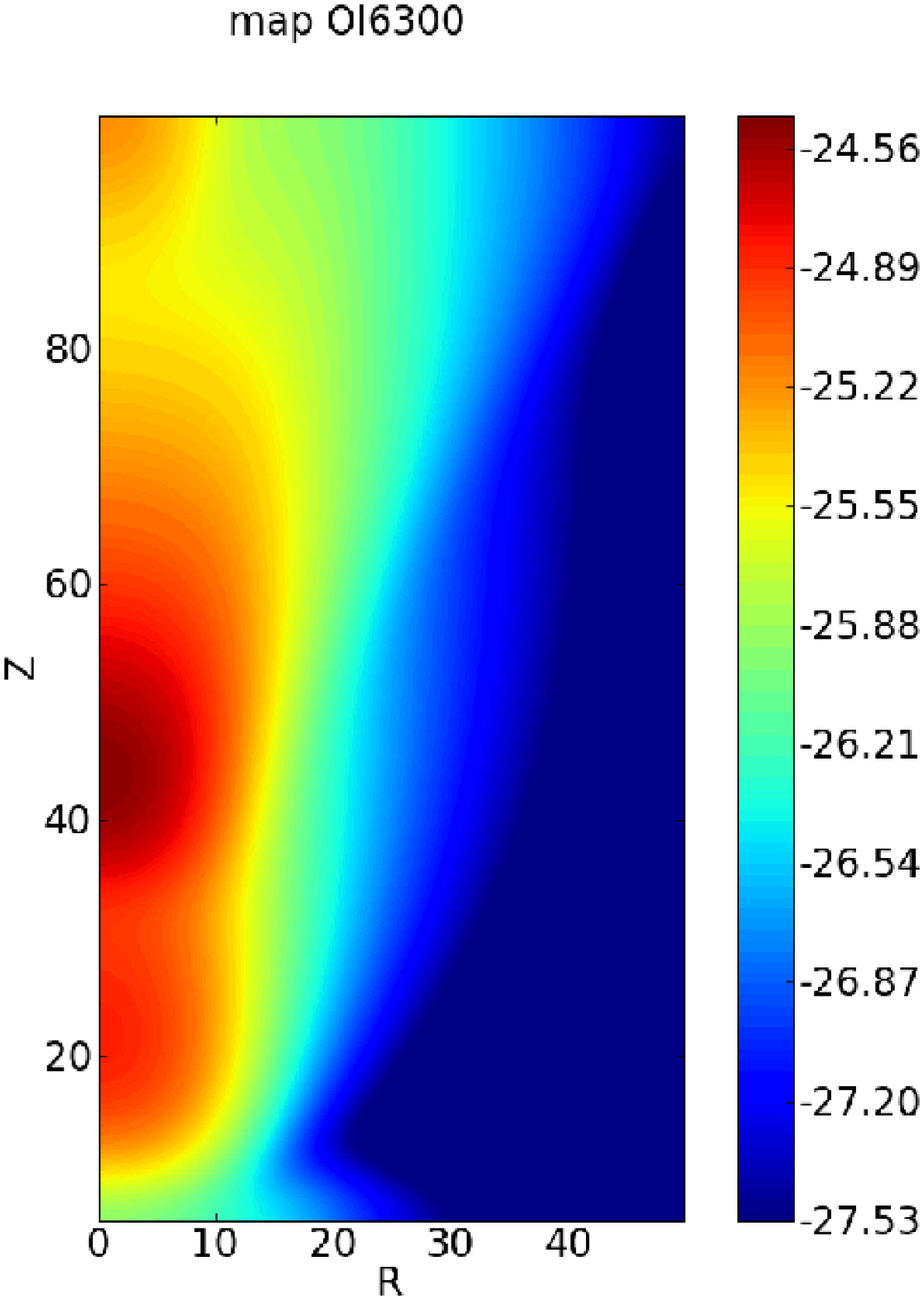}
  \includegraphics[width=0.19\textwidth]{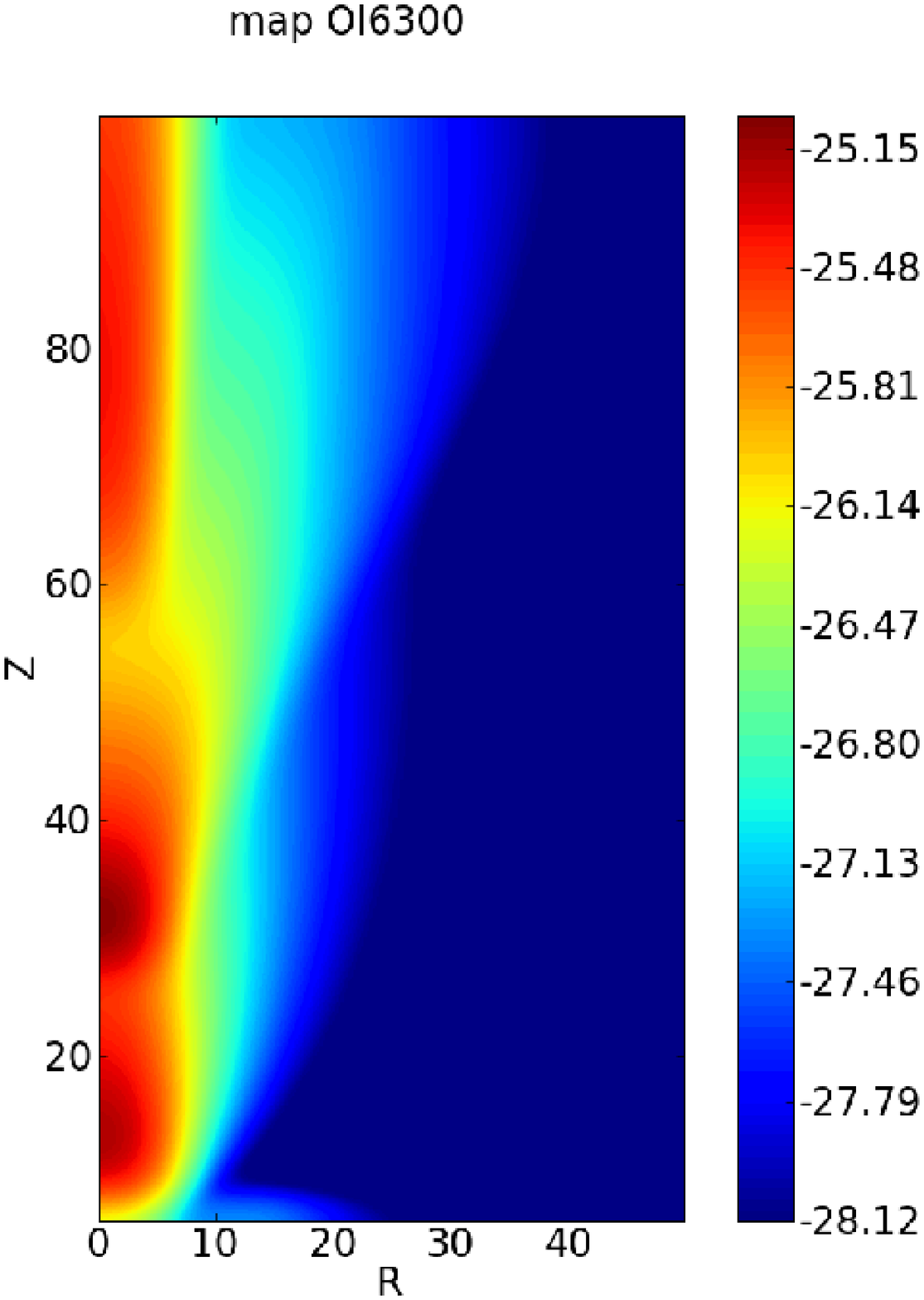}
  \includegraphics[width=0.19\textwidth]{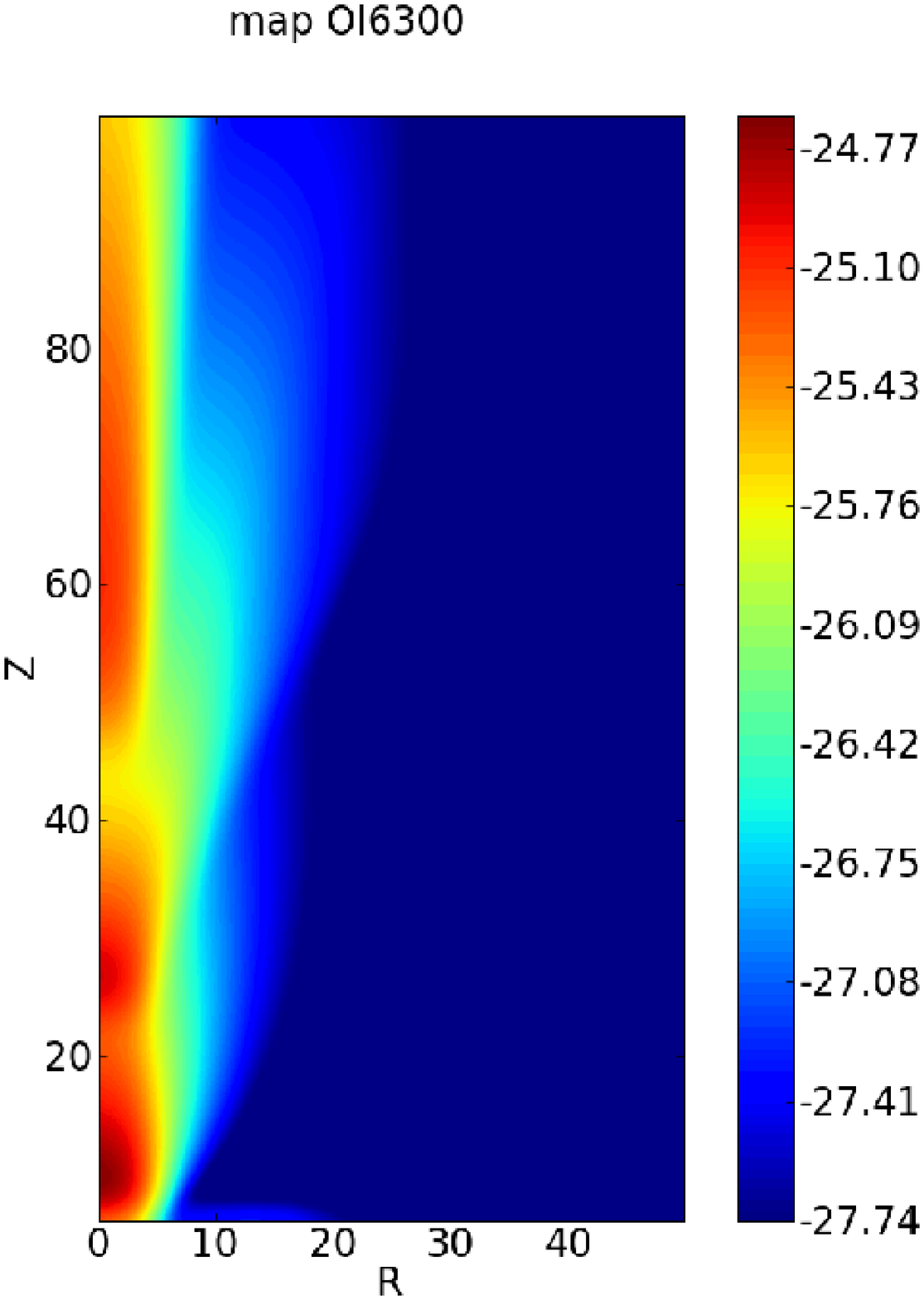}
  \caption{Synthetic emission maps of the [OI] $\lambda$6300 line, convolved 
    with a Gaussian PSF with a FWHM of 15 AU, for model SC1e and runs 
    (500, 600, 0.2), (500, 600, 0.5), (500, 1000, 0.2), (500, 1000, 0.5), 
    (500, 1000, 0.8).}
  \label{Fig_emissmaps_all6}
\end{figure*}
\begin{figure*}[!bt]
  \centering
  \includegraphics[width=0.19\textwidth]{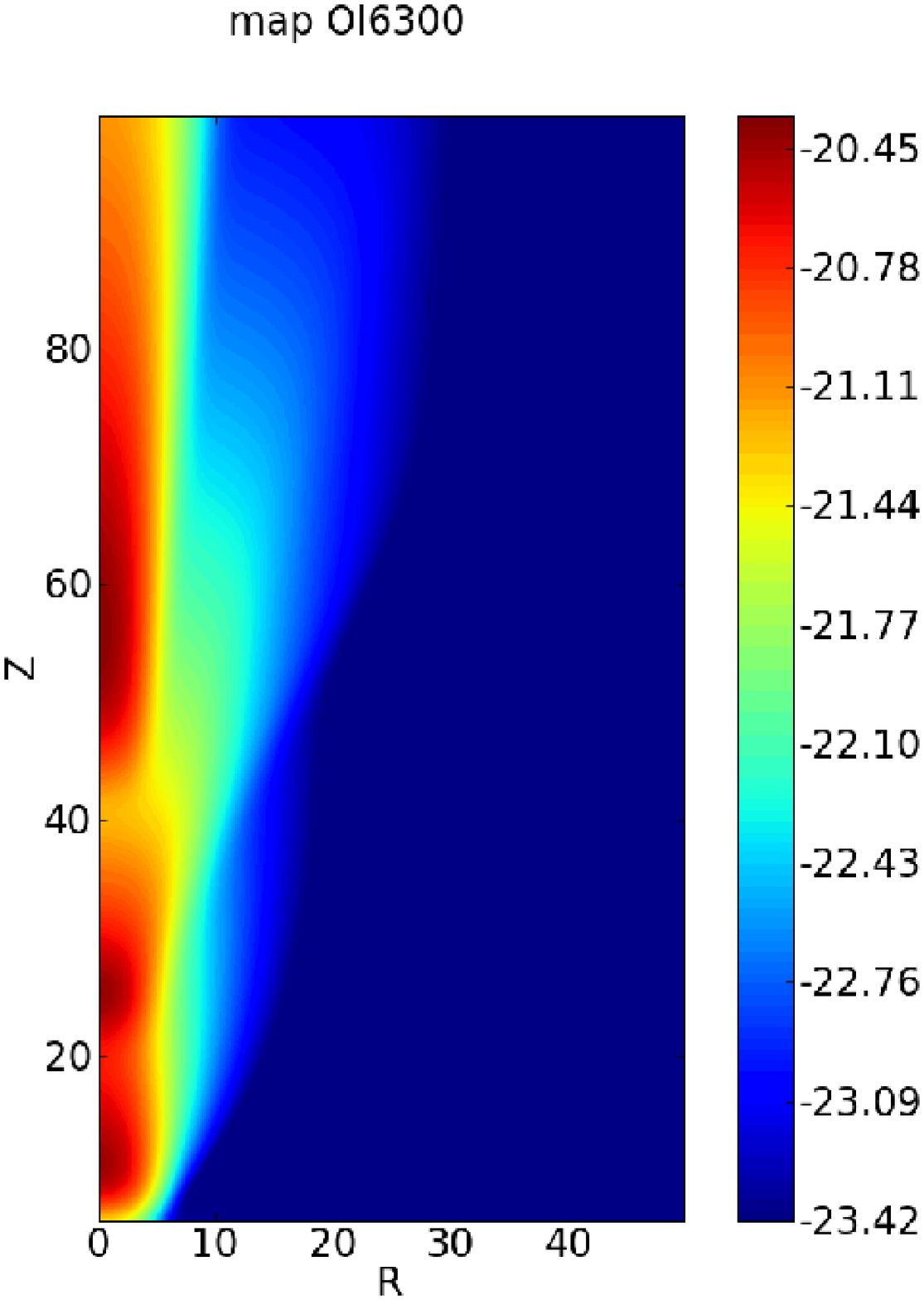}
  \includegraphics[width=0.19\textwidth]{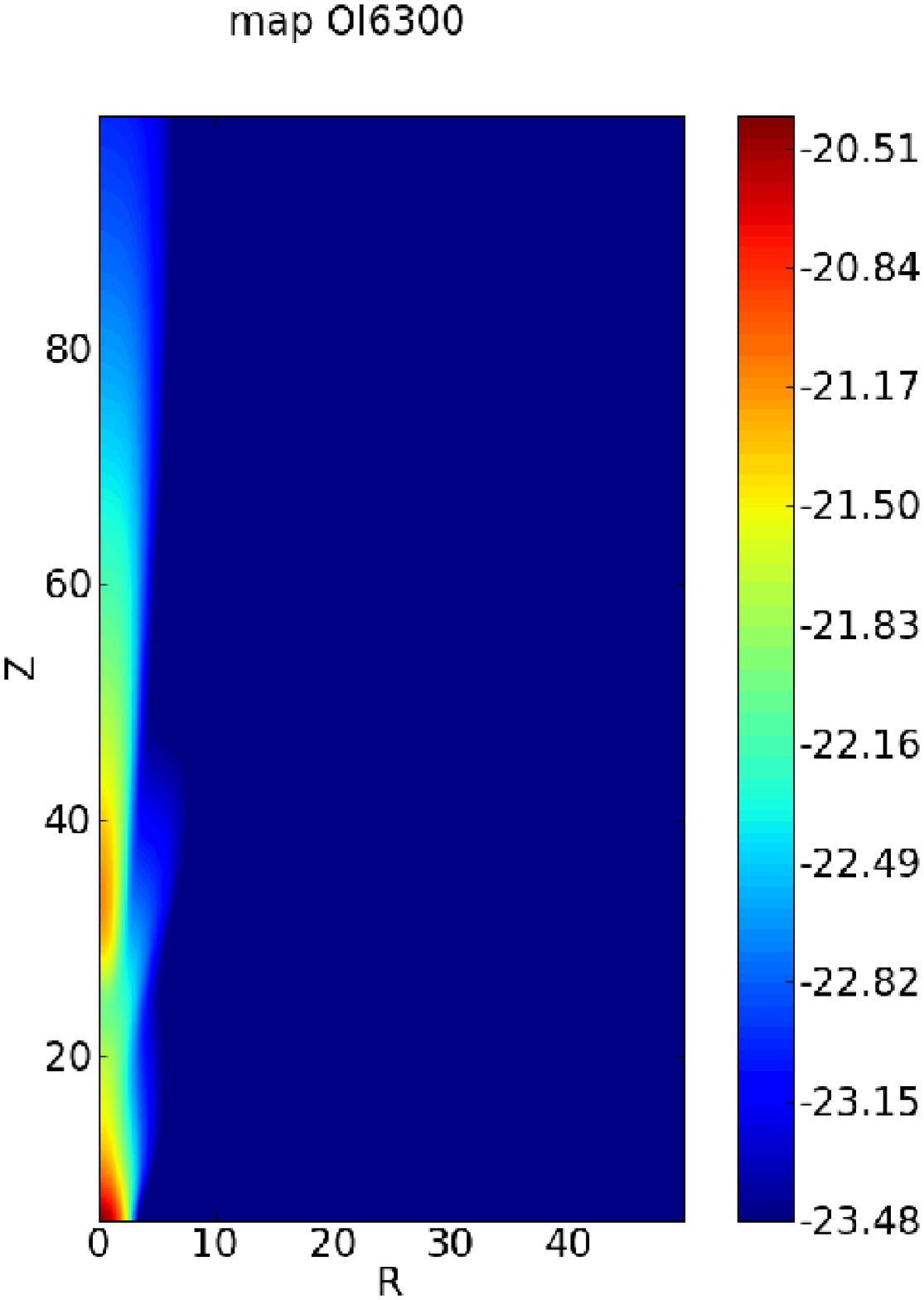}
  \includegraphics[width=0.19\textwidth]{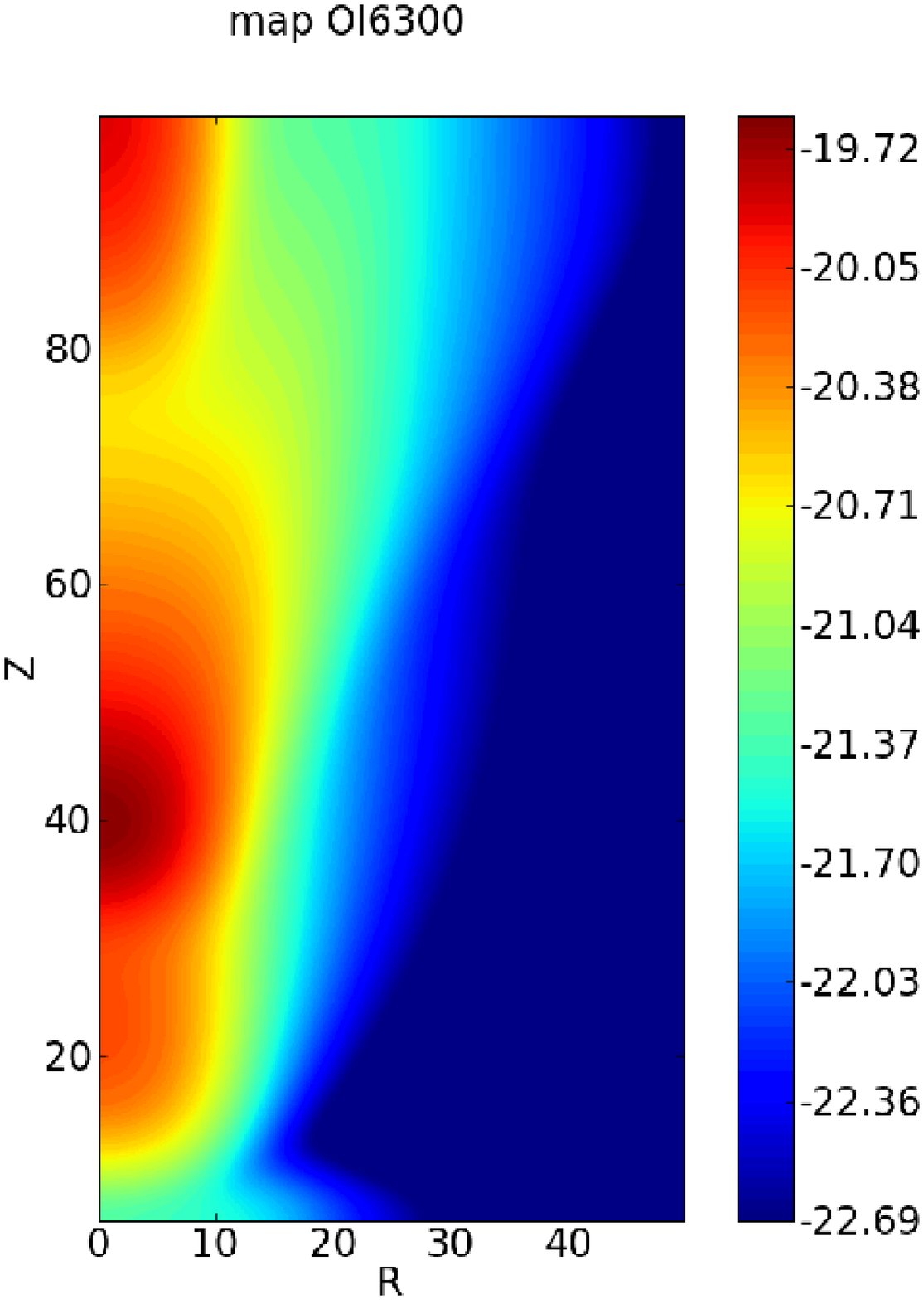}
  \includegraphics[width=0.19\textwidth]{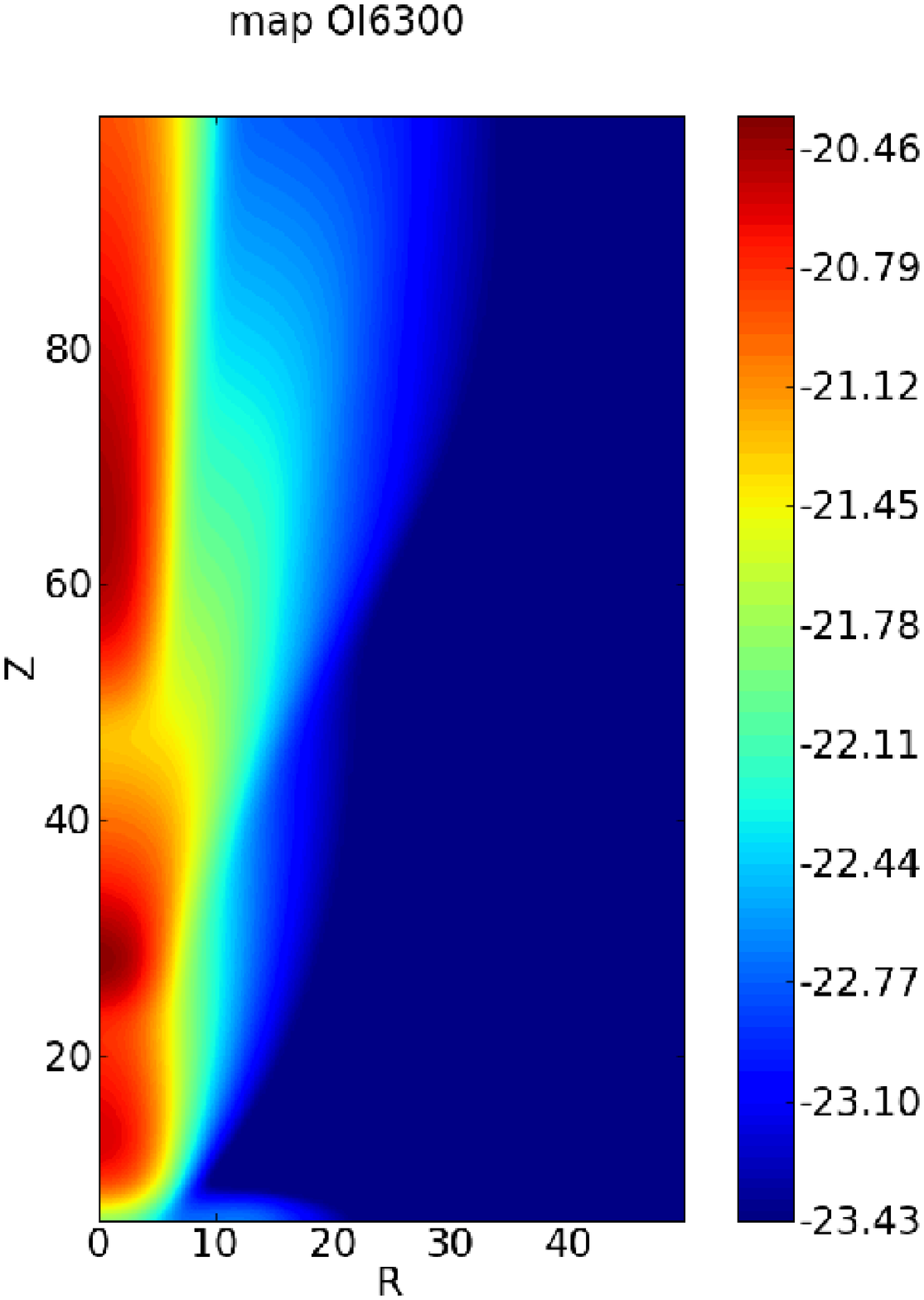}
  \includegraphics[width=0.19\textwidth]{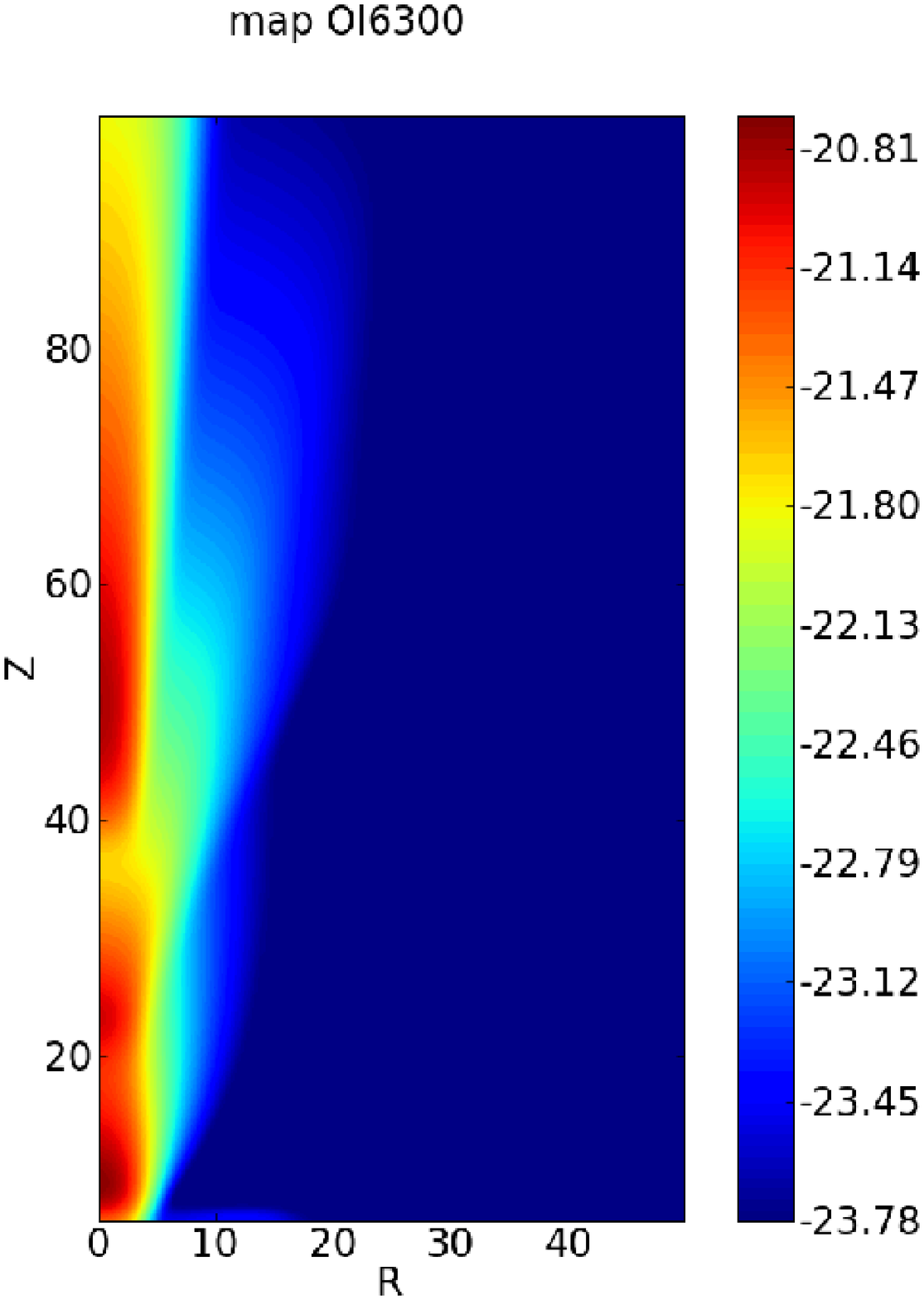}
  \caption{Synthetic emission maps of the [OI] $\lambda$6300 line, convolved 
    with a Gaussian PSF with a FWHM of 15 AU, for model SC1f and runs 
    (500, 600, 0.2), (500, 600, 0.5), (500, 1000, 0.2), (500, 1000, 0.5), 
    (500, 1000, 0.8).}
  \label{Fig_emissmaps_all7}
\end{figure*}
\begin{figure*}[!bt]
  \centering
  \includegraphics[width=0.24\textwidth]{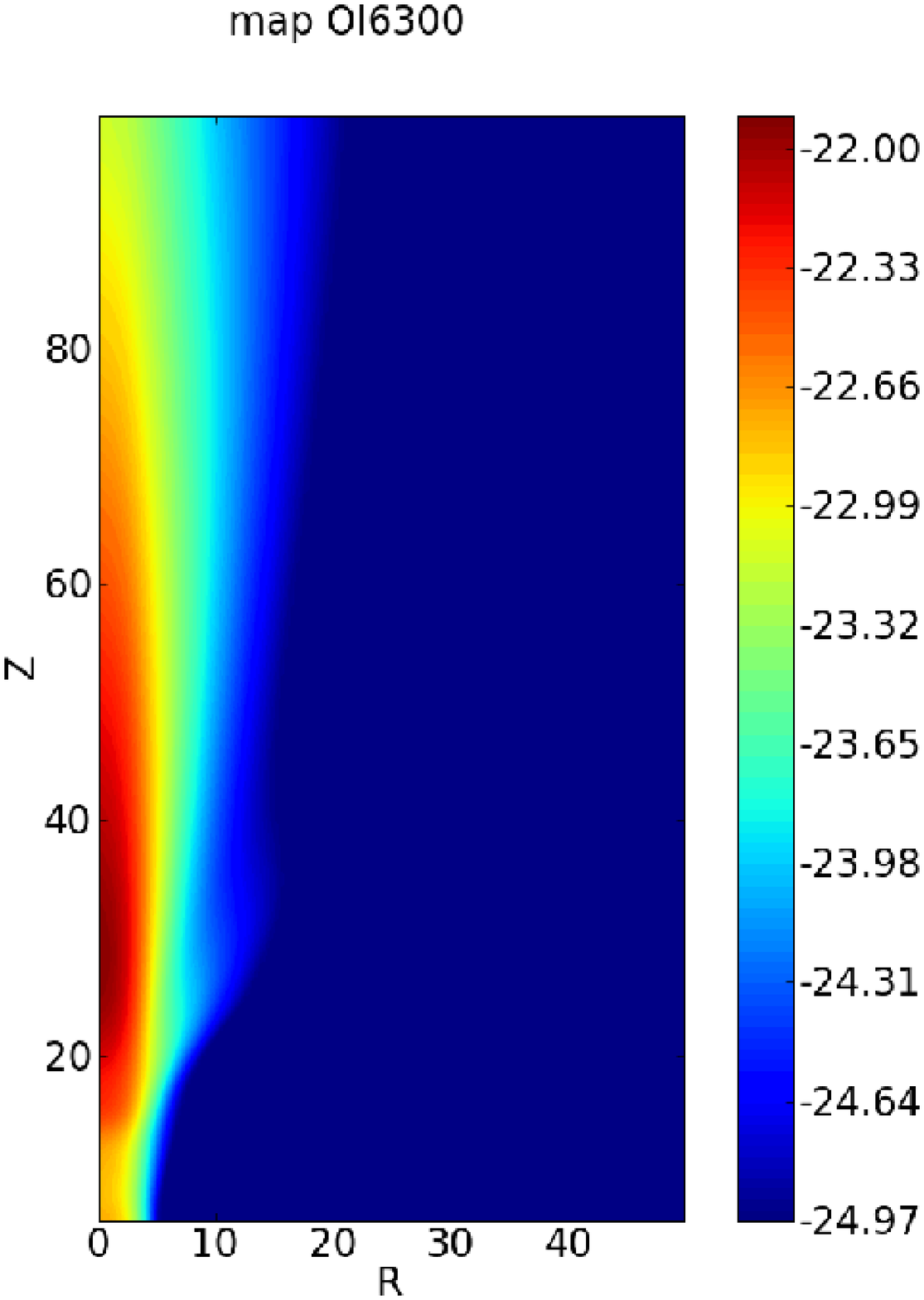}
  \includegraphics[width=0.24\textwidth]{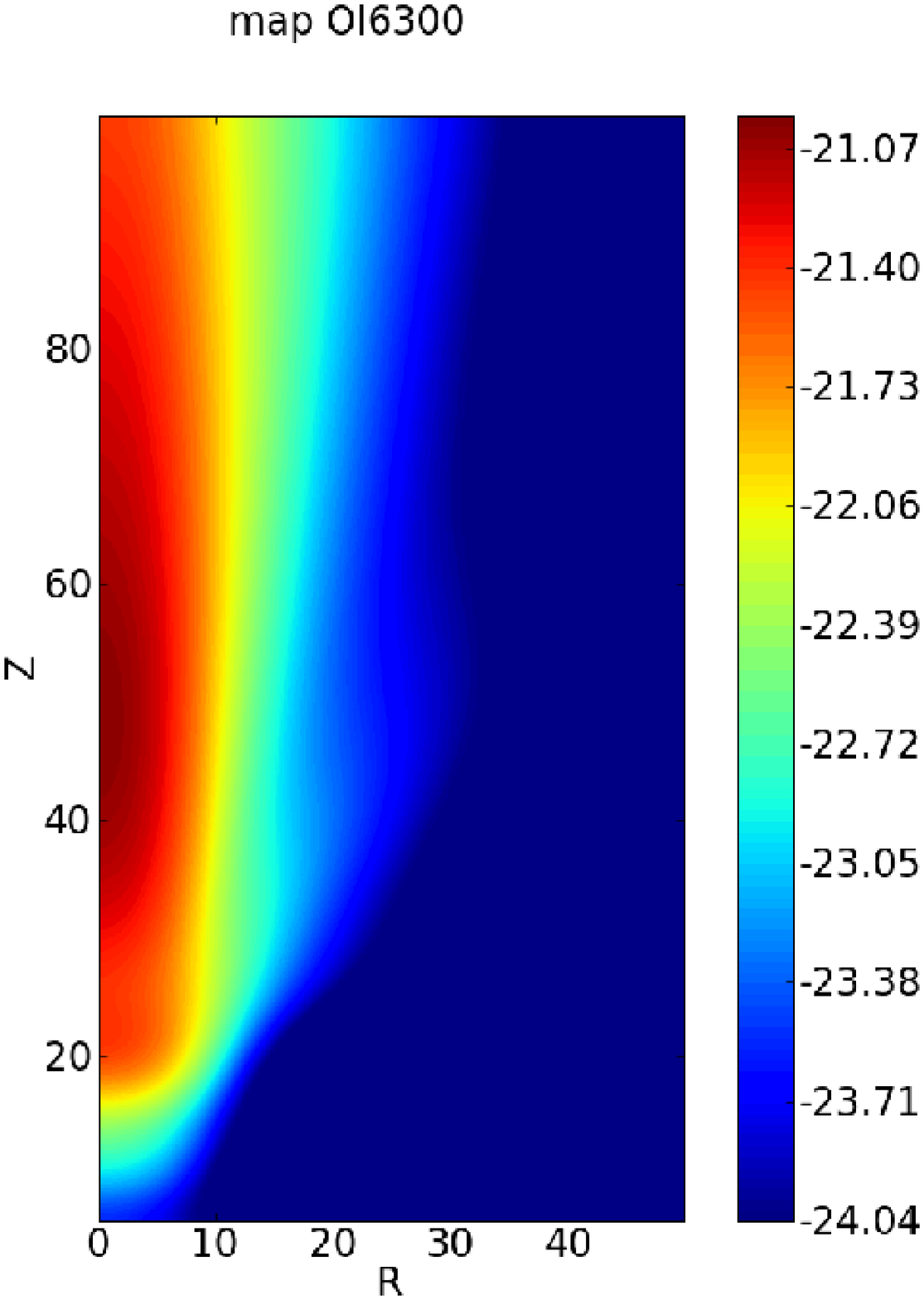}
  \includegraphics[width=0.24\textwidth]{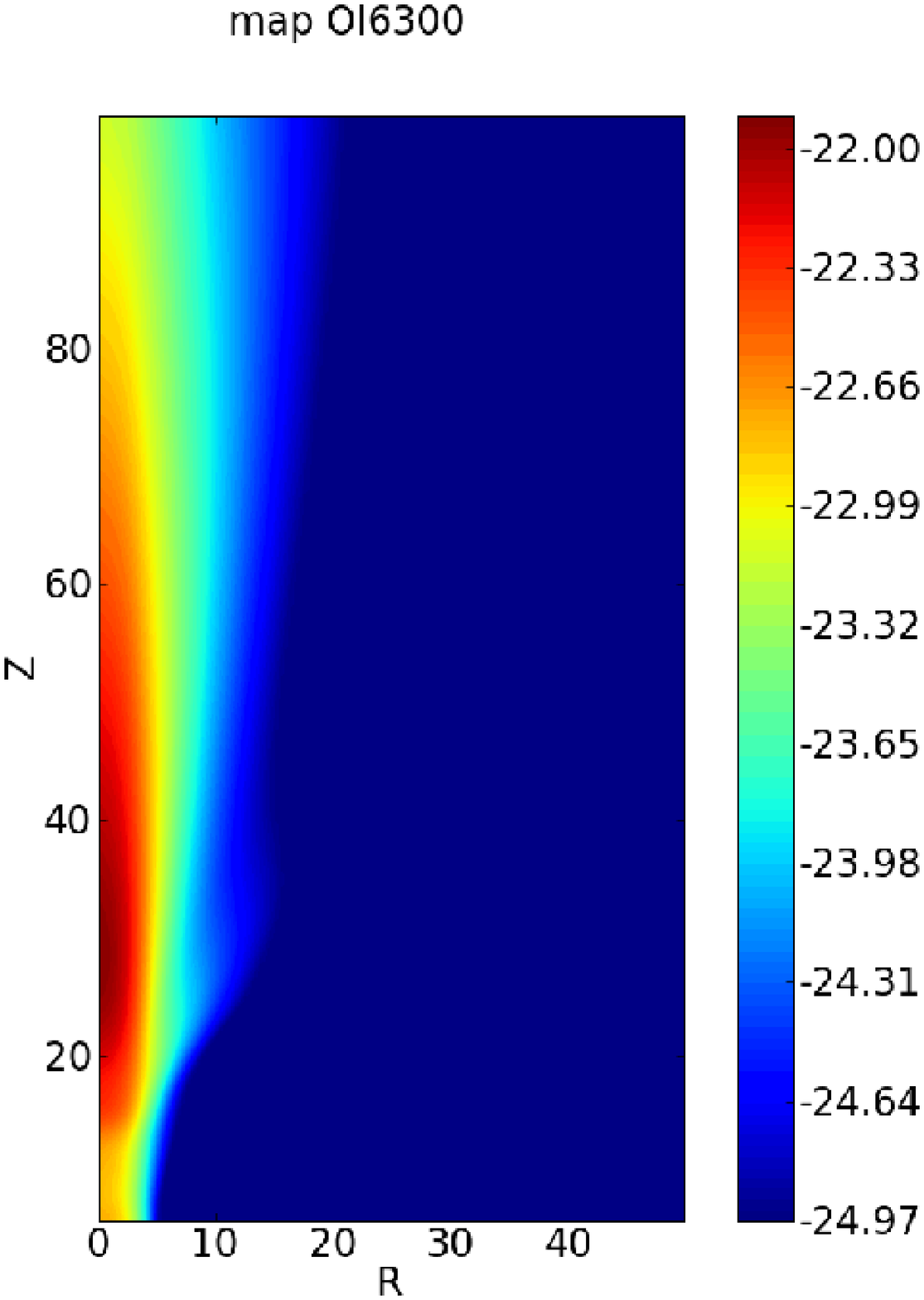}
  \includegraphics[width=0.24\textwidth]{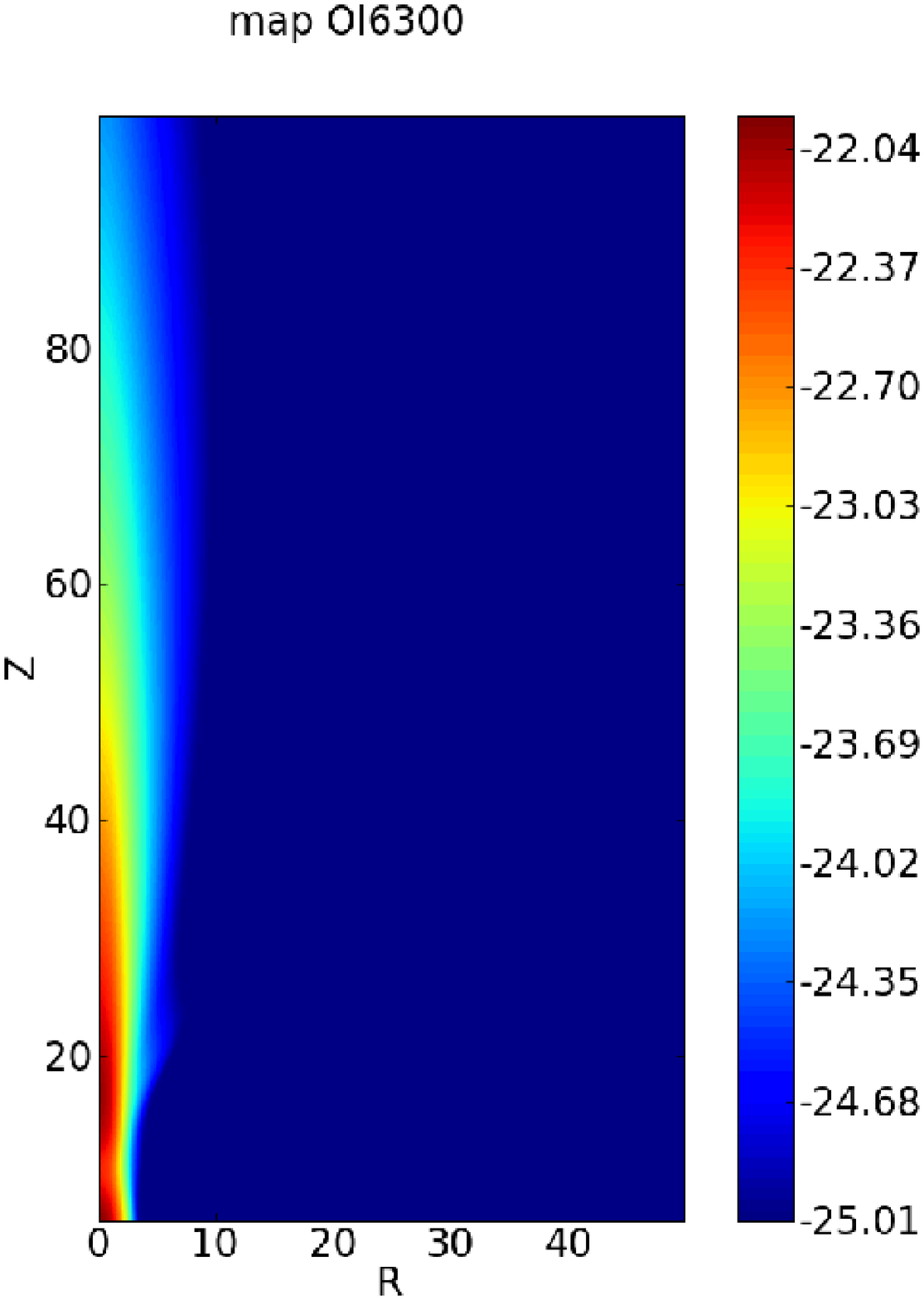}
  \caption{Synthetic emission maps of the [OI] $\lambda$6300 line, convolved 
    with a Gaussian PSF with a FWHM of 15 AU, for model SC1g and runs 
    (500, 600, 0.2), (500, 1000, 0.2), (500, 1000, 0.5), (500, 1000, 0.8).}
  \label{Fig_emissmaps_all8}
\end{figure*}
\begin{figure*}[!bt]
  \centering
  \includegraphics[width=0.24\textwidth]{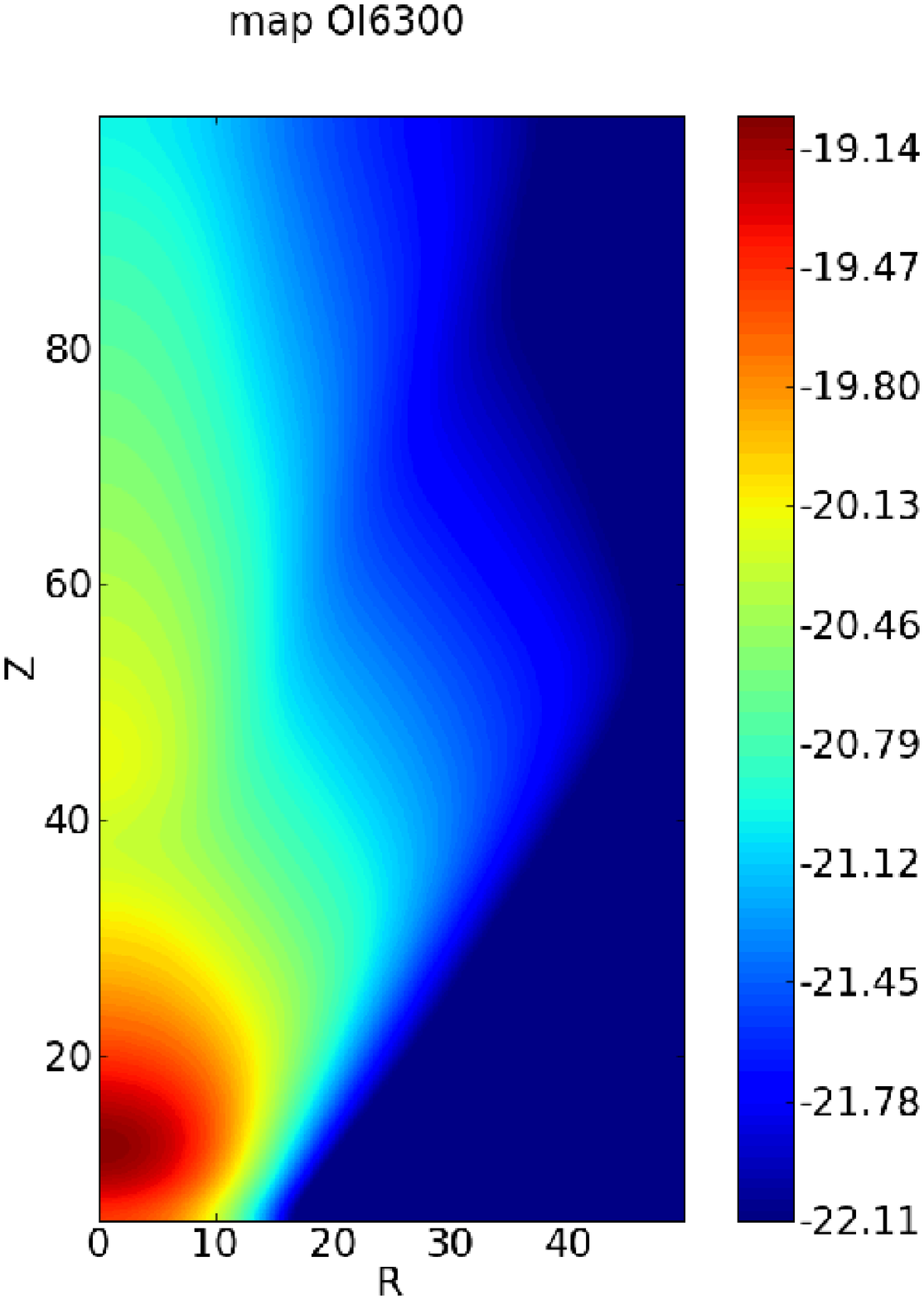}
  \includegraphics[width=0.24\textwidth]{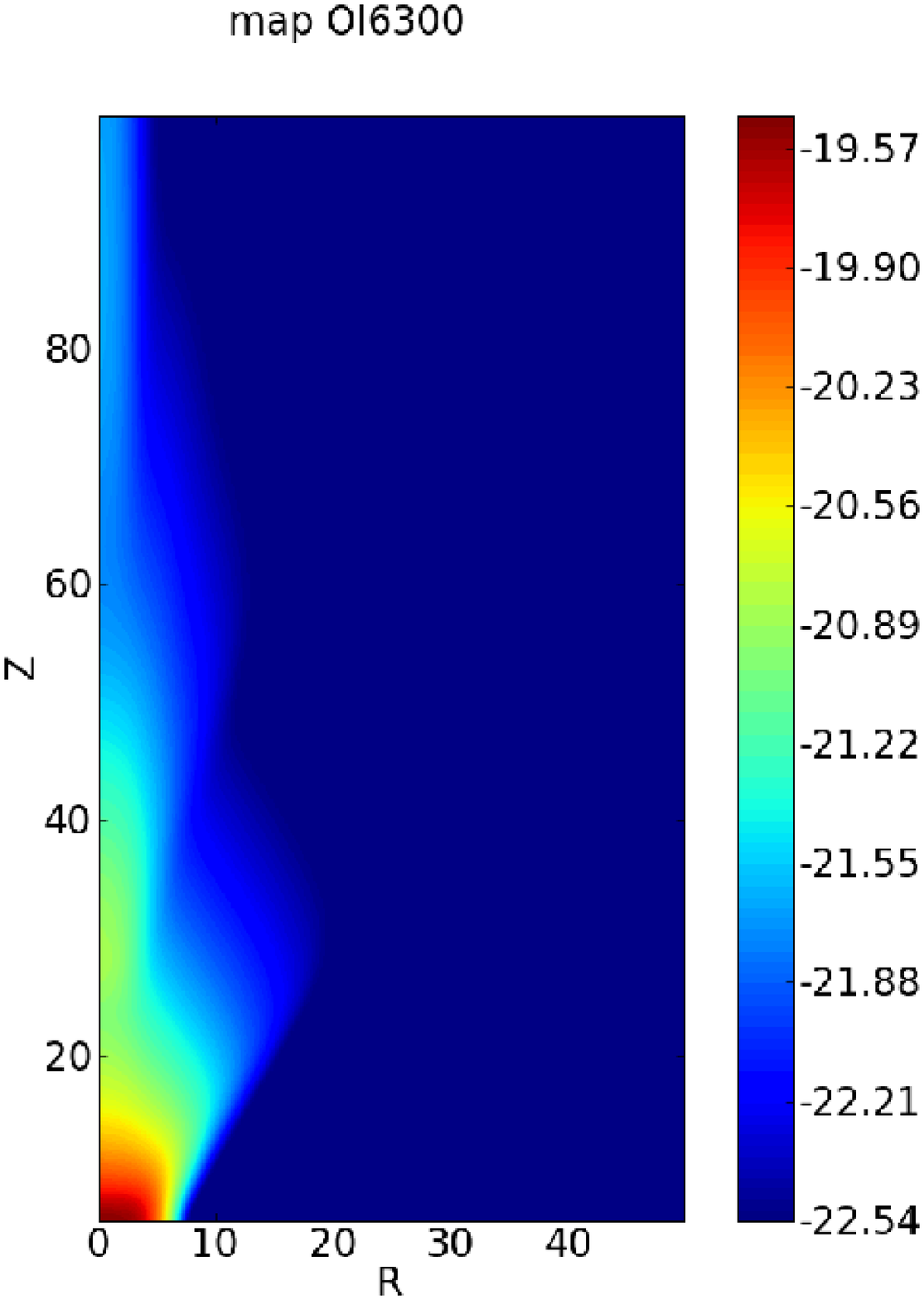}
  \includegraphics[width=0.24\textwidth]{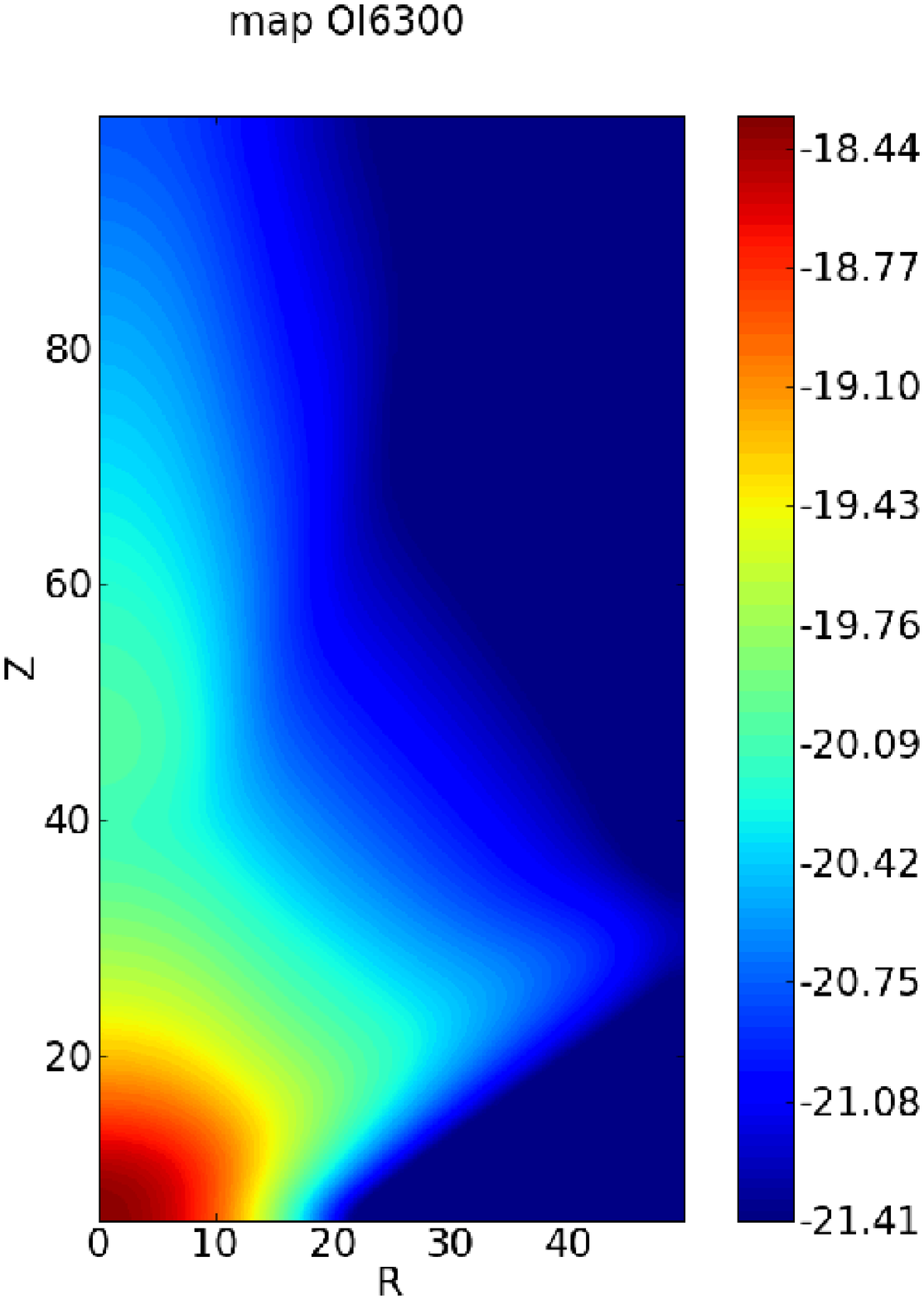}
  \includegraphics[width=0.24\textwidth]{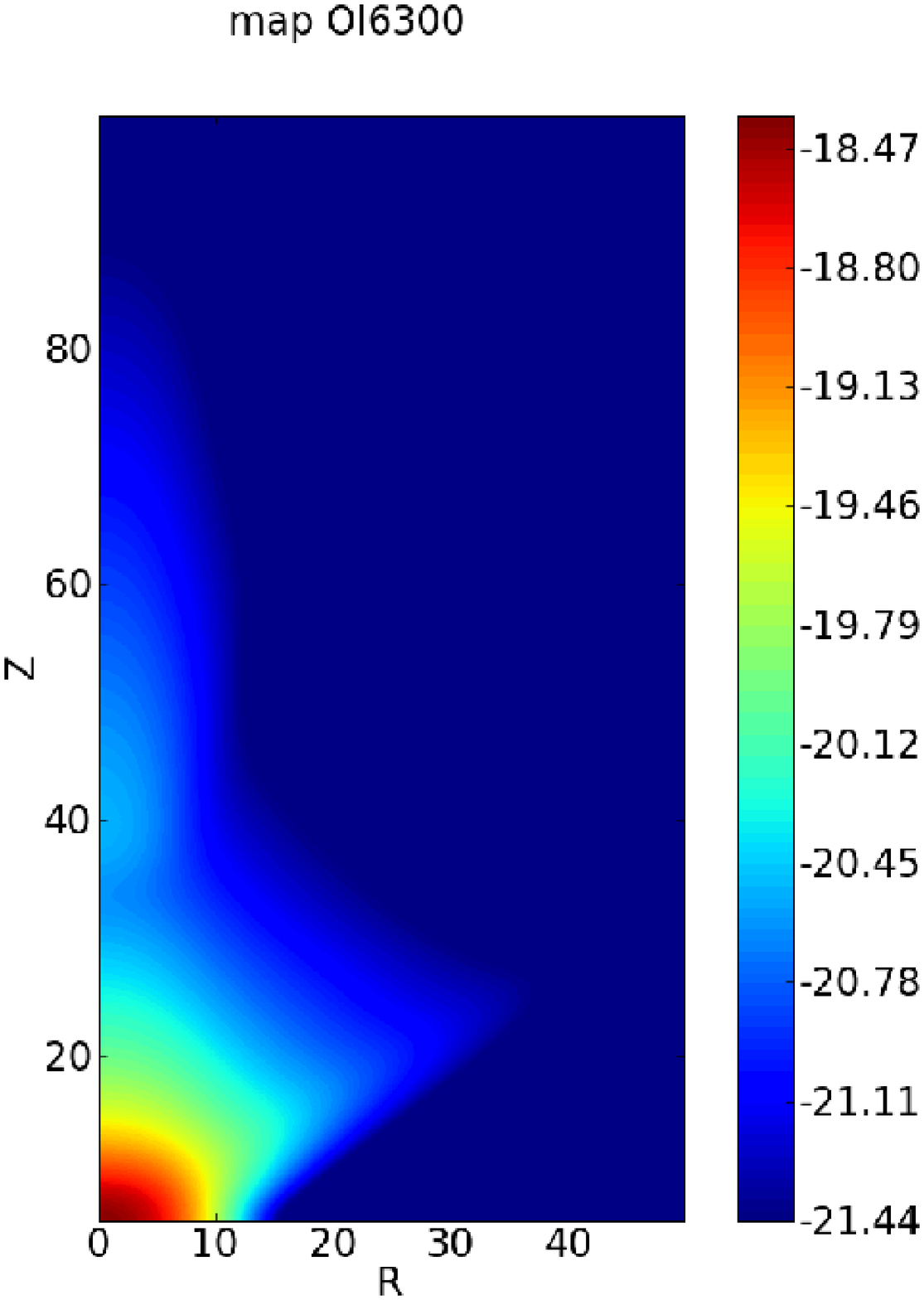}
  \caption{Synthetic emission maps of the [OI] $\lambda$6300 line, convolved 
    with a Gaussian PSF with a FWHM of 15 AU, for model SC2 and runs 
    (500, 600, 0.2), (500, 600, 0.5), (500, 1000, 0.5), (500, 1000, 0.8).}
  \label{Fig_emissmaps_all9}
\end{figure*}
\begin{figure*}[!bt]
  \centering
  \includegraphics[width=0.24\textwidth]{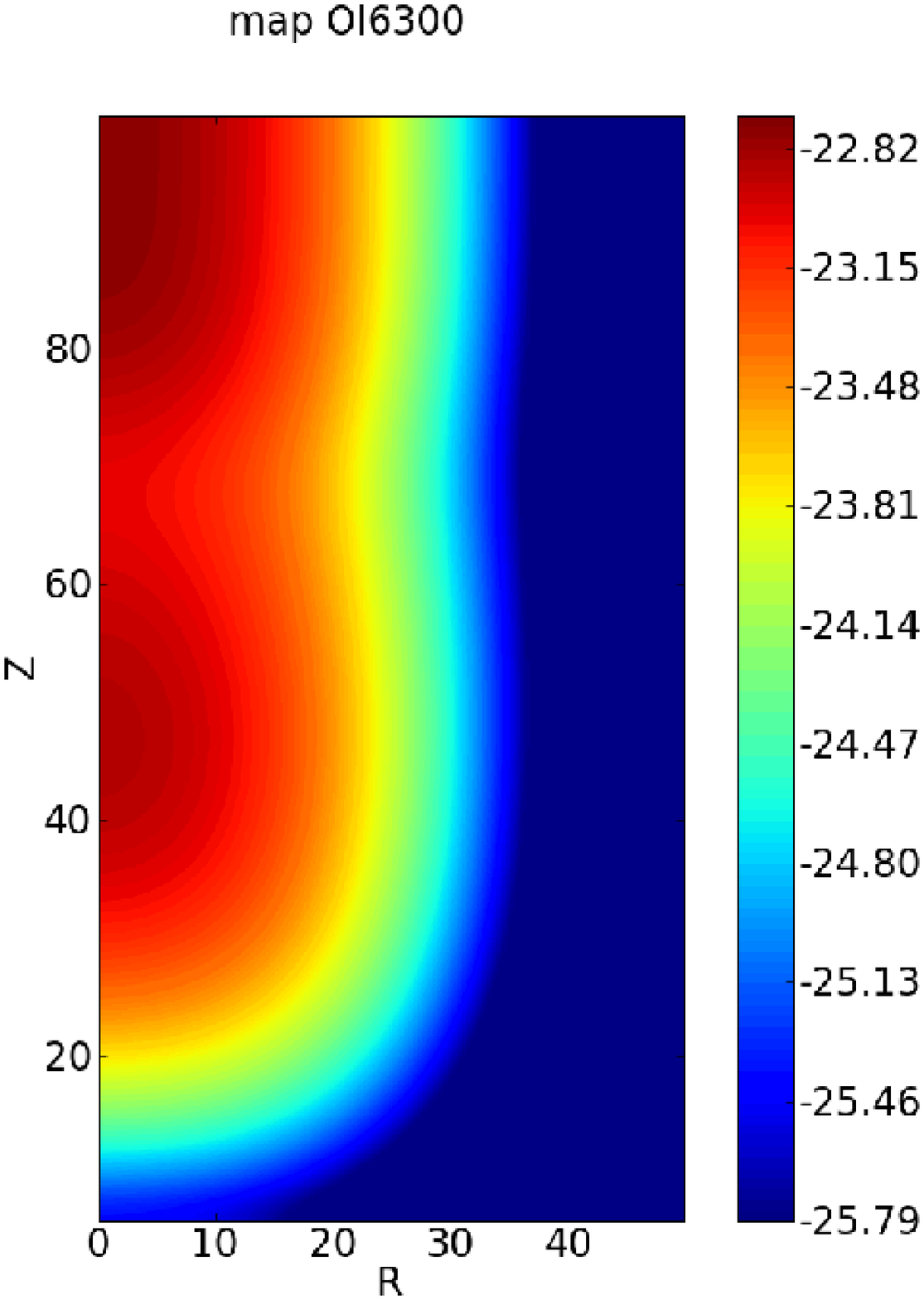}
  \caption{Synthetic emission maps of the [OI] $\lambda$6300 line, convolved 
    with a Gaussian PSF with a FWHM of 15 AU, for model SC3 and run 
    (500, 100, 0.2).}
  \label{Fig_emissmaps_all10}
\end{figure*}
\begin{figure*}[!bt]
  \centering
  \includegraphics[width=0.24\textwidth]{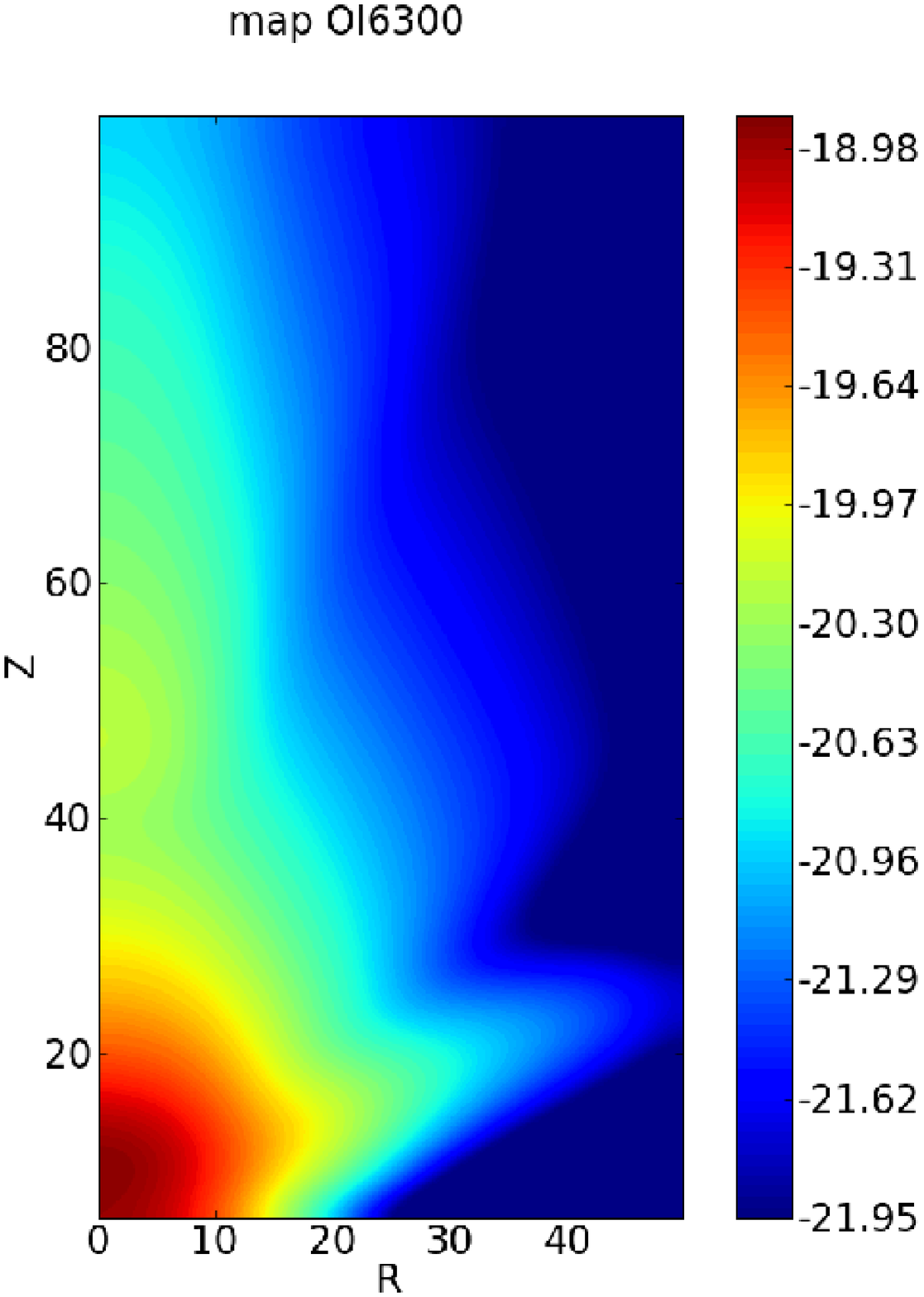}
  \includegraphics[width=0.24\textwidth]{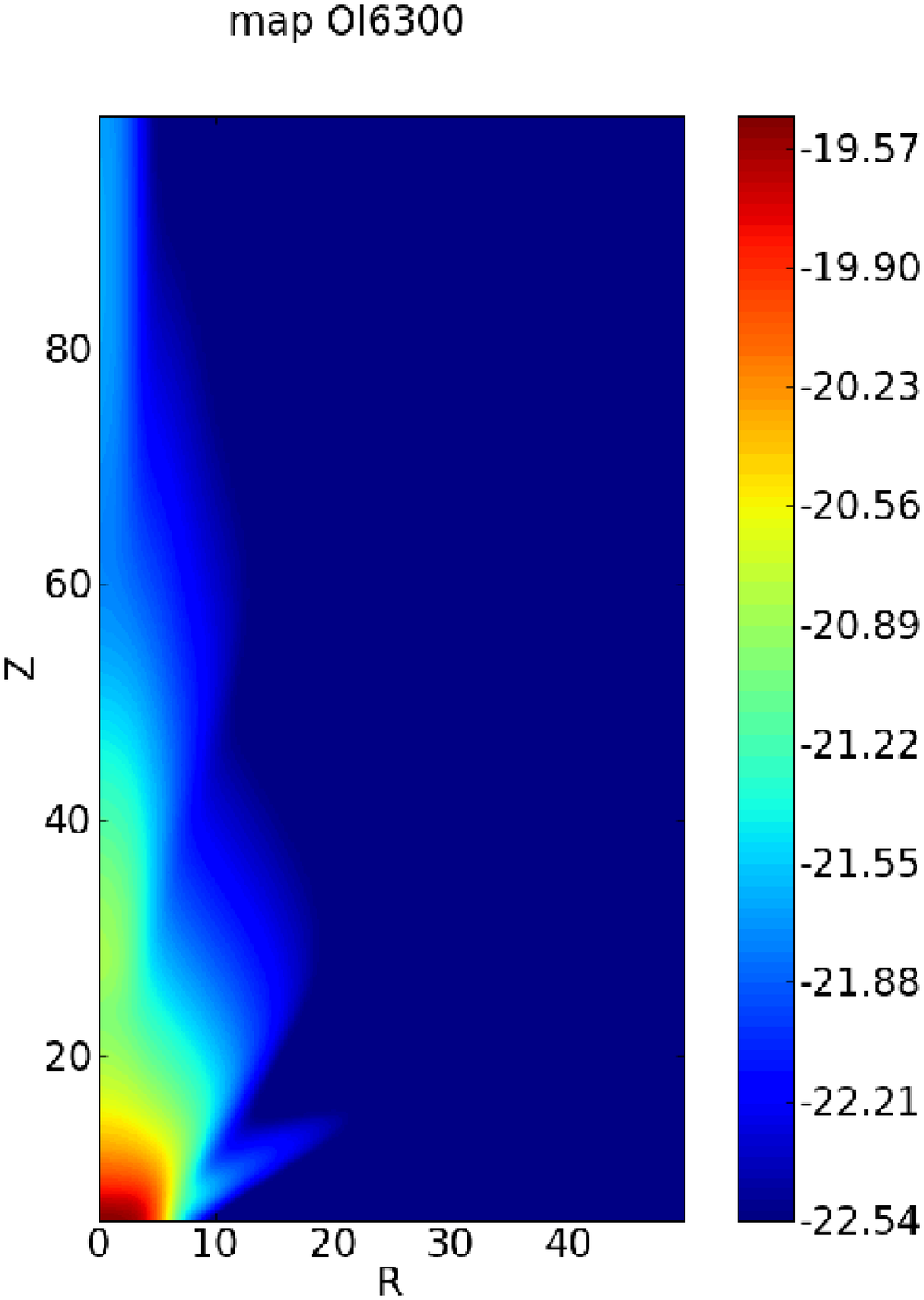}
  \includegraphics[width=0.24\textwidth]{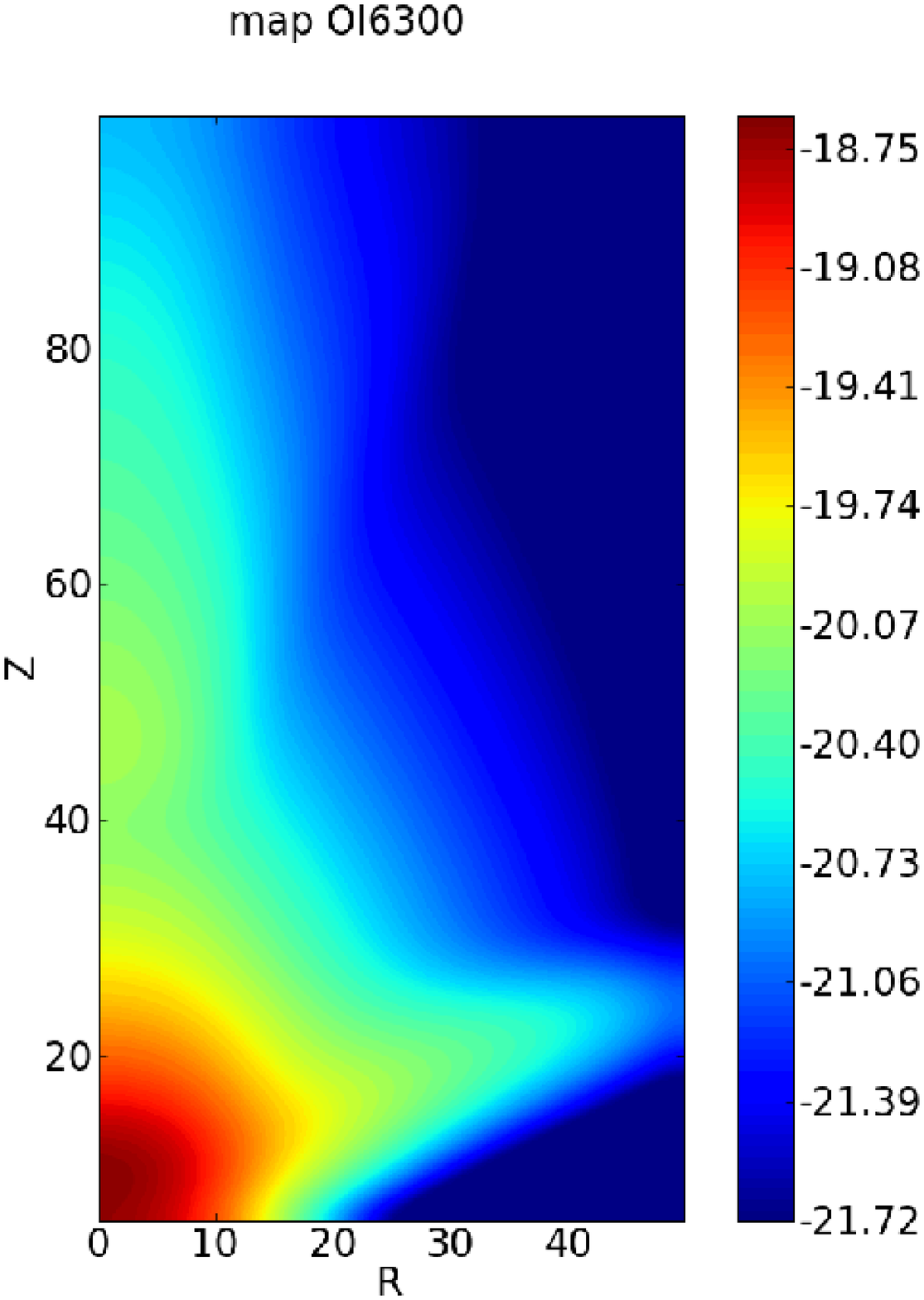}
  \includegraphics[width=0.24\textwidth]{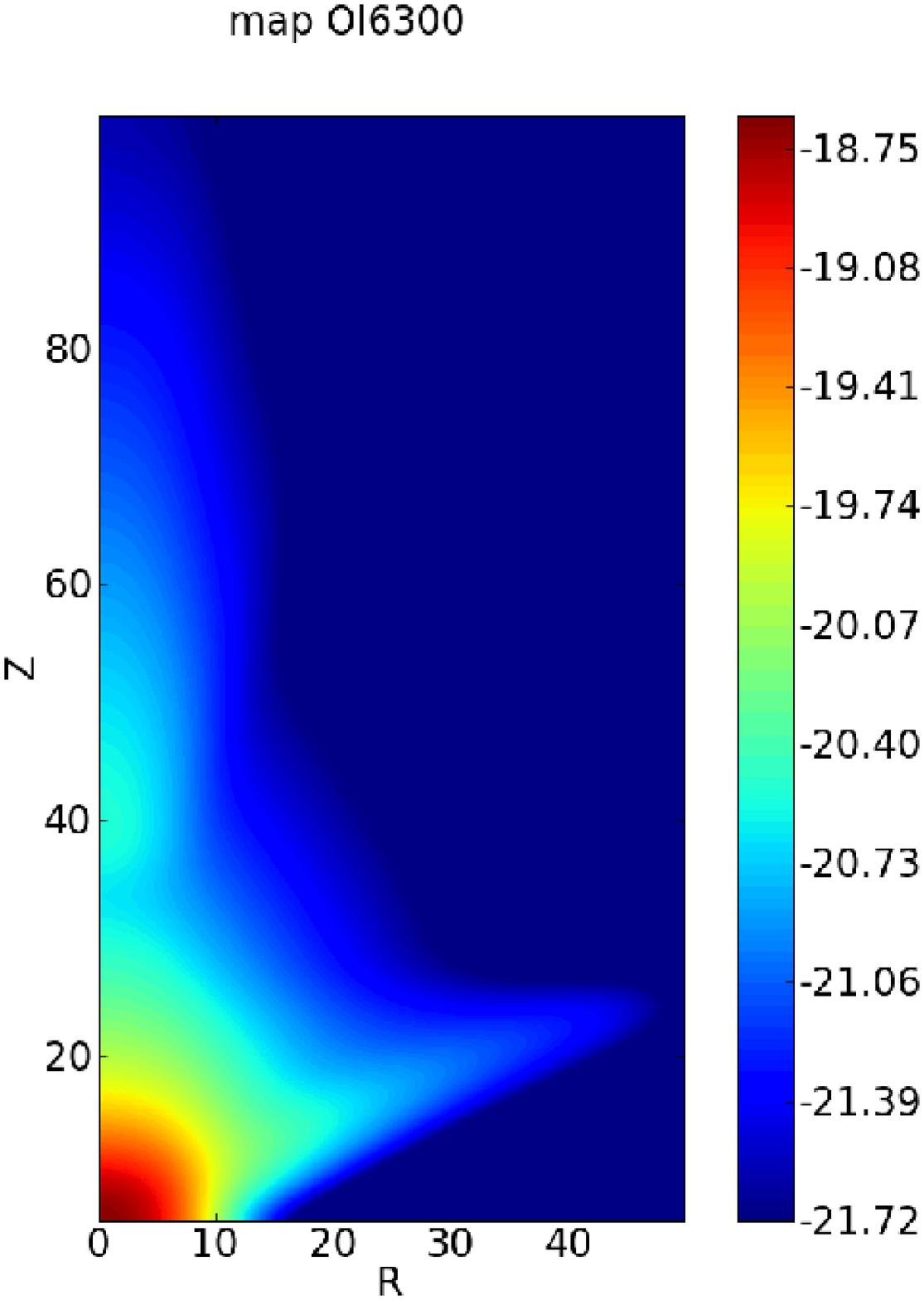}
  \caption{Synthetic emission maps of the [OI] $\lambda$6300 line, convolved 
    with a Gaussian PSF with a FWHM of 15 AU, for model SC4 and runs 
    (500, 600, 0.2), (500, 600, 0.5), (500, 1000, 0.5), (500, 1000, 0.8).}
  \label{Fig_emissmaps_all11}
\end{figure*}

\begin{acknowledgements}
The authors would thank the referee, Sylvie Cabrit, for fruitful 
discussions, suggestions and comments improving this paper. The present work 
was supported in part by the European Community's Marie Curie Actions - Human 
Resource and Mobility within the JETSET (Jet Simulations, Experiments and 
Theory) network under contract MRTN-CT-2004 005592.
\end{acknowledgements}

\end{document}